\documentclass[a4paper,11pt]{article}
\pdfoutput=1 

\usepackage{jheparxiv}

\usepackage[T1]{fontenc} 

\usepackage{tensor}
\usepackage{amsmath}
\usepackage{amssymb}
\usepackage{mathrsfs}
\usepackage{mathtools}
\usepackage{bbm}
\usepackage{graphicx}
\usepackage{stmaryrd}
\usepackage{twistor}
\usepackage{subcaption}
\usepackage{xcolor}
\usepackage{empheq}

\newcommand{\veps}{\varepsilon}
\newcommand{\bo}{\mathbf{o}}
\newcommand{\bi}{\boldsymbol{\iota}}

\renewcommand{\sc}{\mathsf{c}}

\newcommand{\MT}{\mathbb{MT}}
\newcommand{\sa}{\mathsf{a}}
\newcommand{\sH}{\mathsf{H}}
\newcommand{\sG}{\mathsf{G}}

\newcommand{\sh}{\mathsf{h}}
\newcommand{\sF}{\mathsf{F}}
\newcommand{\sg}{\mathsf{g}}

\newcommand{\m}{\mathrm{m}}
\newcommand{\nbar}{\bar{\nabla}}

\title{Graviton scattering on self-dual black holes}

\author[a]{Tim Adamo,}
\author[b]{Giuseppe Bogna,}
\author[b]{Lionel Mason,}
\author[c]{Atul Sharma}

\affiliation[a]{School of Mathematics and Maxwell Institute for Mathematical Sciences,\\
University of Edinburgh, EH9 3FD, UK\vspace{0.1cm}}
\emailAdd{t.adamo@ed.ac.uk}

\affiliation[b]{The Mathematical Institute, University of Oxford,\\ Woodstock Road, Oxford OX2 6GG, United
Kingdom\vspace{0.1cm}}
\emailAdd{giuseppe.bogna@maths.ox.ac.uk}
\emailAdd{lmason@maths.ox.ac.uk}

\affiliation[c]{Center for the Fundamental Laws of Nature \& Black Hole Initiative,\\Harvard University, Cambridge, MA, 02138, USA \vspace{0.1cm}}
\emailAdd{atulsharma@fas.harvard.edu}

\abstract{The computation of gravitational wave scattering on black hole spacetimes is an extremely hard problem, typically requiring approximation schemes that either treat the black hole perturbatively or are only amenable to numerical techniques. In this paper, we consider linearised gravitational waves (or gravitons) scattering on the self-dual analogue of a black hole: namely, the self-dual Taub-NUT metric. Using the hidden integrability of the self-dual sector,  we solve the linearised Einstein equations on these self-dual black hole backgrounds exactly in terms of simple, explicit quasi-momentum eigenstates.  Using a description of the self-dual Taub-NUT metric and its gravitons in terms of twistor theory, we obtain an explicit formula, exact in the background, for the tree-level maximal helicity violating (MHV) graviton scattering amplitude at arbitrary multiplicity, with and without spin. This is obtained from the description of the MHV amplitudes in terms of the perturbation theory of a chiral sigma model whose target is the twistor space of the background. The incorporation of spin effects on these backgrounds is a straightforward application of the Newman-Janis shift. We also demonstrate that the holomorphic collinear splitting functions in the self-dual background are equal to those in flat space so that the celestial symmetry algebra is undeformed.}  

\begin{document}
\maketitle
\flushbottom

\section{Introduction}\label{sec:intro}
The study of gravitational wave scattering in a black hole spacetime is an important but notoriously difficult endeavour. These difficulties manifest themselves in both conceptual and technical ways: the presence of an event horizon means that the usual S-matrix of quantum field theory (QFT) does not exist as a unitary operator evolving incoming states from past infinity to outgoing states at future infinity~\cite{Hawking:1975vcx,Hawking:1976ra}, and external graviton wavefunctions in black hole spacetimes are typically determined by equations which do not admit analytic solutions~\cite{Regge:1957td,Zerilli:1970se,Teukolsky:1972my,Teukolsky:1973ha}. Furthermore, the Feynman rules of perturbative gravity in a black hole background -- already non-polynomial in Minkowski space -- are woefully complicated. While some of the conceptual obstacles in this problem can be overcome or forestalled -- for instance, by defining tree-level graviton scattering amplitudes on the black hole background variationally through the perturbiner formalism~\cite{Arefeva:1974jv,Abbott:1983zw,Jevicki:1987ax,Rosly:1997ap,Selivanov:1999as,Ilderton:2023ifn,Adamo:2023cfp,Kim:2023qbl} -- this does not ameliorate the underlying technical difficulties in the associated calculations.

Yet for all these imposing obstacles, there are equally many reasons to be interested in multi-graviton scattering amplitudes on black hole backgrounds. These amplitudes encode information relevant to various approximation regimes of interest for the gravitational two-body problem (e.g., the self-force expansion)~\cite{Cheung:2023lnj,Kosmopoulos:2023bwc,Cheung:2024byb}, give probes of quantum gravity~\cite{tHooft:1996rdg,Akhoury:2013bia,Betzios:2016yaq,Gaddam:2021zka,He:2023qha,Aoude:2024sve}, and also provide much needed raw data for asymptotically flat holography~\cite{Pasterski:2020pdk,Costello:2022jpg,Costello:2023hmi,Fernandes:2023ibv,Adamo:2024mqn}, where the mass of the black hole provides a scale akin to the cosmological constant in asymptotically anti-de Sitter spaces. Most approaches in these scenarios treat the black hole perturbatively in some way to facilitate concrete calculations~\cite{Luna:2017dtq,Vines:2017hyw,Guevara:2018wpp,Arkani-Hamed:2019ymq,Bautista:2021wfy,Saketh:2022wap,Bautista:2022wjf,Gonzo:2022tjm,Cangemi:2023bpe,Bohnenblust:2024hkw}.

Of course, an analytic approach which treats the black hole background \emph{exactly} (i.e., non-perturbatively) is highly desirable, though currently out of reach for astrophysical black holes such as the Schwarzschild and Kerr metrics. Consequently, it is reasonable to look for simpler `toy' scenarios which treat the background exactly while still capturing the full non-linearity of perturbative general relativity through scattering amplitudes. 

\medskip

In this paper, we consider the \emph{self-dual} version of a black hole metric as such a toy model. This enables explicit calculation of some graviton scattering amplitudes to high multiplicity while treating the background metric exactly, and has other desirable properties such as straightforward incorporation of spin effects.

But what, precisely, is a self-dual black hole? A 4-dimensional complex, holomorphic metric is said to be (vacuum) self-dual if it solves the vacuum Einstein equations and has a purely self-dual Weyl curvature tensor -- this is equivalent to the metric being hyperk\"ahler. As the self-dual and anti-self-dual parts of the Weyl curvature are related by complex conjugation in Lorentzian signature, non-trivial self-dual metrics do not admit Lorentzian-real slices. Thus, astrophysical black holes of the Kerr-Newman family are definitely \emph{not} self-dual.

However, one can consider the one-parameter generalization of the Schwarzschild metric due to Taub, Newman, Unti and Tamburino, known as the Taub-NUT metric~\cite{Taub:1950ez,Newman:1963yy}. This Lorentzian-real metric depends on two parameters, the familiar ADM mass of Schwarzschild ($M$) and a new quantity called the NUT charge ($N$). The metric is only locally asymptotically flat due to wire singularities on the sphere at infinity~\cite{Misner:1963fr}; when $N=0$ it reduces to the Schwarzschild metric. 

Upon complexifying the Taub-NUT metric, it is possible to set $N=-\im\,M$, at which point the metric becomes self-dual with only a single parameter, $M$. This self-dual metric then admits a Euclidean-real slice which is a complete, Ricci-flat  metric on $\R^4$: the famous \emph{self-dual Taub-NUT} (SDTN) gravitational instanton~\cite{Hawking:1976jb,Gibbons:1978tef,Gibbons:1979xm}. Although defined on $\R^4$, Wick-rotated time now takes values in a topologically non-trivial circle bundle. It is this metric which we will study as a toy setting for gravitational scattering on a black hole.

\medskip

It has long been known that the self-dual sector of general relativity is classically integrable~\cite{Penrose:1976js}, and its self-duality is the key property of the SDTN metric which makes it a tractable toy model for black holes (cf., \cite{Crawley:2023brz,Guevara:2023wlr,Araneda:2024cqu,Guevara:2024edh} for other recent studies which have taken this perspective). For instance, we were recently able  to solve the linearised Einstein equations exactly on the SDTN background without the need to separate variables or use partial wave expansions~\cite{Adamo:2023fbj}: the resulting \emph{quasi-momentum eigenstates} generalise the usual plane wave momentum eigenstates of Minkowski space to the curved SDTN metric. Using these wavefunctions, it was possible to compute the two point function of negative helicity gravitons on the SDTN metric, in closed form and exactly in the parameter $M$.

In this paper, we use the twistor description of SDTN, which manifests its hidden integrability, to compute the classical (tree-level) maximal-helicity-violating (MHV) graviton scattering amplitude on SDTN with an arbitrary number of external gravitons. This MHV configuration involves two external gravitons of negative helicity (i.e., gravitons whose anti-self-dual linearised Weyl curvature satisfies the zero-rest-mass equation) and an arbitrary number of positive helicity (i.e., gravitons whose linearised Weyl curvature is purely self-dual), and corresponds to the first non-trivial perturbations away from the integrable self-dual sector of perturbative gravity around the SDTN background~\cite{Mason:2008jy,Adamo:2021bej,Miller:2024oza,Guevara:2025tsm}.

Each external graviton is labelled by a frequency $\omega$ and spatial three momentum $\vec{k}$ which are on-shell ($\omega^2+\vec{k}^2=0$) and form a (necessarily complex, in Euclidean signature) massless 4-momentum $k^{\alpha\dot\alpha}=\kappa^{\alpha}\tilde{\kappa}^{\dot\alpha}$ parametrized by $\omega$ and a point $(z,\bar{z})$ on the complexified sphere. In addition, the topologically non-trivial nature of the SDTN time coordinate means that gravitons can also carry a topological charge, but configurations of arbitrary topological charge can be obtained from `minimal' quasi-momentum eigenstates where the charge is fixed by the frequency through simple differential operations~\cite{Adamo:2023fbj}. Schematically, the MHV graviton amplitude in this case is given by (neglecting overall numerical factors and powers of the gravitational coupling constant):
\begin{multline}\label{schematicMHV}
\cM_n=\delta\!\left(\sum_{i=1}^{n}\omega_i\right)\frac{\la1\,2\ra^6}{\la1\,3\ra^2\,\la2\,3\ra^2}\sum_{t\geq0} \int_{\R^3}\d^{3}\vec{x}\,\left(1+\frac{2M}{r}\right)\int_{(S^2)^t}\prod_{\m=1}^{t}\varpi_\m\,\mathcal{D}_{\m}\,\left|\cH[t]\right|\Big|_{\veps=0} \\
\times\e^{\im\vec{k}\cdot\vec{x}}\,\prod_{i\in\mathfrak{n}_+}\left(\frac{r}{1+|\zeta|^2}\right)^{2M\omega_i}\left(\bar{\zeta}\,z_i+1\right)^{4M\omega_i}\,\prod_{j\in\mathfrak{n}_-}\left(\frac{r}{1+|\zeta|^2}\right)^{-2M\omega_i}\left(\zeta-z_i\right)^{-4M\omega_j}\,,
\end{multline}
where gravitons 1 and 2 have negative helicity and $\mathfrak{n}_{\pm}$ are the sets of positive/negative frequency gravitons. In this expression, $r$ is a radial coordinate on $\R^3$, $\zeta\in \C$ a stereographic coordinate
on the corresponding spheres, 
\be\label{totspmom}
\vec{k}:=\sum_{i=1}^{n}\vec{k}_i\,,
\ee
is the total spatial momentum of the external gravitons, $\la i\,j\ra:=\kappa_i^{\alpha}\,\kappa_{j\,\alpha}$ and the sum over $t\geq0$ introduces additional integrals over $S^2$ with
\be\label{spheremeasure}
\varpi_\m:=\frac{\d\zeta_\m\wedge\d\bar{\zeta}_\m}{(1+|\zeta_\m|^2)^2}\,,
\ee
denoting the K\"ahler measure on each of these copies of the 2-sphere with stereographic coordinates $(\zeta_\m,\bar{\zeta}_\m)$.

The matrix $\cH[t]$ is a $(n+t-3)\times(n+t-3)$ matrix whose entries are constructed from the kinematic data of the external gravitons, the SDTN background geometry and a set of $t$ formal parameters $\{\veps_1,\ldots,\veps_t\}$ such that the determinant $|\cH[t]|$ is a polynomial in each $\veps_\m$ for $\m=1,\ldots,t$. This polynomial is then acted upon by certain differential operators $\{\mathcal{D}_\m\}$ with respect to these formal parameters, and then evaluated at $\veps_1=\veps_2=\cdots=\veps_t=0$, extracting only the coefficients in this polynomial of a particular order. The sum over $t\geq0$ corresponds to the presence of explicit tail contributions to the amplitude, generated by the external gravitons scattering off the SDTN background itself. 

\medskip

Obtaining such a formula, which is \emph{exact} in the SDTN background, would be impossible with traditional background field theory methods. The key is a two-dimensional, chiral conformal field theory (CFT), known as the twistor sigma model~\cite{Adamo:2021bej}, which describes the hyperk\"ahler structure of the SDTN metric as well as its perturbations. Using the twistor sigma model, it is possible to recast the computation of the MHV graviton amplitude in terms of a tree-level, connected correlation function in a 2d CFT -- a quantity which \emph{can} be calculated straightforwardly.

The formula \eqref{schematicMHV} can be shown to reduce to Hodges' well-known expression for the MHV graviton scattering amplitude on Minkowski space~\cite{Hodges:2012ym} upon taking the flat limit of the SDTN background. Furthermore, the result trivially extends to a spinning SDTN background by means of the Newman-Janis trick~\cite{Newman:1965tw}. The integrals appearing in the formula can be performed explicitly for low numbers of external points, to give completely closed-form expressions. For general numbers of external gravitons, it is also possible to examine the leading holomorphic collinear limits between positive helicity external gravitons; remarkably, the collinear splitting functions are un-deformed from flat space.

\medskip

\paragraph{Structure of the paper:} In Section \ref{sec:class_mech}, we review the self-dual Taub-NUT space and its geometry, as well as the incorporation of spin. Section \ref{sec:twistor_space} develops the twistor description of the SDTN metric via Penrose's non-linear graviton construction. We then introduce the quasi-momentum eigenstate solutions to the linearised Einstein equations on SDTN and their description in twistor space. In Section~\ref{sec:mhv_amplitudes}, we derive the MHV graviton amplitudes on SDTN using the twistor sigma model description of its hyperk\"ahler geometry. Some basic properties of the formula, such as its explicit evaluation for low numbers of external gravitons and the flat space limit, are explored. Section~\ref{sec:celestial_ope} computes the holomorphic collinear limits of this MHV amplitude formula, confirming that the resulting splitting functions are equivalent to those in flat space. Finally, Section~\ref{sec:concl} concludes with a brief discussion of interesting future directions.

The appendices include additional material for the interested reader. Appendix~\ref{subsec:other_twistor_spaces} discusses other presentations of the SDTN twistor space which have appeared in the literature over the years. Appendix~\ref{app:GHmets} demonstrates how a generalisation of the presentation of the SDTN metric used throughout the paper can be used to describe \emph{all} Gibbons-Hawking gravitational instantons in Euclidean signature. Appendix~\ref{app:2-point} provides details regarding how the twistor-based formalism of this paper reproduces the graviton 2-point function computed in our earlier work from a purely spacetime perspective.

    
\section{Self-dual Taub-NUT and its hyperk\"ahler structure}\label{sec:class_mech}

The Taub-NUT metric (viewed as a complexified Lorentzian metric solving the vacuum Einstein equations) depends on two parameters: the ADM mass $M$ and NUT parameter $N$~\cite{Taub:1950ez,Newman:1963yy}. When $N\to0$, the metric reduces to the Schwarzschild black hole, but for $N\neq0$ the metric generically has `wire singularities' on the celestial sphere at infinity. However, in the configuration where the mass and NUT parameter are related by $N=-\im\,M$, the metric becomes a complete, Ricci-flat and self-dual -- or \emph{hyperk\"ahler} gravitational instanton -- metric on the Euclidean real slice $\R^4$~\cite{Hawking:1976jb,Gibbons:1978tef,Gibbons:1979xm}. 

In this section, we review the geometry of the self-dual Taub-NUT (SDTN) metric. We also emphasize the hyperk\"ahler structure of the SDTN metric, and its straightforward extension to the `spinning' self-dual Kerr-Taub-NUT (SDKTN) metric. Our focus will be on the Euclidean-real metric, although we will also comment on the relation to other signatures. The reader who is already familiar with these notions may wish to simply skim this section, to absorb the notation and conventions that will be made use of through the rest of the paper.


\subsection{The SDTN geometry}

To begin, we fix notation conventions (broadly following~\cite{Adamo:2017qyl}). Let $x^a=(x^0,x^1,x^2,x^3)$ be holomorphic Cartesian coordinates on complexified Minkowski space $\M$ with holomorphic metric $(\d x^0)^2-(\d x^1)^2-\cdots -(\d x^3)^2$. The Euclidean real slice $\R^4\subset\M$ corresponds to letting $x^0\in\R$ and $x^1,x^2,x^3\in\im\R$, the set of imaginary numbers. Re-labeling $x^0\to t$ and $\im\,x^i\to x^i$ for $i=1,2,3$, the Cartesian coordinates $(t,x^1,x^2,x^3)$ on $\R^4$ can be encoded in the 2-spinor variables
	\be
	x^{\alpha\dot\alpha}=\frac{1}{\sqrt{2}}\left(\begin{array}{cc}
		t+ \im x^3& \im x^1+ x^2  \\
		\im x^1- x^2& t-\im x^3 
	\end{array}\right)\,.
	\ee
The 2-spinor indices are raised and lowered using the 2-dimensional Levi-Civita symbols $\epsilon_{\alpha\beta}$, $\epsilon_{\dot\alpha\dot\beta}$, etc., so that the flat metric on $\R^4$ is $\d s^2_{\R^4}=\epsilon_{\alpha\beta}\,\epsilon_{\dot\alpha\dot\beta}\,\d x^{\alpha\dot\alpha}\,\d x^{\beta\dot\beta}$.  It will be useful to employ the usual notation for 2-spinor invariants: $\la a\,b\ra:=a^{\alpha}b_{\alpha}=\epsilon^{\alpha\beta}a_{\beta}b_{\alpha}$, and $[a\,b]:=~a^{\dot\alpha}b_{\dot\alpha}=\epsilon^{\dot\alpha\dot\beta}a_{\dot\beta}b_{\dot\alpha}$.

In these variables, let $T^{\alpha\dot\alpha}\coloneqq\text{diag}(1,1)/\sqrt{2}$ denote the unit, future-pointing `time-like' vector\footnote{We sometimes abuse terminology by referring to quantities in Euclidean signature with Lorentzian nomenclature. In this case, a Euclidean `time-like' vector is one which would truly be time-like under Wick rotation to Lorentzian signature, and future-pointing means pointing in the positive direction of Euclidean time.}. This vector can be used to project the coordinates onto the spatial 3-slices by defining
	\be
\vec{x}^{\alpha\beta}\coloneqq\,\epsilon_{\dot\alpha\dot\beta}\,T^{(\alpha|\dot\alpha|}\,x^{\beta)\dot\beta}=\frac{1}{2}\left(\begin{array}{cc}
		\im x^1+x^2&-\im x^3  \\
		-\im x^3&-\im x^1+x^2 
	\end{array}\right)\,,\label{eq:3d-x_from_4d-x}
	\ee
so that $\vec{x}^{\alpha\beta}$ are coordinates on $\R^3$, whilst $t=T^{\alpha\dot\alpha} x_{\alpha\dot\alpha}$ is the coordinate in the direction of $T^{\alpha\dot\alpha}$. The inverse relation
	\be
x^{\alpha\dot\alpha}=t\,T^{\alpha\dot\alpha}-2\vec{x}^{\alpha\beta}\,T_\beta{}^{\dot\alpha}\,,\label{eq:4d-x_from_3d-x}
	\ee
enables us to pass between the $x^{\alpha\dot\alpha}$ and $(t,\vec{x}^{\alpha\beta})$ coordinates. 

Finally, it will be useful to fix constant dyads on the bundle of undotted and dotted spinors. We take these dyads to be $\iota^\alpha=\tilde\iota^{\dot\alpha}\coloneqq(1,0)$ and $o^\alpha=\tilde o^{\dot\alpha}\coloneqq(0,-1)$, normalized so that $\la \iota \,o\ra=1=[\tilde\iota\,\tilde o]$.

\medskip

The Euclidean self-dual Taub-NUT (SDTN) space $\CM$ is (topologically) Euclidean space $\R^4$ equipped with the metric 
\begin{equation}
    \d s^2=\left(1+\frac{2M}{r}\right)^{-1}\left(\d t-2M(1-\cos\theta)\d\phi\right)^2+\left(1+\frac{2M}{r}\right)\left(\d r^2+r^2\d\Omega_2^2\right)\,.\label{eq:metric}
\end{equation}
Here, $(r,\theta,\phi)$ are the standard spherical polar coordinates on $\R^3\setminus\{0\}\cong\R_{+}\times S^2$, defined in terms of the spatial coordinates $\vec{x}^{\alpha\beta}$ in the obvious way. The coordinate ranges are $r\in(0,\infty)$, $(\theta,\phi)\in S^2$, and $t\sim t+8\pi M$, with $\d\Omega_2^2$ the round metric on the 2-sphere parametrized by $\theta$ and $\phi$. 

The metric \eqref{eq:metric} is a complete, Ricci-flat and self-dual metric on $\R^4$. This can be verified in many equivalent ways, but perhaps the cleanest is to compute the Newman-Penrose scalars~\cite{Newman:1961qr} of the metric, finding~\cite{Kinnersley:1969zza}
\be\label{NPscal}
\widetilde{\Psi}_2=-\frac{2\,M}{(r+2\,M)^3}\,, \qquad \Psi_2=0\,,
\ee
with $\widetilde{\Psi}_2$ the only non-vanishing component of the Riemann curvature tensor. This immediately shows that the metric is vacuum (as the Ricci curvature vanishes), type D (as only $\widetilde{\Psi}_2\neq0$) and self-dual (as $\Psi_2$=0). The apparent singularity at $r=0$ is the Euclidean analogue of a black hole event horizon: it is a coordinate singularity which the metric can be extended across by a change of coordinates. 

Note that the metric \eqref{eq:metric} is manifestly in Gibbons-Hawking form. Recall that Gibbons-Hawking metrics~\cite{Hawking:1976jb,Gibbons:1978tef,Gibbons:1979xm} are defined by a pair $(V,A)$, where $V$ is a scalar function and $A$ is a 1-form, related by the abelian monopole equation 
\begin{equation}
    \d V=\star_3\,\d A\,,
\end{equation}
for $\star_3$ the Hodge star on $\R^3$. For the SDTN metric, this pair is given by
\begin{equation}
    V=1+\frac{2M}{r}\,,\qquad A=2M(1-\cos\theta)\d\phi:=2\,M\,\sa\,,\label{eq:bogol_pair}
\end{equation}
where $\sa$ is the gauge potential of a magnetic monopole. The metric \eqref{eq:metric} is then equal to
\be\label{GHmet}
\d s^2=\frac{(\d t-A)^2}{V}+V\,\d\vec{x}^2\,,
\ee
which is precisely of Gibbons-Hawking form.

As the `true' singularity at $r=-2M$ is not on the Euclidean manifold, one might worry that the SDTN metric is a poor toy model for a black hole. However, in \emph{split} signature, the SDTN metric has an event horizon, where the signs of the metric encoding the Kleinian `causal structure' are interchanged, behind which the metric can be continued up to a true singularity~\cite{Crawley:2021auj,Crawley:2023brz,Adamo:2023fbj,Easson:2023ytf}. In this sense, we are justified in viewing SDTN as an integrable toy model of a black hole.   

\medskip

In the 2-spinor variables, the SDTN metric \eqref{eq:metric} can be written as $\d s^2=\epsilon_{\alpha\beta}\,\epsilon_{\dot\alpha\dot\beta}\,e^{\alpha\dot\alpha}\,e^{\beta\dot\beta}$ in terms of the tetrad
\begin{equation}\label{SDTNtetrad}
    e^{\alpha\dot\alpha}=\left(1+\frac{2M}{r}\right)^{-1/2}T^{\alpha\dot\alpha}\left(\d t-2M(1-\cos\theta)\,\d\phi\right)-2\left(1+\frac{2M}{r}\right)^{1/2}\d \vec{x}^{\alpha\beta}\,T_\beta{}^{\dot\alpha}\,.
\end{equation}
From this tetrad, a basis of closed anti-self-dual (ASD) 2-forms on $\CM$ is given by
\begin{equation}
    \begin{aligned}
        \Sigma^{\alpha\beta}&=e^{\alpha\dot\alpha}\wedge e^\beta{}_{\dot\alpha}\\&=2\left(\d t-2M(1-\cos\theta)\,\d\phi\right)\wedge\d \vec{x}^{\alpha\beta}+2\left(1+\frac{2M}{r}\right)\d \vec{x}^{\alpha\gamma}\wedge\d \vec{x}^\beta{}_\gamma\,,
    \end{aligned}\label{eq:asd_2-forms}
\end{equation}
which will prove useful in later computations as the undotted spin connection is flat in this frame.


\subsection{Hyperk\"ahler structure of SDTN}

As the SDTN metric \eqref{eq:metric} is self-dual and Ricci flat, it has holonomy $\SU(2)$ and is thus a hyperk\"ahler (HK) manifold. A HK manifold has a 2-sphere's worth of complex structures which are K\"ahler with respect to the metric. This structure can be manifested very cleanly for SDTN following LeBrun's construction~\cite{lebrun1991complete}.

Let $(y,z)$ be complex coordinates on $\C^2$ and consider the K\"ahler potential
\begin{equation} \Omega=8M(u^2+v^2)+2(u^4+v^4)\,,\label{eq:kahler_potential}
\end{equation}
where $u$, $v$ are two real-valued, positive functions defined implicitly by
\begin{equation}
    |y|=\e^{(u^2-v^2)/4M}u\,,\qquad |z|=uv\,.
\end{equation}
The associated K\"ahler metric (obtained by implicit differentiation of the K\"ahler potential) is
\begin{equation}
    \d s^2=4\left(1+\frac{2M}{u^2+v^2}\right)|\d z|^2+16M^2\left(1+\frac{2M}{u^2+v^2}\right)^{-1}\left|\frac{\d y}{y}-\frac{v^2}{u^2+v^2}\frac{\d z}{z}\right|^2\,.\label{eq:kahler_metric}
\end{equation}
Introducing Cartesian coordinates $\vec{x}^{i}=(x^1,x^2,x^3)$, $r=|\vec x|$ on $\R^3$ and performing the diffeomorphism
\begin{equation}
    y=\sqrt{\frac{r+x^3}{2}}\e^{-\im(t+\im x^3)/4M}\,,\qquad z=\frac{1}{2}(x^1+\im x^2)\,,\label{yztoreal}
\end{equation}
one finds that
\begin{equation}
    u=\sqrt{\frac{r+x^3}{2}}\,,\qquad v=\sqrt{\frac{r-x^3}{2}}\,.
\end{equation}
In these new coordinates, the K\"ahler metric \eqref{eq:kahler_metric} becomes
\begin{equation}
    \d s^2=\left(1+\frac{2M}{r}\right)^{-1}\left(\d t-\frac{2M}{r}\frac{x^1\d x^2-x^2\d x^1}{r+x^3}\right)^2+\left(1+\frac{2M}{r}\right)((\d x^1)^2+(\d x^2)^2+(\d x^3)^2)\,,
\end{equation}
which is precisely the SDTN metric \eqref{eq:metric} upon passing to spherical coordinates.

Remarkably, the K\"ahler potential \eqref{eq:kahler_potential} itself is extremely simple when written in the Gibbons-Hawking coordinates:
\be\label{sdtnkahler}
\Omega = r^2+(x^3)^2 + 8Mr\,.
\ee
However, observe that these coordinates are \emph{not} holomorphic/anti-holomorphic with respect to the associated complex structure. So although the potential \eqref{sdtnkahler} is simple, taking the appropriate derivatives to arrive at the metric is not.

\medskip

At this stage, one may note that by working with a single K\"ahler potential, we have chosen a single complex structure, rather than the 2-sphere's worth associated with a HK structure. The existence of this $S^2$ of complex structures is guaranteed by checking that $\Omega$ satisfies Plebanski's `first heavenly equation'~\cite{Plebanski:1975wn}, and in fact the choice of complex structure is generic. Indeed, orientation- and metric-compatible complex structures on SDTN are in one-to-one correspondence with points on the unit sphere, parametrized by $\vec{n}\in\R^3$, $\vec{n}\,^2=1$~\cite{lebrun1991complete}. 

To see this, let $z$ be a complex coordinate on the plane $\vec{n}\cdot\vec{x}=0$, and $y=u\,\e^{-\im(t+\im\vec{n}\cdot\vec{x})/4M}$. Any such coordinates are related to the coordinates $(y,z)$ defined in \eqref{yztoreal} by $\SO(3)$ rotations of $\vec{x}$, which rotate the K\"ahler potential \eqref{sdtnkahler} into
\be
\Omega = r^2+(\vec{n}\cdot\vec{x})^2 + 8Mr\,,\qquad\vec{n}\in S^2\,.\label{eq:kahlerpotential}
\ee
As the SDTN metric is invariant under $\SO(3)$ rotations of $\vec{x}$, it follows that the SDTN metric will be recovered from any of this $S^2$ family of K\"ahler potentials. 


\subsection{From SDTN to self-dual Kerr-Taub-NUT}
\label{sec:spin}

All astrophysical black holes have some angular momentum, making the Kerr metric the archetypal astrophysically relevant black hole solutions. It has long been known that the Kerr metric can be obtained from the (non-spinning) Schwarzschild solution by a complex coordinate transformation, known as the Newman-Janis shift~\cite{Newman:1965tw}. In essence, this works by shifting the spatial coordinates of the Schwarzschild metric into the complex along the mass-rescaled spin vector $\vec{a}$ of the desired Kerr metric; ensuring that the resulting metric is Lorentzian-real requires a non-linear superposition of this shift and its complex conjugate which is somewhat non-trivial.

While understanding of the Newman-Janis shift has improved over the years due to a variety of new perspectives (cf., \cite{Talbot:1969bpa,Drake:1998gf,Keane:2014sta,Erbin:2014aya,Adamo:2014baa,Erbin:2014aja,Erbin:2016lzq,Rajan:2016zmq,Arkani-Hamed:2019ymq,Emond:2020lwi,Beltracchi:2021ris,Adamo:2023fbj,Kim:2024mpy}), it is fair to say that the non-linear nature of the reality conditions needed to obtain a Lorentzian-real metric after this complex shift make it a tool of limited applicability for gravitational scattering in black hole backgrounds. However, in the self-dual sector, this non-linearity disappears and the self-dual Kerr-Taub-NUT metric (cf., \cite{Carter:1968ks,Carter:1968rr,Miller:1973hqu}) is obtained by a complex \emph{linear} translation of the SDTN metric.

To see this, observe that any (Lorentzian-real) vacuum solution to the Einstein equations which is algebraically special of type D has a Weyl tensor which can be written in 2-spinors as~\cite{Penrose:1984uia}
\be\label{DWs1}
\Psi_{\alpha\beta\gamma\delta}=6\,\Psi_2\,\bo_{(\alpha}\,\bo_{\beta}\,\bi_{\gamma}\,\bi_{\delta)}\,,
\ee
where $\Psi_{\alpha\beta\gamma\delta}$ is the anti-self-dual Weyl spinor and $\{\bo_{\alpha},\bi_{\alpha}\}$ is a normalised spinor dyad aligned with the two degenerate principal null directions. The scalar $\Psi_2$ is the only non-vanishing Newman-Penrose scalar for the curvature of a vacuum type D metric. The SD Weyl curvature is obtained by complex conjugation.

Now, it can be shown that any such vacuum type D spacetime admits a valence-two ASD Killing spinor $\chi_{\alpha\beta}$, obeying $\nabla_{(\alpha}{}^{\dot\alpha}\chi_{\beta\gamma)}=0$ which is related to the Weyl curvature by~\cite{Walker:1970un,Hughston:1972qf,Hughston:1973,Jeffryes:1984a,Penrose:1986uia}:
\be\label{R2KS}
\chi_{\alpha\beta}=\frac{\bo_{(\alpha}\,\bi_{\beta)}}{(6\,\Psi_2)^{1/3}}\,, \qquad \Psi_{\alpha\beta\gamma\delta}=C\,\frac{\chi_{(\alpha\beta}\,\chi_{\gamma\delta)}}{(\chi^{\rho\sigma}\,\chi_{\rho\sigma})^{5/2}}\,,
\ee
where $C$ is a numerical constant. For the Schwarzschild and Kerr metrics, one finds that \eqref{R2KS} is realised for $\chi_{\alpha\beta}=\vec{x}_{\alpha\beta}+\im\,\vec{a}_{\alpha\beta}$ and $C=M$. 

Compatibility between this imaginary shift along the spin vector for $\Psi_{\alpha\beta\gamma\delta}$ and its complex conjugate $\bar{\Psi}_{\dot\alpha\dot\beta\dot\gamma\dot\delta}$ is responsible for the non-linear reality conditions of the Newman-Janis shift at the level of the metric. However, for SDTN and self-dual Kerr-Taub-NUT (SDKTN) it follows that $\Psi_{\alpha\beta\gamma\delta}=0$, so that the `spinning' self-dual metric can be obtained by a simple shift $\vec{x}^{\alpha\beta}\to\vec{x}^{\alpha\beta}-\im\,\vec{a}^{\alpha\beta}$. With Euclidean reality conditions, this leads to a real-valued Euclidean SDKTN metric in Euclideanised Kerr coordinates.

It is in this sense that SDKTN can be viewed as a simple complex diffeomorphism of SDTN, a fact which has been observed several times before, in Euclidean as well as split signature~\cite{Crawley:2023brz,Desai:2024fgr}. The important takeaway here is that to compute quantities like wavefunctions or scattering amplitudes on SDKTN, it suffices to first perform the computation in `pure' SDTN and then simply translate $\vec{x}\to\vec{x}-\im\vec{a}$ at the end to obtain the desired quantities in SDKTN.


\section{Twistor theory for self-dual Taub-NUT}\label{sec:twistor_space}

The famous non-linear graviton theorem of Penrose~\cite{Penrose:1976js} states that any vacuum self-dual four-manifold corresponds to a twistor space $\CPT$, obtained by a complex deformation of the twistor space of flat space, and equipped with an integrable complex structure and a four-parameter family of holomorphic curves obeying certain consistency conditions. Hence, there is a twistor description of the self-dual Taub-NUT metric, and there have been several formulations introduced over the years~\cite{Sparling:1976,Hitchin:1979rts,Galperin:1985de}.

In this Section, we review the twistor description of SDTN which will be used throughout the paper. After describing the twistor space and how the SDTN metric is reconstructed from the twistor data, we show how solutions to the linearised Einstein equations on SDTN are encoded in the cohomology of its twistor space. 


\subsection{The SDTN twistor space}

The twistor space $\PT$ of complexified Minkowski space, $\M$, is the open subset of $\P^3$ obtained by removing a projective line. In particular, let $Z^A=(\mu^{\dot\alpha},\lambda_\alpha)$ be homogeneous coordinates on $\P^3$, considered up to projective rescalings $Z^A\sim r\,Z^A$ for all $r\in\C^*$, with $[Z^A]$ the projective equivalence class corresponding to a point in $Z\in\P^3$. Twistor space is given by
\begin{equation}
    \PT=\{[Z^A]\in\P^3\,|\,\lambda_\alpha\neq0\}\,.
\end{equation}
This means that $\PT$ fibres holomorphically over $\P^1$, the projection being simply $Z^A\mapsto\lambda_\alpha$, with $[\lambda_{\alpha}]$ serving as homogeneous coordinates on the $\P^1$ base. There is then a two-set open cover of $\PT=U_0\cup U_1$ given by
\begin{equation}
    U_0=\{Z\in\PT\,|\,\lambda_0\neq 0\}\,,\qquad U_1=\{Z\in\PT\,|\,\lambda_1\neq 0\}\,,
\end{equation} 
induced by the natural open covering of $\P^1$. Twistor space is also equipped with a holomorphic, $\mathcal{O}(2)$-valued\footnote{The bundles $\mathcal{O}(n)\to\PT$ are defined as the pullbacks of the bundles $\mathcal{O}(n)\to\P^1$ by the holomorphic projection $\PT\to\P^1$.} symplectic 2-form $\Sigma$ on the fibres of the fibration $\PT\to\P^1$
\begin{equation}
\Sigma=\d\mu^{\dot\alpha}\wedge\d\mu_{\dot\alpha}\,,\label{eq:symplectic_form}
\end{equation}
together with a dual $\mathcal{O}(-2)$-valued Poisson bracket
\begin{equation}
    \{\cdot\,,\cdot\}=\epsilon^{\dot\alpha\dot\beta}\mathcal{L}_{\dot\alpha}\wedge\mathcal{L}_{\dot\beta}\,,
\end{equation}
where $\mathcal{L}_{\dot\alpha}$ denotes the Lie derivative along $\p/\p\mu^{\dot\alpha}$. 

Flat space $\M$ is realized as the moduli space of linear, holomorphic Riemann surfaces in $\PT$, referred to as \emph{twistor lines}. Explicitly, the twistor line corresponding to any point $x^{\alpha\dot\alpha}\in\M$ is given by
\begin{equation}\label{incidence}
    X\cong\P^1\quad \colon\quad \left\{Z\in\PT\,|\,\mu^{\dot\alpha}=x^{\alpha\dot\alpha}\,\lambda_\alpha\right\}\,.
\end{equation}
From this, it is easy to see that the normal bundle to any twistor line is $\cN_X\cong~\cO(1)\oplus\cO(1)$.

To single out a particular real slice of $\M$, $\PT$ must be equipped with some reality structure. For instance, to single out the Euclidean real slice $\R^4\subset\M$, $\PT$ is equipped with an anti-holomorphic involution $\sigma:Z^{A}\to\hat{Z}^{A}$ which acts as
\begin{equation}\label{qconj}
\hat{Z}^{A}=(-\bar{\mu}^{\dot{1}},\,\bar{\mu}^{\dot{0}},\,-\bar{\lambda}_{1},\,\bar{\lambda}_0)\,.
\end{equation}
This `quaternionic conjugation' operation squares to minus the identity (so it has no fixed points), and acts as the antipodal map on the $\P^1$ base of twistor space. It is easy to show that the twistor lines defined by \eqref{incidence} that are preserved by $\sigma$ correspond to real $x\in\R^4$.

\medskip

A key result in twistor theory is that curved, self-dual metrics can be obtained by performing complex deformations of the `flat' twistor space $\PT$. In particular, the \emph{non-linear graviton theorem}~\cite{Penrose:1976js,Atiyah:1978wi} establishes a one-to-one correspondence between self-dual, Ricci-flat Riemannian metrics\footnote{Here, and throughout, we implicitly assume that we are considering suitably convex regions of such a self-dual Ricci-flat Riemannian manifold.} and `curved' twistor spaces $\CPT$ obtained by a complex deformation of $\PT$ which:
\begin{enumerate}
 \item preserve the holomorphic fibration over $\P^1$,
 \item admit a holomorphic $\cO(2)$-valued symplectic form $\Sigma$ on the fibres,
 \item are equipped with an anti-holomorphic involution $\sigma$, and
 \item have a 4 (real) dimensional family of rational holomorphic curves which are invariant under $\sigma$ and have normal bundle $\cO(1)\oplus\cO(1)$.
\end{enumerate}
Just as in the flat model, points in the self-dual `spacetime' correspond to the holomorphic rational curves in twistor space, and the metric can be reconstructed from the holomorphic data on $\CPT$.

The condition that $\CPT$ be a complex deformation of $\PT$ means that the complex structure of $\CPT$ is encoded in an anti-holomorphic Dolbeault operator which is locally of the form $\nbar=\dbar+V$, for $V\in\Omega^{0,1}(\PT,T^{1,0}\PT)$, which is integrable:
\begin{equation}
    \dbar V+\frac{1}{2}[V,\,V]=0\,,\label{eq:integrability}
\end{equation}
where $[V,V]$ is the Lie bracket. The further requirements of a holomorphic fibration over $\P^1$ and a holomorphic weighted symplectic form on the fibres means that $V$ must be a Hamiltonian vector field
\be\label{hamvec}
V=\left\{\sh,\,\cdot\right\}\,, \qquad \sh\in\Omega^{0,1}(\PT,\cO(2))\,,
\ee
with respect to the weighted symplectic form.

\medskip

Thus, to describe the twistor space of SDTN, it suffices to specify the appropriate $\sh$ valued in $\cO(2)$. In order to do this, it will be convenient to introduce the notation
\begin{equation}\label{convents}
    \mu^-\coloneqq[\mu\,\tilde \iota]\,,\qquad\mu^+\coloneqq[\mu\,\tilde o]\,,\qquad \eta\coloneqq \mu^+\mu^-\,,\qquad \bar e^0\coloneqq\frac{\la\hat{\lambda}\,\d\hat{\lambda}\ra}{\la \lambda\,\hat\lambda\ra^2}\,,
\end{equation}
where $\mu^{\pm}$ are the decomposition of $\mu^{\dot\alpha}$ with respect to the dotted spinor dyad $\{\tilde{o}_{\dot\alpha},\,\tilde{\iota}_{\dot\alpha}\}$, $\hat{\lambda}_{\alpha}=(-\overline{\lambda_1},\overline{\lambda_0})$ and $\bar{e}^0$ trivializes the canonical bundle of $\P^1$. Now, define the $\cO(2)$-valued Hamiltonian~\cite{Galperin:1985de}
\begin{equation}
    \sh\coloneqq\frac{\eta^2\,\bar e^0}{4M}\in\Omega^{0,1}(\PT,\mathcal{O}(2))\,.\label{eq:hamiltonian}
\end{equation}
$M$ is a parameter that will be identified with the `mass' of the SDTN metric. 

The complex deformation associated with this Hamiltonian is
\begin{equation}\label{eq:beltrami}
    V=\{\sh,\,\cdot\}=\frac{\eta\,\bar e^0}{2M}\wedge\left(\mu_+\cL_+ -\mu_-\cL_-\right)\,,
\end{equation}
where $\cL_\pm$ denotes the Lie derivative along $\p/\p\mu^\pm$. The obstruction to integrability \eqref{eq:integrability} of the complex structure $\nbar$ in $\P^3$ is then
\begin{equation}
    \dbar \sh+\frac{1}{2}\{\sh,\sh\}=\frac{\pi^2\,\eta^2}{M}\,\bar\delta^{2}(\lambda)\,,
\end{equation}
where 
\be\label{holdelt}
\bar{\delta}^2(\lambda):=\frac{1}{(2\pi\im)^2}\bigwedge_{\alpha=0,1}\dbar\left(\frac{1}{\lambda_{\alpha}}\right)\,,
\ee
is the $(0,2)$-distribution that acts as a holomorphic delta function. Since the complex deformation must preserve the holomorphic fibration over $\P^1$, which has holomorphic homogeneous coordinates $[\lambda_{\alpha}]$, it follows that $\lambda_{\alpha}\neq 0$ on $\CPT$. Therefore, $\nbar^2=0$ as required\footnote{It is interesting to contrast this with the analogous twistor space for the Eguchi-Hanson metric~\cite{Bittleston:2023bzp}, where the complex structure is induced by a source on the $\mu^{\dot\alpha}=0$ locus, which \emph{is} in the twistor space.}.

In the complex structure \eqref{eq:beltrami}, $\lambda_\alpha$ are still holomorphic coordinates (as the holomorphic fibration over $\P^1$ is undeformed), but the $\mu^{\dot\alpha}$ are no longer holomorphic. Nevertheless, one can construct local holomorphic coordinates on $U_0$ and $U_1$ straightforwardly. First, observe that
\begin{equation}
    \bar\nabla\mu^{\pm}=\pm\mu^\pm\,\frac{\eta\,\bar e^0}{2M}\,,
\end{equation}
so that although $\nbar\mu^{\pm}\neq0$, the combination $\eta=\mu^+\,\mu^-$ \emph{is} still holomorphic: $\nbar\eta=0$. Then holomorphic coordinates on the fibres of $\CPT\to\P^1$ are given by
\begin{equation}
    \rho^\pm=\mu^\pm\,\exp\!\left(\pm\frac{\eta\, f(\lambda)}{2M}\right)\,,
\end{equation}
where $f$ is defined in the two patches by
\begin{equation}
    f(\lambda)=-\frac{1}{\la\lambda\,\hat\lambda\ra}\begin{cases}
        \hat\lambda_0/\lambda_0\,,&\lambda_\alpha\in U_0\\
        \hat\lambda_1/\lambda_1\,,&\lambda_\alpha\in U_1
    \end{cases}\,.
\end{equation}
In particular, $\dbar f=-\bar e^0$, from which the holomorphicity of $\rho^\pm$ follows.

The holomorphic coordinates $\rho^\pm$ can be understood as taking values in $\mathcal{O}(1)\otimes L^{\pm 1}$, where $L$ is a line bundle over the total space of $\mathcal{O}(2)\to\P^1$ with transition function
\begin{equation}
    \phi_{10}=\exp\left(\frac{\eta}{2M\,\lambda_0\,\lambda_1}\right)\,,\label{eq:transition_function}
\end{equation}
over $U_0\cap U_1$. Then $\CPT$ is the sub-bundle of $\mathcal{O}(1)\otimes(L\oplus L^{-1})$ defined by
\begin{equation}
    \rho^+\,\rho^-=\eta\,.
\end{equation}
This relation is useful when comparing this twistor space with other twistor constructions of SDTN in the literature -- see Appendix~\ref{subsec:other_twistor_spaces}.


\subsection{Reconstructing the SDTN metric}

The holomorphic symplectic form on the fibres of $\CPT\to\P^1$ is now
\begin{equation}
    \Sigma=2\,\d\rho^-\wedge\d \rho^+\,.
\end{equation}
It can be checked that $\Sigma$ is a global section of the exterior square of the conormal bundle: that is, it changes from $U_0$ to $U_1$ only by terms proportional to $\d\lambda_\alpha$. Furthermore, $\Sigma$ coincides with the symplectic form on $\PT$ \eqref{eq:symplectic_form}, again up to terms proportional to $\d\lambda_\alpha$. Thus, one can continue to work with the homogeneous coordinates $(\mu^{\dot\alpha},\lambda_{\alpha})$ at the level of the symplectic form.

To reconstruct the spacetime metric, one must first solve for the holomorphic rational curves in twistor space corresponding to points in the four-manifold. These curves are given by $\cO(1)$-valued maps $\sF^{\dot\alpha}(x,\lambda)$ satisfying
\begin{equation}
\left.\dbar\right|_{X}\sF^{\dot\alpha}=\left.\frac{\p\sh}{\p\mu_{\dot\alpha}}\right|_{X}\,,\label{eq:eq_for_twistor_lines}
\end{equation}
where $\dbar|_{X}$ is the Dolbeault operator along the twistor line $X\cong\P^1$. Decomposing $\sF^{\dot\alpha}$ as $\sF^{\dot\alpha}=-\sF^-\tilde o^{\dot\alpha}+\sF^+\tilde\iota^{\dot\alpha}$ and inserting the Hamiltonian \eqref{eq:hamiltonian}, one obtains
\begin{equation}\label{holcurve}
    \left.\dbar\right|_{X}\sF^\pm=\pm \sF^\pm\,\frac{\sF^+\,\sF^-}{2M}\,\bar e^0\,,
\end{equation}
as the equation determining twistor curves in $\CPT$.

To solve this equation, introduce the spinors 
\be\label{chispinors}
\chi^\alpha_+\coloneqq \sqrt{\frac{r}{1+\zeta\bar\zeta}}\,(\bar\zeta,-1)\,,\qquad \chi^\alpha_-\coloneqq \sqrt{\frac{r}{1+\zeta\bar\zeta}}\,(1,\zeta)\,.
\ee
where $\zeta=\e^{\im\phi}\tan\frac{\theta}{2}$ and $\bar\zeta=\e^{-\im\phi}\tan\frac{\theta}{2}$ are stereographic coordinates on $S^2$. These spinors are related to the coordinates on the `spatial' slice $\R^3$ as 
\be\label{2spinident}
\vec{x}^{\alpha\beta}=\im\,\chi_+^{(\alpha}\,\chi_-^{\beta)}\,,
\ee
and can be used to construct solutions to \eqref{holcurve}:
\be\label{SDTNcurves}
\sF^\pm(x,\lambda)=\la \chi_\pm\,\lambda\ra\,\exp\!\left[\mp\frac{\im}{4M}\left(t+2\frac{\vec{x}^{\alpha\beta}\,\lambda_\alpha\,\hat\lambda_\beta}{\la\lambda\,\hat\lambda\ra}\right)\right]\,.
\ee
This follows by direct calculation using the identity
\begin{equation}
\left.\dbar\right|_{X}\left(\frac{\vec{x}^{\alpha\beta}\,\lambda_\alpha\,\hat\lambda_\beta}{\la\lambda\,\hat\lambda\ra}\right)=\im\la\chi_+\,\lambda\ra\,\la\chi_-\,\lambda\ra\,.
\end{equation}
Consequently, the map
\begin{multline}
    \sF^{\dot\alpha}(x,\lambda)=\tilde\iota^{\dot\alpha}\,\langle \chi_+\,\lambda\rangle\,\exp\!\left[-\frac{\im}{4M}\left(t+2\frac{\vec{x}^{\alpha\beta}\,\lambda_\alpha\,\hat\lambda_\beta}{\langle \lambda\,\hat\lambda\rangle}\right)\right] \\
    -\tilde o^{\dot\alpha}\,\langle\chi_-\,\lambda\rangle\,\exp\!\left[\frac{\im}{4M}\left(t+2\frac{\vec{x}^{\alpha\beta}\,\lambda_\alpha\,\hat\lambda_\beta}{\langle\lambda\,\hat\lambda\rangle}\right)\right]\,.\label{eq:twistor_lines}
\end{multline}
describes holomorphic rational curves in $\CPT$.

It is straightforward to check that these curves are invariant under the anti-holomorphic involution $\sigma:(\mu^{\dot\alpha},\lambda_\alpha)\mapsto(\hat\mu^{\dot\alpha},\hat\lambda_\alpha)$ when $\vec{x}^{\alpha\beta}=\hat{\vec{x}}^{\alpha\beta}$ and $t\in\R$, as required for Euclidean reality conditions.
That is, a point $(\mu^{\dot\alpha},\lambda_\alpha)$ lies on a twistor line if and only if the conjugate point $(\hat\mu^{\dot\alpha},\hat\lambda_\alpha)$ lies on the same line as well. This happens because $\hat{\tilde\iota}^{\dot\alpha}=-\tilde o^{\dot\alpha}$ and $\hat\chi_+^\alpha=-\chi_-^\alpha$. 

For $\vec{x}^{\alpha\beta}\neq0$, $(\vec{x}^{\alpha\beta},t)$ provide coordinates on a circle bundle over $\R^3-0$, the fibre coordinate being $t\in S^1$ of radius $8\pi M$. When $\vec{x}^{\alpha\beta}=\vec{0}$, there is a unique twistor curve $\sF^{\dot\alpha}(t,\vec{0},\lambda)=0$, which just corresponds to adding a single point as the fibre over the origin of $\R^3$. This means that the 4-manifold topology is \emph{still} $\R^4$, although we will see that the metric is not flat. 

\medskip

Let $\CM$ denote the 4-dimensional moduli space of twistor curves (which we have just established to have topology $\R^4$) and let $p:\CM\times\P^1\to\CPT$ be the projection from the projectivised, un-dotted spin bundle of $\CM$ to twistor space, given by $p:(x^{\alpha\dot\alpha},\lambda_{\alpha})\mapsto (\sF^{\dot\alpha}(x,\lambda),\lambda_{\alpha})$. To reconstruct the metric on $\CM$, one computes the pullback of the symplectic form $\Sigma$~\cite{Gindikin:1986}. Using the holomorphic curves \eqref{eq:twistor_lines},
\begin{equation}
    \begin{aligned}
        p^*\Sigma&=-2\lambda_\alpha\,\lambda_\beta\left[\d \chi_+^\alpha\wedge\d \chi_-^\beta+\frac{\d \vec{x}^{\alpha\beta}}{4M}\wedge\left(\d t+2\,\frac{\d \vec{x}^{\gamma\delta}\,\lambda_\gamma\,\hat\lambda_\delta}{\langle\lambda\,\hat\lambda\rangle}\right)\right]\mod\d \lambda_\alpha\,,
    \end{aligned}
\end{equation}
where `mod $\d\lambda_{\alpha}$' means up to terms proportional to differential forms pointing along the $\P^1$ directions of $\CM\times\P^1$.

Now, using the identity
\begin{equation}
\d \vec{x}^{\alpha\beta}\wedge\d \vec{x}^{\gamma\delta}=-\frac{\epsilon^{\alpha\gamma}}{2}\,\d \vec{x}^{\beta\eta}\wedge\d \vec{x}^\delta{}_\eta-\frac{\epsilon^{\beta\delta}}{2}\,\d \vec{x}^{\alpha\eta}\wedge\d \vec{x}^\gamma{}_{\eta}\,,
\end{equation}
it follows that $p^{*}\Sigma$ can be rewritten as
\begin{equation}
    p^*\Sigma=\frac{\lambda_\alpha\,\lambda_\beta}{4M}\,\Sigma^{\alpha\beta}(x)\: \mod\d\lambda_\alpha\,,
\end{equation}
for
\begin{equation}
    \Sigma^{\alpha\beta}(x)=-8M\,\d \chi_+^{(\alpha}\wedge\d\chi_-^{\beta)}+2\,\d t\wedge\d \vec{x}^{\alpha\beta}+2\,\d \vec{x}^{\alpha\gamma}\wedge\d \vec{x}^\beta{}_{\gamma}\,.
\end{equation}
These are precisely the ASD 2-forms of the SDTN metric \eqref{eq:asd_2-forms}! To see this, note that in terms of the magnetic monopole 1-form $\sa$ appearing in \eqref{eq:bogol_pair}
\begin{equation}
    \d \chi_\pm^\alpha=\left(\frac{\vec{x}^{\beta\gamma}\,\d \vec{x}_{\beta\gamma}}{r^2}\mp\frac{\im\,\sa}{2}\right)\chi_\pm^\alpha-\frac{\im}{r^2}\,\chi_\mp^\alpha\,\chi_\pm^\beta\,\chi_\pm^\gamma\,\d \vec{x}_{\gamma\delta}\,,
\end{equation}
so that 
\begin{equation}
    \d \chi_+^{(\alpha}\wedge\d\chi_-^{\beta)}=\frac{1}{2}\,\d \vec{x}^{\alpha\beta}\wedge \sa-\frac{1}{2r}\,\d \vec{x}^{\alpha\gamma}\wedge\d \vec{x}^\beta{}_\gamma\,,
\end{equation}
upon which the equivalence with \eqref{eq:asd_2-forms} is immediate.

This implies that the exterior derivative $\d_{x}\sF^{\dot\alpha}$ -- where $\d_x$ denotes the exterior derivative with $\lambda$ held constant -- is proportional to $e^{\alpha\dot\alpha}\,\lambda_{\alpha}$ for a tetrad $e^{\alpha\dot\alpha}$, up to a frame rotation:
\begin{equation}\label{framedef}
    \d_x\sF^{\dot\alpha}=\sH^{\dot\alpha}{}_{\dot\beta}(x,\lambda)\,e^{\alpha\dot\beta}\,\lambda_\alpha\,,
\end{equation}
where $\sH^{\dot\alpha}{}_{\dot\beta}(x,\lambda)$ is valued in SL$(2,\C)$. The matrix $\sH^{\dot\alpha}{}_{\dot\beta}(x,\lambda)$ acts as a holomorphic frame for the bundle $\cN_X\otimes\cO(-1)$ over the twistor curve $X$, where $\cN_X\cong\cO(1)\oplus\cO(1)$ is the normal bundle of the curve in $\CPT$.

For the SDTN tetrad \eqref{SDTNtetrad} in our chosen coordinate system, one finds
\begin{equation}
    \begin{aligned}
    \sH^{\dot\alpha}{}_{\dot\beta}(x,\lambda)&=\frac{\im\, T_{\beta\dot\beta}\,\tilde o^{\dot\alpha}}{2M\,\sqrt{V}}\left(\frac{\langle\lambda\,\chi_-\rangle}{\langle\lambda\,\hat\lambda\rangle}\,\hat\lambda^\beta-\frac{2M}{r}\,\chi_-^\beta\right)\exp\!\left[\frac{\im}{4M}\left(t+2\,\frac{\vec{x}^{\gamma\delta}\,\lambda_\gamma\,\hat\lambda_\delta}{\langle\lambda\,\hat\lambda\rangle}\right)\right]\\&\quad+\frac{\im\, T_{\beta\dot\beta}\,\tilde \iota^{\dot\alpha}}{2M\,\sqrt{V}}\left(\frac{\langle\lambda\,\chi_+\rangle}{\langle\lambda\,\hat\lambda\rangle}\,\hat\lambda^\beta-\frac{2M}{r}\,\chi_+^\beta\right)\exp\!\left[-\frac{\im}{4M}\left(t+2\,\frac{\vec{x}^{\gamma\delta}\,\lambda_\gamma\,\hat\lambda_\delta}{\langle\lambda\,\hat\lambda\rangle}\right)\right]\,.\label{eq:dotted_frame}
    \end{aligned}
\end{equation}
It is a straightforward (albeit somewhat tedious) calculation to verify that this frame obeys \eqref{framedef} as well as $\sH_{\dot\gamma\dot\alpha}(x,\lambda)\,\sH^{\dot\gamma}{}_{\dot\beta}(x,\lambda)=\epsilon_{\dot\alpha\dot\beta}$.

\medskip

It is worth noting that in addition to the special case of the SDTN metric, this basic twistor construction can be easily generalised to describe \emph{any} Gibbons-Hawking gravitational instanton~\cite{Gibbons:1978tef,Gibbons:1979xm}. See Appendix~\ref{app:GHmets} for the details.


\subsection{Quasi-momentum eigenstates}

Gravitons are solutions of the linearised vacuum Einstein equations, in this case around the SDTN metric. For generic curved spacetimes, the linearised Einstein equations are a complicated coupled system of partial differential equations which cannot be solved exactly. For astrophysical black holes, it is a remarkable fact that (when written as equations for the radiative Newman-Penrose scalars of the gravitational perturbation) the linearised Einstein equations are separable, leaving only a single non-trivial radial equation of Schr\"odinger type~\cite{Teukolsky:1972my,Teukolsky:1973ha}. However, this Teukolsky equation does not generally lend itself to analytical solutions, and in any case describes only the radial modes of the gravitational perturbation, which must be combined with spheroidal harmonics in a partial wave sum to give the graviton wavefunction. 

A standard, separable Teukolsky-like description of gravitational perturbations exists for all Taub-NUT metrics~\cite{Bini:2002kp,Bini:2003sy,Bini:2003bm} -- and indeed any type D Einstein metric~\cite{Araneda:2018ezs} -- and is known to simplify dramatically at the self-dual point~\cite{Araneda:2024cqu}. However, it turns out that in this case one can dispense with separation of variables and partial wave sums to solve the linearised Einstein equations directly~\cite{Adamo:2023fbj}. This is possible thanks to the relationship between the linearised Einstein equations on SDTN and the charged Killing spinor equations in a background self-dual dyon electromagnetic field. 

The resulting \emph{quasi-momentum eigenstates} have an array of desirable properties: they are specified by a massless 4-momentum, have a smooth flat space limit (where they reproduce the usual momentum eigenstates), are everywhere smooth on the celestial sphere and vanish at the origin (i.e., the Euclidean `event horizon'). The Penrose transform~\cite{Penrose:1969ae,Hitchin:1980hp,Eastwood:1981jy} guarantees that these quasi-momentum eigenstates can be generated from cohomological data on the SDTN twistor space. 

Here, we review the quasi-momentum eigenstates on SDTN for massless scalars as well as negative and positive helicity gravitational perturbations, showing in each case what the corresponding representative is on twistor space.


\subsubsection{Massless scalars} 

Quasi-momentum eigenstates are described by a (complex) null momentum $k^{\alpha\dot\alpha}=\kappa^\alpha\tilde\kappa^{\dot\alpha}$ and a discrete charge\footnote{In~\cite{Adamo:2024xpc}, this charge was denoted by $m$; here, we use $q$ to avoid confusion with $M$, the `mass' of the SDTN metric.} $2q\in\Z$, chosen such that $q\pm 2M\omega\in\Z_{\geq0}$, where $\omega$ is the energy of the state and $\Z_{\geq0}$ is the set of non-negative integers. This implies that the energy $\omega$ is quantized in units of $2M$; this is an expected feature since the Euclidean time of the SDTN metric is periodic. The on-shell momentum can be parametrized as
\begin{equation}
    \kappa^\alpha=(1,z)\,,\qquad\tilde\kappa^{\dot\alpha}=\frac{\sqrt{2}\,\omega}{1+z\tilde z}\,(1,\tilde z)\,,\label{eq:spinor-helicity1}
\end{equation}
for $(z,\tilde z)$ affine coordinates on the complexified celestial sphere $\P^1\times\P^1$ which parameterize the `direction' of the momentum. Note that $\tilde{z}\neq\bar{z}$ for complex momenta (and the momentum must necessarily be complex in Euclidean signature in order to be null). 

In terms of the quantum numbers $q$, $\kappa^\alpha$ and $\tilde\kappa^{\dal}$, scalar quasi-momentum eigenstates are given by \cite{Adamo:2023fbj}
\begin{equation}\label{scalarqme}
    \begin{aligned}
    \phi^{(q)}(x)&\coloneqq\la\chi_+\,\kappa\ra^{q+2M\omega}\,\la\chi_-\,\kappa\ra^{q-2M\omega}\,\e^{\im k\cdot x}\\&=\left(\frac{r}{1+\zeta\bar\zeta}\right)^{q}(\zeta-z)^{q-2M\omega}\,(\bar\zeta z+1)^{q+2M\omega}\,\e^{\im k\cdot x}\,.
    \end{aligned}
\end{equation}
Observe that these wavefunctions are regular on the celestial sphere, thanks to the requirement that $q\pm2M\omega$ are non-negative integers. Among these solutions, there are distinguished \emph{minimal} states, which are regular as $r\to0$ and have the slowest possible growth as $r\to\infty$. These correspond to $q=2M|\omega|$, with the positive frequency minimal state given by
\be\label{pfminscal}
\phi^{+}(x)=\left(\frac{r}{1+\zeta\bar{\zeta}}\right)^{2M\omega}\left(\bar{\zeta}\,z+1\right)^{4M\omega}\,\e^{\im\,k\cdot x}\,,
\ee
and the negative frequency minimal state given by
\be\label{nfminscal}
\phi^{-}(x)=\left(\frac{r}{1+\zeta\bar{\zeta}}\right)^{-2M\omega}\left(\zeta-z\right)^{-4M\omega}\,\e^{\im\,k\cdot x}\,.
\ee
These minimal states are universal, in the sense that any other quasi-momentum eigenstate can be obtained by acting on them with an appropriate differential operator in momentum space
\be\label{min2qmom}
\phi^{(q)}(x)=\left(\frac{1+z\,\tilde{z}}{2\,\omega}\,\frac{\partial}{\partial\tilde{z}}\right)^{q-2M\omega}\phi^{+}(x)=
\left(\frac{1+z\,\tilde{z}}{2\,\omega}\,\frac{\partial}{\partial\tilde{z}}\right)^{q+2M\omega}\phi^{-}(x)\,,
\ee
for either positive or negative frequency fields, respectively.

As the scalar field traverses the curved SDTN metric, the on-shell 4-momentum is effectively `dressed' by the background. This is a familiar feature of solutions to background-coupled equations of motion (cf., \cite{Wolkow:1935zz,Seipt:2017ckc,Adamo:2017nia}) which can also be expressed cleanly in the spinor-helicity formalism~\cite{Adamo:2019zmk}; in the case of self-dual background fields, one expects that only the dotted momentum spinor will be dressed by the background~\cite{Adamo:2020syc,Adamo:2020yzi,Adamo:2021hno,Adamo:2022mev,Adamo:2023fbj,Adamo:2024xpc}. Indeed, a straightforward calculation shows that
\be\label{dressmom1}
\d \phi^{(q)}(x)=\im\,\kappa_{\alpha}\,\tilde{K}^{(q)}_{\dot\alpha}(x)\,\phi^{(q)}(x)\,e^{\alpha\dot\alpha}\,,
\ee
where the background-dressed dotted momentum spinor is
\be\label{dressedspinor}
\tilde{K}^{(q)}_{\dot\alpha}(x)=\frac{1}{\sqrt{V}}\left(\tilde\kappa_\dal+(q+2M\omega)\,\frac{\chi_{+\gamma}\,T^\gamma{}_\dal}{r\,\langle\kappa\,\chi_+\ra}-(q-2M\omega)\,\frac{\chi_{-\gamma}\,T^\gamma{}_\dal}{r\,\langle\kappa\,\chi_-\ra}\right)\,.
\ee
This background dressing can be conveniently expressed in terms of a dressing matrix
\begin{equation}\label{dmatrix1}
    \tilde K_\dal^{(q)}(x)=\tilde\kappa_{\dot\beta}\,\sG^{\dot\beta}{}_\dal(x;k,q)\,,
\end{equation}
for
\begin{multline}\label{dmatrix2}
    \sG^\dal{}_{\dot\beta}(x;k,q)=\frac{1}{\sqrt{V}}\left(\delta^\dal{}_{\dot\beta}+(q+2M\omega)\,\frac{\kappa_\alpha\,\chi_{+\beta}\, T^{\alpha\dal}\,T^\beta{}_{\dot\beta}}{\omega\, r\,\langle \kappa\,\chi_+\rangle}\right. \\
    \left.-(q-2M\omega)\,\frac{\kappa_\alpha\,\chi_{-\beta}\, T^{\alpha\dal}\,T^\beta{}_{\dot\beta}}{\omega\, r\,\langle \kappa\,\chi_-\rangle}\right)\,.
\end{multline}
It is straightforward to show that this dressing matrix is unimodular and satisfies
\be\label{dmatrix3}
\sG^\dal{}_{\dot\beta}(x;k,q)\,\epsilon_{\dot\alpha\dot\gamma}\,\sG^{\dot\gamma}{}_{\dot\delta}(x;k,q)=\epsilon_{\dot\beta\dot\delta}\,,\qquad\d(\kappa_\alpha\,\sG^\dal{}_{\dot\beta}(x;k,q)\,e^{\alpha\dot\beta})=0\,,
\ee
for $e^{\alpha\dot\alpha}$ the SDTN tetrad \eqref{SDTNtetrad}.

\medskip

Now, the Penrose transform ensures that any solution of the massless scalar wave equation on SDTN can be represented by a cohomology class in $H^{0,1}_{\nbar}(\CPT,\cO(-2))$, the Dolbeault cohomology group with respect to $\nbar$ defined by the Hamiltonian \eqref{eq:beltrami}. The twistor representative for the quasi-momentum eigenstate \eqref{scalarqme} is
\begin{equation}
\Phi^{(q)}(Z)=\int_{\C^*}\d s\,s\,\bar\delta^{2}(s\lambda-\kappa)\left(s\,\mu^+\right)^{q+2M\omega}\left(s\mu^-\right)^{q-2M\omega}\,\exp\!\left(-\xi\,s^2\eta\right)\,,\label{eq:scalar_twistor_rep}
\end{equation}
where
\be\label{xidef}
\xi:=\frac{\sqrt{2}\,[\bar\kappa\,\tilde\kappa]}{\la\kappa\,\hat\kappa\ra}\,,
\ee
for
\begin{equation}
    \hat\kappa^\alpha=(-\bar z,\,1)\,,\qquad\bar\kappa^{\dot\alpha}=(1,\,\bar z)\,.\label{eq:spinor-helicity2}
\end{equation}
It follows immediately that $\Phi^{(q)}$ is a cohomology class of the appropriate weight. Indeed, the scale parameter $s$ has the opposite projective scale to $(\mu^{\dot\alpha},\lambda_{\alpha})$, so the measure $\d s\,s$ ensures that $\Phi^{(q)}$ has homogeneity weight $-2$. Furthermore, since both $\nbar\mu^{\pm}$ and $\bar{\delta}^2(s\,\lambda-\kappa)$ have $(0,1)$-form components proportional to $\bar{e}^0$ and $\bar{e}^0\wedge\bar{e}^0=0$, it follows that $\nbar\Phi^{(q)}=0$.

To see that \eqref{eq:scalar_twistor_rep} does indeed give rise to the quasi-momentum eigenstate \eqref{scalarqme}, one first pulls the representative back to the holomorphic rational curve in twistor space corresponding to the point $x\in\CM$. Using the formulae \eqref{SDTNcurves} for the twistor curves, one finds
\begin{multline}\label{scalrepline}
\left.\Phi^{(q)}\right|_{X}=\la\chi_+\,\kappa\ra^{q+2M\omega}\,\la\chi_-\,\kappa\ra^{q-2M\omega}\,\frac{\la \iota\,\kappa\ra}{\la\iota\,\lambda\ra}\,\bar\delta(\la \lambda\,\kappa\ra)\\
\times\,\exp\!\left[\im\,\omega\, t+\im\, \vec{x}^{\alpha\beta}\left(2\omega\,\frac{\kappa_\alpha\,\hat\kappa_\beta}{\la\kappa\,\hat\kappa\ra}+\frac{\sqrt{2}\,[\bar\kappa\,\tilde\kappa]\,\kappa_\alpha\,\kappa_\beta}{\la\kappa\,\hat\kappa\ra}\right)\right]\,.
\end{multline}
Here, we have used the definition of $\xi$ from \eqref{xidef}, the support\footnote{This holomorphic delta function has support where $\lambda_{\alpha}\propto\kappa_{\alpha}$. In factors which are homogeneous of weight zero in $\lambda_{\alpha}$ (e.g., the argument of the exponential in $\Phi^{(q)}|_{X}$), one can then simply replace $\lambda_{\alpha}$ with $\kappa_\alpha$. Similarly, this implies that $\hat{\lambda}_{\alpha}\propto\hat{\kappa}_{\alpha}$, and since $\Phi^{(q)}$ has homogeneity zero in $\hat{\lambda}_{\alpha}$, we can replace $\hat{\lambda}_{\alpha}=\hat{\kappa}_{\alpha}$ everywhere.} of the remaining holomorphic delta function $\bar{\delta}(\la\lambda\,\kappa\ra)$, and $\iota_{\alpha}$ appears as part of a Jacobian upon performing the $s$-integral. The exponential can be further simplified by using \eqref{eq:spinor-helicity1} and \eqref{eq:spinor-helicity2} to deduce that
\begin{equation}
\omega=\la\kappa|T|\tilde\kappa]\,,\qquad 2\omega\,\kappa^{(\alpha}\hat\kappa^{\beta)}+\sqrt{2}\,[\bar\kappa\,\tilde\kappa]\,\kappa^\alpha\,\kappa^\beta=-2\,\la\kappa\,\hat\kappa\ra\,\kappa^{(\alpha}\,T^{\beta)\dal}\,\tilde\kappa_\dal\,,
\end{equation}
so that \eqref{eq:4d-x_from_3d-x} implies
\begin{equation}
    \omega t+ \vec{x}^{\alpha
    \beta}\,\left(2\omega\,\frac{\kappa_\alpha\,\hat\kappa_\beta}{\la\kappa\,\hat\kappa\ra}+\frac{\sqrt{2}\,[\bar\kappa\,\tilde\kappa]\,\kappa_\alpha\,\kappa_\beta}{\la\kappa\,\hat\kappa\ra}\right)=k\cdot x\,.
\end{equation}
With this identity, the quasi-momentum eigenstate scalar field on SDTN is recovered by the usual Penrose integral formula
\begin{equation}\label{scalarintegral}
\phi^{(q)}(x)=\int_{X}\D\lambda\wedge\left.\Phi^{(q)}\right|_{X}\,,
\end{equation}
with $\D\lambda:=\la\lambda\,\d\lambda\ra$ and the integral performed trivially against the remaining holomorphic delta function in \eqref{scalrepline}.

The Penrose transform implies an important relationship between the holomorphic frame \eqref{eq:dotted_frame} arising on twistor space and the dressing frame \eqref{dmatrix2} for the dotted momentum spinor. Observe that, by \eqref{framedef}
\be\label{frameident1}
\d\phi^{(q)}(x)=e^{\alpha\dot\beta}\,\int_X\D\lambda\wedge\sH^{\dot\alpha}{}_{\dot\beta}(x,\lambda)\,\lambda_{\alpha}\,\left.\frac{\partial\Phi^{(q)}}{\partial\mu^{\dot\alpha}}\right|_{X}\,, 
\ee
but upon comparison with \eqref{dressmom1} this implies that
\be\label{frameident2}
\sH^{\dot\alpha}{}_{\dot\beta}(x,\lambda)\,\lambda_{\alpha}\,\left.\frac{\partial\Phi^{(q)}}{\partial\mu^{\dot\alpha}}\right|_{X}=\im\,\kappa_{\alpha}\,\tilde{K}^{(q)}_{\dot\beta}(x)\,\left.\Phi^{(q)}\right|_{X}\,.
\ee
In particular, this means that derivatives of the twistor quasi-momentum eigenstate representatives, contracted with the frame $\sH^{\dot\alpha}{}_{\dot\beta}$ can be replaced by contractions of the 4-momentum with the dressing matrix $\sG^{\dot\alpha}{}_{\dot\beta}$. In other words, these frame-contracted derivatives are effectively exponential in nature. This will have important consequences in our later calculations of scattering amplitudes.


\subsubsection{Gravitational perturbations}

On SDTN, gravitational perturbations -- or gravitons -- can still be decomposed into positive and negative helicity, although this decomposition is no longer symmetric (like in Minkowski space) due to the chirality of the background metric~\cite{Mason:2008jy}. Positive helicity gravitons can still be characterised by metric perturbations $g_{ab}\to g_{ab}+h_{ab}$ such that $h_{ab}$ solves the linearised Einstein equations and has purely self-dual linearised Weyl tensor. However, it is not possible to define negative helicity gravitons in a similar fashion, as a generic infinitesimal diffeomorphism on the self-dual background will give rise to a non-vanishing piece of self-dual linear Weyl curvature. 

Thus, we define positive helicity gravitons to be those metric perturbations $h_{ab}$ such that their corresponding linear Weyl tensor obeys $\psi_{\alpha\beta\gamma\delta}=0$ (i.e., the linear Weyl curvature is purely self-dual), while negative helicity gravitons are those metric perturbations $h_{ab}$ whose linear anti-self-dual Weyl tensor obeys $\nabla^{\alpha\dot\alpha}\psi_{\alpha\beta\gamma\delta}=0$. In both cases, the equations of motion can be solved in terms of quasi-momentum eigenstates akin to the scalars discussed above, with corresponding representatives on twistor space.

\medskip

\paragraph{Negative helicity gravitons:} The quasi-momentum eigenstate of a negative helicity graviton is expressed at the level of the corresponding zero-rest-mass field, namely, the linearised anti-self-dual Weyl spinor:
\be\label{nhelqme}
\begin{split}
\psi^{(q)}_{\alpha\beta\gamma\delta}(x)&=\kappa_{\alpha}\,\kappa_{\beta}\,\kappa_{\gamma}\,\kappa_{\delta}\,\la\chi_+\,\kappa\ra^{q+2M\omega}\,\la\chi_{-}\,\kappa\ra^{q-2M\omega}\,\e^{\im\,k\cdot x} \\
&=\kappa_{\alpha}\,\kappa_{\beta}\,\kappa_{\gamma}\,\kappa_{\delta}\left(\frac{r}{1+\zeta\,\bar{\zeta}}\right)^{q}\,(\bar{\zeta}\,z+1)^{q+2M\omega}\,(\zeta-z)^{q-2M\omega}\,\e^{\im\,k\cdot x}\,.
\end{split}
\ee
Minimal states $\psi^{\pm}_{\alpha\beta\gamma\delta}$ for positive/negative frequency again correspond to $q=2M|\omega|$, and have the slowest possible growth in $r$ and are regular at the origin. Generic quasi-momentum eigenstates can be obtained from the minimal ones by the same relations \eqref{min2qmom} as in the scalar case.

The Penrose transforms dictates that these negative helicity gravitons are represented by cohomology classes in $H^{0,1}_{\nbar}(\CPT,\cO(-6))$, and a suitable representative is obtained by simply modifying the projective weight of the scalar representative:
\be\label{nhelrep}
\tilde{h}^{(q)}(Z)=\int_{\C^{*}}\d s\,s^{5}\,\bar{\delta}^{2}(s\,\lambda-\kappa)\,\left(s\,\mu^+\right)^{q+2M\omega}\,\left(s\,\mu^-\right)^{q-2M\omega}\,\exp\!\left(-\xi\,s^2\,\eta\right)\,,
\ee
with $\xi$ given by \eqref{xidef} as before. To see that this representative gives rise to the wavefunction \eqref{nhelqme} on SDTN, one simply evaluates the integral formula
\be\label{nhelint1}
\psi^{(q)}_{\alpha\beta\gamma\delta}(x)=\int_{X}\D\lambda\wedge\lambda_{\alpha}\,\lambda_{\beta}\,\lambda_{\gamma}\,\lambda_{\delta}\,\left.\tilde{h}^{(q)}\right|_{X}\,,
\ee
following the same steps as in the scalar case.

It is also possible to describe the perturbation to anti-self-dual spin connection corresponding to such a negative helicity graviton. Let the triplet of 1-forms $\gamma_{\alpha\beta}$ denote a linear perturbation to the (vanishing) ASD spin connection of SDTN. This perturbation is related to the zero-rest-mass field by
\begin{equation}
    \d \gamma_{\alpha\beta}=\psi_{\alpha\beta\gamma\delta}\,\Sigma^{\gamma\delta}\,.\label{eq:gamma-eom}
\end{equation}
For the quasi-momentum eigenstate \eqref{nhelqme}, the corresponding spin connection perturbation is given by
\be
 \gamma_{\alpha\beta}^{(q)}=\frac{2\im}{[\tilde{a}\,\tilde\kappa]}\,\kappa_\alpha\,\kappa_\beta\,\kappa_\gamma\,\tilde{a}_{\dot\delta}\,\sG^{\dot\delta}{}_{\dot\gamma}(x;k,q)\,\phi^{(q)}\,e^{\gamma\dot\gamma}\,,\label{eq:linearized_asd_spin_connection}
\ee
for $\tilde{a}_{\dot\alpha}$ a fixed reference spinor amounting to a gauge choice for the spin connection. It follows that this satisfies \eqref{eq:gamma-eom} upon using the identities \eqref{dmatrix3}. 

The Penrose transform also extends to this potential description of the negative helicity gravitons~\cite{Mason:2008jy}. Here, $\gamma_{\alpha\beta}$ is represented by
\be\label{scPtrans}
\tilde{b}\in H^{0,1}_{\nbar}(\CPT,\Omega^{1,0}\otimes\cO(-4))\,,
\ee
defined modulo
\begin{equation}\label{btilfree}
    \tilde b\sim \tilde b+\p c_{-4}+c_{-6}\wedge\D\lambda\,,
\end{equation}
where $c_{-k}\in\Omega^{0,1}(\CPT,\cO(-k))$, with the corresponding integral formula being
\be\label{gammaint}
\gamma_{\alpha\beta}(x)=\int_{X}\D\lambda\wedge\lambda_{\alpha}\,\lambda_{\beta}\,\left.\tilde{b}\right|_{X}\,.
\ee
This is related to the more standard Penrose transform for the negative helicity zero-rest-mass field by using the freedom \eqref{btilfree} to fix the gauge
\be\label{btilgf}
\tilde{b}=\tilde{b}_{\dot\alpha}\,\d\mu^{\dot\alpha}\,,
\ee
upon which the integral formula \eqref{gammaint} becomes
\be\label{gammint2}
\gamma_{\alpha\beta\gamma\dot\gamma}(x)=\int_{X}\D\lambda\wedge\lambda_{\alpha}\,\lambda_{\beta}\,\lambda_{\gamma}\,\sH^{\dot\alpha}{}_{\dot\gamma}(x,\lambda)\,\left.\tilde{b}_{\dot\alpha}\right|_{X}\,,
\ee
and we can identify
\be\label{btilgf2}
\tilde{h}=\epsilon^{\dot\alpha\dot\beta}\,\frac{\partial\tilde{b}_{\dot\alpha}}{\partial\mu^{\dot\beta}}\,,
\ee
as the cohomology class in $H^{0,1}_{\nbar}(\CPT,\cO(-6))$.

\medskip

\paragraph{Positive-helicity gravitons:} In contrast to the negative helicity case, positive helicity gravitons can be described directly at the level of a metric perturbation. In particular, positive helicity gravitons on SDTN can be described in terms of a spin-raising operator acting on massless scalars, as any undotted spinor field on SDTN is covariantly constant (since the ASD spin connection vanishes). This is essentially a linearisation of Plebanski's construction of any SD vacuum metric from a scalar potential~\cite{Plebanski:1975wn}. For quasi-momentum eigenstates, the corresponding positive helicity metric perturbation is given by~\cite{Adamo:2023fbj}
\be\label{phelqme}
h^{(q)}_{\alpha\dot\alpha\beta\dot\beta}(x)=\beta_{\alpha}\,\beta_{\beta}\,\beta_{\gamma}\,\beta_{\delta}\,\nabla^{\gamma}{}_{\dot\alpha}\,\nabla^{\delta}_{\dot\beta}\phi^{(q)}(x)\,,
\ee
where $\beta_{\alpha}$ is an arbitrarily-chosen constant spinor amounting to a choice of lightfront gauge for the metric perturbation.

The Penrose transform states that these positive helicity gravitons should be described by cohomology classes $h\in H^{0,1}(\CPT,\cO(2))$. Given such an $h$ the field on SDTN is recovered by restriction to twistor curves:
\be\label{phelPT1}
h|_{X}=\dbar|_{X}j(x,\lambda)\,,
\ee
for some function $j$ of homogeneity $+2$ on $\P^1$, as $H^{0,1}(\P^1,\cO(2))$ is trivial. Holomorphicity of $h$ on $\CPT$ then implies that $\lambda^{\alpha}\nabla_{\alpha\dot\alpha}j$ is holomorphic on $\P^1$, which in turn means that
\be\label{phelPT2}
\lambda^{\alpha}\,\nabla_{\alpha\dot\alpha}j(x,\lambda)=\lambda^{\alpha}\,\lambda^{\beta}\,\lambda^{\gamma}\,\varphi_{\dot\alpha\alpha\beta\gamma}(x)\,,
\ee
for $\varphi_{\dot\alpha\alpha\beta\gamma}(x)$ a field on $\CM$ which is totally symmetric in its undotted spinor indices. This acts as a potential for a metric perturbation
\begin{equation}\label{phelPT3}
h_{\alpha\dal\beta\dot\beta}=\nabla^\gamma{}_{(\dal}\varphi_{\dot\beta)\alpha\beta\gamma}\,,
\end{equation}
which is easily seen to be self-dual, and hence positive helicity.

For the quasi-momentum eigenstates, one takes the twistor representative
\begin{equation}\label{phelrep}
    h^{(q)}=\int_{\C^*}\frac{\d s}{s^3}\,\bar\delta^{2}(s\lambda-\kappa)\,\left(s\mu^+\right)^{q+2M\omega}\left(s\mu^-\right)^{q-2M\omega}\,\exp\!\left(-\xi\,s^2\,\eta\right)\,.
\end{equation}
Following the procedure \eqref{phelPT1} -- \eqref{phelPT2} gives the potential
\be\label{phelrep2}
\varphi_{\dal\alpha\beta\gamma}^{(q)}(x)=\im\,\frac{\beta_\alpha\,\beta_\beta\,\beta_\gamma}{\langle \beta\,\kappa\rangle^3}\,\tilde{K}_\dal^{(q)}(x)\,\phi^{(q)}(x)\,,
\ee
and the corresponding metric perturbation
\be\label{phelrep3}
h^{(q)}_{\alpha\dot\alpha\beta\dot\beta}(x)=-\frac{\beta_{\alpha}\,\beta_{\beta}\,\beta_{\gamma}\,\beta_{\delta}}{\la\beta\,\kappa\ra^{4}}\nabla^{\gamma}{}_{\dot\alpha}\nabla^{\delta}_{\dot\beta}\phi^{(q)}(x)\,,
\ee
which is of the desired form.


\section{MHV scattering on self-dual Taub-NUT}\label{sec:mhv_amplitudes}

Equipped with the twistor description of SDTN and its gravitational perturbations, we now turn to dynamics; that is, the computation of tree-level graviton scattering amplitudes on the SDTN metric. Our particular focus will be on \emph{maximal helicity violating} (MHV) amplitudes, which involve two negative helicity and arbitrarily many positive helicity external gravitons. In a helicity grading of the tree-level S-matrix, this MHV configuration is the first non-trivial set of amplitudes as one moves away from the (classically integrable) self-dual sector of the theory. While computing even these amplitudes on a curved metric such as SDTN would be practically impossible with traditional background field methods, twistor theory enables us to obtain an explicit formula for the MHV amplitude at arbitrary multiplicity.

After a brief review of the generating functional of MHV amplitudes, we show that computing the MHV amplitudes boils down to computing connected, tree-level correlation functions in a certain 2d CFT (a \emph{twistor sigma model}) which constitutes a variational principle for the holomorphic curves in twistor space~\cite{Adamo:2021bej}. The computation of this correlator can then be operationalised using methods from algebraic combinatorics, leading to a remarkably compact formula for the MHV amplitudes, whose basic features are then analyzed.


\subsection{MHV generating functional and twistor sigma model}

A tree-level graviton MHV amplitude has two negative helicity external gravitons and arbitrarily many positive helicity gravitons. On any self-dual background metric, these positive helicity gravitons can be viewed equivalently as linear \emph{self-dual} perturbations to the background metric. Consequently, a generating functional for MHV amplitudes on any self-dual background is given by the classical two-point function for negative helicity gravitons on a generic self-dual background~\cite{Mason:2008jy,Adamo:2021bej,Adamo:2022mev}. The MHV amplitude is then obtained by expanding this generic self-dual background in terms of self-dual perturbations (the positive helicity gravitons) on the desired self-dual scattering background.

Identifying this generating functional is particularly simple in Plebanski's chiral formulation of general relativity~\cite{Plebanski:1977zz}. Here, the classical action is
\be\label{Plebact}
S[e,\Gamma]=\int_{M}\Sigma^{\alpha\beta}\wedge\left(\d\Gamma_{\alpha\beta}+\kappa^2\,\Gamma_{\alpha}{}^{\gamma}\wedge\Gamma_{\gamma\beta}\right)\,,
\ee
where $\kappa^2=16\pi G$, $\Sigma^{\alpha\beta}=e^{\alpha\dot\alpha}\wedge e^{\beta}{}_{\dot\alpha}$ is the triplet of ASD 2-forms associated to the vierbein $e^{\alpha\dot\alpha}$ on $M$ and $\Gamma_{\alpha\beta}$ is the ASD spin connection. Despite this action's apparent chirality (depending only on the ASD spin connection), its equations of motion are equivalent to the vacuum Einstein equations, and \eqref{Plebact} differs from the Einstein-Hilbert action only by a topological term~\cite{Plebanski:1977zz,Frauendiener:1990,Capovilla:1991qb,Krasnov:2009pu,Sharma:2021pkl}. One remarkable consequence of the Plebanski formulation is that it immediately makes it clear that general relativity admits a perturbative expansion around the self-dual sector: $\kappa^2\to0$ in \eqref{Plebact} yields an action whose classical equations of motion $\d\Sigma^{\alpha\beta}=0=\d\Gamma^{\alpha\beta}$ are equivalent to the vacuum self-duality equations.

This perspective also makes the form of the MHV generating functional manifest. This generating functional should be the portion of the classical action which is bi-linear in ASD perturbations on a non-linear SD background. If $\gamma_{1,2}^{\alpha\beta}$ are two perturbations to the ASD spin connection on $M$ -- now a SD, Ricci flat (i.e., hyperk\"ahler) manifold -- then the MHV generating functional is
\be\label{MHVgen1}
\mathcal{G}(1,2)=\int_{M}\Sigma^{\alpha\beta}\wedge\gamma_{1\,\alpha}{}^{\gamma}\wedge\gamma_{2\,\gamma\beta}\,,
\ee
where $\Sigma^{\alpha\beta}$ are the triplet of ASD 2-forms on $M$.

To recover a MHV amplitude from this generating functional, one writes the metric on $M$ as 
\be\label{MHVexpand1}
g_{M}=g_{\CM}+\sum_{i=3}^{n}\veps_i\,h_i\,,
\ee
where $g_{\CM}$ is the metric on the desired SD scattering background $\CM$, $\{h_i\}$ are positive helicity/SD gravitons on $\CM$ and $\{\veps_i\}$ are formal parameters. The $n$-point MHV amplitude is then given by
\be\label{MHVexpand2}
\cM^{\mathrm{MHV}}_{n}=\left.\frac{\partial^{n-2}\mathcal{G}(1,2)}{\partial\veps_3\cdots\partial\veps_n}\right|_{\veps_3=\cdots=\veps_n=0}\,,
\ee
that is, the piece of the generating functional which is multi-linear in the positive helicity gravitons on the scattering background. Note that because both $M$ and $\CM$ are vacuum SD, $\gamma_{1,2}^{\alpha\beta}$ represent negative helicity gravitons on \emph{both}.

While the generating functional \eqref{MHVgen1} gives a beautiful geometric interpretation of MHV scattering on any SD background, it is practically difficult to work with. For starters, the generating functional is not manifestly gauge invariant: the negative helicity particles enter at the level of their corresponding spin connection perturbations (i.e., as potentials rather than zero-rest-mass fields), whereas any resulting amplitude must be gauge invariant. How this gauge invariance for the amplitude emerges from the expansion of the generating functional is not obvious. Secondly -- and perhaps more importantly -- it is not at all clear how to operationalise the perturbative expansion of \eqref{MHVgen1} to obtain the MHV amplitude in practice. 

\medskip

Remarkably, these two problems can be solved simultaneously~\cite{Adamo:2021bej}. As we are interested in extracting scattering amplitudes, let us consider the negative helicity gravitons in \eqref{MHVgen1} to be quasi-momentum eigenstates of the form \eqref{eq:linearized_asd_spin_connection} with on-shell (undressed) 4-momenta and charges $k_1^{\alpha\dot\alpha}=\kappa_1^{\alpha}\tilde{\kappa}_1^{\dot\alpha}$, $q_1$ and $k_2^{\alpha\dot\alpha}=\kappa_2^{\alpha}\tilde{\kappa}_2^{\dot\alpha}$, $q_2$, respectively. Now, introduce a new set of coordinates $y^{\alpha\dot\alpha}=(y_1^{\dot\alpha},y_{2}^{\dot\alpha})$ on $M$ adapted to the spinor dyad $(\kappa_1^{\alpha},\kappa_2^{\alpha})$, defined by 
\be\label{ycoords}
y_i^\dal=\kappa_{i\,\alpha}\,x^{\alpha\dal}
-\im\, \frac{q_i+2M\omega_i}{\omega_i}\,\kappa_{i\,\alpha}\,T^{\alpha\dal}\log\la i\,\chi_+\ra-\im\, \frac{q_i-2M\omega_i}{\omega_i}\,\kappa_{i\,\alpha}T^{\alpha\dal}\log\la i\,\chi_-\ra\,,
\ee
for $i=1,2$ and $\la i\,a\ra:=\kappa_i^{\alpha}\,a_{\alpha}$ for any spinor $a_{\alpha}$. 

These coordinates have the two-fold advantage of locally solving for the closure of the dressing matrix and rendering the scalar quasi-momentum eigenstates of the two negative helicity fields as a pure exponential. In particular, one can show that
\be\label{dy}
\d y_{i}^{\dot\alpha}=\sG^{\dot\alpha}{}_{\dot\beta}(x;k_i,q_i)\,\kappa_{i\,\beta}\,e^{\beta\dot\beta}\,,
\ee
and
\be\label{y-qme}
\phi_i^{(q_i)}(y)=\e^{\im\,[y_i\,i]}\,,
\ee
abbreviating $[\tilde{a}\,i]:=\tilde{a}^{\dot\alpha}\,\tilde{\kappa}_{i\,\dot\alpha}$ for any spinor $\tilde{a}^{\dot\alpha}$. In particular, this means that the negative helicity perturbations to the ASD spin connection can be written as
\be\label{y-spinconn}
\gamma_{i}^{\alpha\beta}(y)=2\im\,\frac{[\tilde{a}_i\,\d y_i]}{[\tilde{a}_i\,i]}\,\kappa_i^{\alpha}\,\kappa_i^{\beta}\,\e^{\im\,[y_i\,i]}\,.
\ee
Furthermore, \eqref{dy} combined with the unimodularity property \eqref{dmatrix3} implies that 
\be\label{y-Sigma}
\d y_{i}^{\dot\alpha}\wedge \d y_{i\,\dot\alpha}=\kappa_{i\,\alpha}\,\kappa_{i\,\beta}\,\Sigma^{\alpha\beta}\,,
\ee
for each of $i=1,2$. The remaining projection of $\Sigma^{\alpha\beta}$ onto the spinor dyad can then be written as
\be\label{y-Sigma12}
\kappa_{1\,\alpha}\,\kappa_{2\,\beta}\,\Sigma^{\alpha\beta}=\d y_1^{\dot\alpha}\wedge\d y_2^{\dot\beta}\,\Omega_{\dot\alpha\dot\beta}(y)\,,
\ee
for some $\Omega_{\dot\alpha\dot\beta}$. Now, the hyperk\"ahler property of both $M$ and $\CM$ ensures that $\d\Sigma^{\alpha\beta}=0$, and thus there exists some scalar $\Omega(y)$ such that locally
\be\label{y-Potential}
\Omega_{\dot\alpha\dot\beta}=\la1\,2\ra\,\frac{\partial^2\Omega}{\partial y_1^{\dot\alpha}\,\partial y_2^{\dot\beta}}\,.
\ee
This $\Omega$ is precisely the first Plebanski scalar for the hyperk\"ahler structure~\cite{Plebanski:1975wn}.

Implementing this at the level of the generating functional \eqref{MHVgen1} gives
\be\label{MHVgen2}
\begin{split}
\mathcal{G}(1,2)&=-4\,\frac{\la 1\,2\ra}{[\tilde{a}_1\,1]\,[\tilde{a}_2\,2]}\int_{M}\kappa_{1\,\alpha}\,\kappa_{2\,\beta}\,\Sigma^{\alpha\beta}\wedge[\tilde{a}_1\,\d y_1]\wedge[\tilde{a}_2\,\d y_2]\,\e^{\im\,([y_1\,1]+[y_2\,2])}\\
&=-4\,\frac{\la 1\,2\ra^2}{[\tilde{a}_1\,1]\,[\tilde{a}_2\,2]}\int_{M}\d y_1^{\dot\alpha}\wedge\d y_2^{\dot\beta}\,\frac{\partial^2\Omega}{\partial y_1^{\dot\alpha}\,\partial y_2^{\dot\beta}}\,\wedge[\tilde{a}_1\,\d y_1]\wedge[\tilde{a}_2\,\d y_2]\,\e^{\im\,([y_1\,1]+[y_2\,2])} \\
&=-\frac{\la1\,2\ra^4}{[\tilde{a}_1\,1]\,[\tilde{a}_2\,2]}\int_{M}\d^{2}y_1\wedge\d^{2}y_2\,\tilde{a}_1^{\dot\alpha}\,\tilde{a}_2^{\dot\beta}\,\frac{\partial^2\Omega}{\partial y_1^{\dot\alpha}\,\partial y_2^{\dot\beta}}\,\e^{\im\,([y_1\,1]+[y_2\,2])}\,.
\end{split}
\ee
Now, one can integrate-by-parts twice (once with respect to $y_1^{\dot\alpha}$ and once with respect to $y_2^{\dot\beta}$) to obtain
\be\label{MHVgen3}
\mathcal{G}(1,2)=\la1\,2\ra^4\int_M\d^2y_1\wedge\d^2y_2\,\Omega\,\e^{\im\,([y_1\,1]+[y_2\,2])}\,,
\ee
with no boundary term contributions, since $\CM$, $M$ and the wavefunctions are asymptotically flat. In Appendix \ref{app:2-point}, we show that this formula reproduces our previous result for the 2-point amplitude \cite{Adamo:2023fbj}.

This expression \eqref{MHVgen3} for the generating functional is now manifestly gauge invariant: all dependence on the spinors $\tilde{a}_{1,2}^{\dot\alpha}$ has dropped out. This resolves the first difficulty associated with extracting a (gauge-invariant) scattering amplitude from the MHV generating functional but does not make the actual perturbative expansion in terms of positive helicity gravitons on $\CM$ appear any easier. At this point, the twistor description of the background plays a crucial role.

\medskip

As $M$ is simply a deformation of the SDTN space $\CM$ by self-dual radiative data (i.e., a collection of positive helicity gravitons), it also has a twistor description via the non-linear graviton theorem. Consequently, the twistor space of $M$ is described by a weighted Hamiltonian of the form 
\be\label{Mham}
\sh+\sum_{i=3}^{n}\veps_i\,h_i\equiv \sh+h\,,
\ee
where $\sh$ is the SDTN Hamiltonian \eqref{eq:hamiltonian} and each $h_i$ is a class in $H^{0,1}_{\nbar}(\CPT,\cO(2))$ of the form \eqref{phelrep}. Holomorphic curves corresponding to points in $M$ are then described by maps
\be\label{defholc1}
F^{\dot\alpha}(x,\lambda)=\sF^{\dot\alpha}(x,\lambda)+m^{\dot\alpha}(x,\lambda)\,,
\ee
where $\sF^{\dot\alpha}(x,\lambda)$ given by \eqref{eq:twistor_lines} describes the holomorphic twistor curves of SDTN. Continuing to denote holomorphic curves in the SDTN twistor space $\CPT$ by $X$, let $\mathcal{X}$ denote holomorphic curves in the twistor space of $M$. The deformed holomorphic curves, defined by \eqref{defholc1}, must then satisfy
\be\label{defholc2}
\dbar m^{\dot\alpha}=\left.\frac{\partial h}{\partial\mu_{\dot\alpha}}\right|_{\cX}+\left.\frac{\partial\sh}{\partial\mu_{\dot\alpha}}\right|_{\cX}-\left.\frac{\partial\sh}{\partial\mu_{\dot\alpha}}\right|_{X}\,,
\ee
where the $\dbar$-operator is understood to be the one along the curve and we have used the equation \eqref{eq:eq_for_twistor_lines} for the holomorphic curves in the twistor space of SDTN.

To ensure that the deformation $m^{\dot\alpha}(x,\lambda)$ does not introduce any new moduli (i.e., that the curves $\cX$ still form a 4-dimensional family), one must impose boundary conditions on $F^{\dot\alpha}$. We do this in a way which is compatible with the coordinates \eqref{ycoords}, setting
\be\label{ybcond}
F^{\dot\alpha}(x,\kappa_1)=\sF^{\dot\alpha}(x,\kappa_1)=y_1^{\dot\alpha}\,, \qquad F^{\dot\alpha}(x,\kappa_2)=\sF^{\dot\alpha}(x,\kappa_2)=y_2^{\dot\alpha}\,.
\ee
This is equivalent to saying that the deformation $m^{\dot\alpha}(x,\lambda)$ has zeros at $\lambda_{\alpha}=\kappa_{1,2\,\alpha}$, which removes any additional moduli associated with the deformation.

Now, the differential equation \eqref{defholc2} for the holomorphic curves in the deformed twistor space can be obtained as the Euler-Lagrange equations of the action~\cite{Adamo:2021bej,Adamo:2022mev}
\be\label{TSmodel1}
S[m]=\frac{1}{\hbar}\int_{\P^1}\frac{\D\lambda}{\la\lambda\,1\ra^2\,\la\lambda\,2\ra^2}\left([m\,\dbar m]+2\,h|_{\cX}+2\left[\sh|_{\cX}-\sh|_{X}-\left.\frac{\partial\sh}{\partial\mu^{\dot\alpha}}\right|_{X}\,m^{\dot\alpha}\right]\right)\,,
\ee
where $\hbar$ is a formal parameter and for any quantity $g(Z)$ on twistor space it is understood that
\be\label{curvepullbacks}
g|_{\cX}=g(\sF+m,\lambda)\,, \qquad g|_{X}=g(\sF,\lambda)\,.
\ee
Note that this action functional constitutes a well-posed variational problem thanks to the boundary conditions \eqref{ybcond}.

Now, by expanding all insertions of the SDTN Hamiltonian $\sh|_{\cX}$ in powers of $m^{\dot\alpha}$, the action can be written as
\be\label{TSmodel2}
S[m]=\frac{1}{\hbar}\int_{\P^1}\frac{\D\lambda}{\la\lambda\,1\ra^2\,\la\lambda\,2\ra^2}\left([m\,\dbar m]+2\,h|_{\cX}+\sum_{p=2}^{4}\frac{2}{p!}\left.\frac{\partial^{p}\sh}{\partial\mu^{\dot\alpha_1}\cdots\partial\mu^{\dot\alpha_p}}\right|_{X}\,m^{\dot\alpha_1}\cdots m^{\dot\alpha_p}\right)\,,
\ee
with the expansion of $\sh(\sF+m,\lambda)$ terminating at quartic order due to the explicit form of the SDTN Hamiltonian \eqref{eq:hamiltonian}.

This action defines a classical, chiral 2d conformal field theory (CFT) on the Riemann sphere with defects at $\lambda=\kappa_{1,2}$. It governs holomorphic rational maps to twistor space, and as such is referred to as a \emph{twistor sigma model}. The key fact about this twistor sigma model is that its on-shell value encodes the Plebanski scalar $\Omega$ of the hyperk\"ahler manifold $M$. More precisely~\cite{Adamo:2021bej,Adamo:2022mev,Sharma:2022arl}
\be\label{TSM-PS}
\Omega=\Omega_{\mathrm{SDTN}}-\left.\frac{\hbar}{4\pi\im}\,S[m]\right|_{\mathrm{on-shell}}\,,
\ee
where $\Omega_{\mathrm{SDTN}}$ is the Plebanski scalar of the SDTN metric and $S[m]|_{\mathrm{on-shell}}$ denotes the twistor sigma model action \eqref{TSmodel2} evaluated on solutions of its equations of motion \eqref{defholc2}.

\medskip

At first, this may seem like a mere curiosity, but in fact it resolves the issue of understanding how to perturbatively expand the generating functional \eqref{MHVgen3} to obtain an explicit formula for the tree-level MHV graviton amplitudes on the SDTN background. Indeed, \eqref{TSM-PS} means that this generating functional is equivalent to
\be\label{MHVgen4}
\mathcal{G}(1,2)=-\frac{\hbar\,\la1\,2\ra^4}{4\pi\im}\int_{M}\d^2 y_1\,\d^2 y_{2}\,\e^{\im\,([y_1\,1]+[y_2\,2])}\,S\big|_{\mathrm{on-shell}}\,,
\ee
and the perturbative expansion of $M$ in positive helicity gravitons on $\CM$ is now equivalent to classical expansion of the on-shell twistor sigma model action in the $\{h_i\}$ (i.e., the representatives of the positive helicity gravitons on $\CPT$). 

In other words, the $n$-point MHV amplitude is controlled by the piece of the on-shell twistor sigma model action which is of order $n-2$ in $h$ and linear in each of the $h_3,\ldots,h_n$. Now, multi-linear pieces of on-shell actions are computed by tree-level connected Feynman diagrams: in particular, the multi-linear piece of interest here is given by
\be\label{FeynCorr1}
\left.\frac{\partial^{n-2}S[m]|_{\mathrm{on-shell}}}{\partial\veps_3\cdots\partial\veps_n}\right|_{\veps_3=\cdots=\veps_n=0}=\left\la\prod_{i=3}^{n}V_i\right\ra^{\mathrm{conn.,\,tree}}_{\mathrm{SDTN}}\,,
\ee
where the quantity on the right-hand-side of this equation is the connected, tree-level (i.e., $O(\hbar^0)$ in this case) correlation function of vertex operators
\be\label{VO}
V_i=\int_{\P^1}\frac{\D\lambda_i}{\la1\,\lambda_i\ra^2\,\la2\,\lambda_i\ra^2}\,h_i(\sF+m,\lambda_i)\,,
\ee
in the 2d CFT 
\begin{multline}\label{SDTN-SM}
S_{\mathrm{SDTN}}[m]=\int_{\P^1}\frac{\D\lambda}{\la1\,\lambda\ra^2\,\la2\,\lambda\ra^2}\left[m^{\dot\alpha}\left(\epsilon_{\dot\beta\dot\alpha}\,\dbar+\frac{\partial^2 \sh}{\partial\mu^{\dot\alpha}\partial\mu^{\dot\beta}}(\sF,\lambda)\right)m^{\dot\beta}\right. \\
+\frac{\bar{e}^0}{24M}\left([\tilde{o}\,m]^2\,[\tilde{\iota}\,m]\,[\tilde{\iota}\,\sF]+[\tilde{\iota}\,m]^2\,[\tilde{o}\,m]\,[\tilde{o}\,\sF]+3\,[\tilde{\iota}\,m]^2\,[\tilde{o}\,m]^2\right)\Bigg]\,,
\end{multline}
on the Riemann sphere.

The terms in the second line of \eqref{SDTN-SM}, which can be thought of as explicit `background' terms associated with SDTN, mean that this 2d CFT is not free. However, as we are only interested in computing tree-level (or semi-classical) correlation functions \eqref{FeynCorr1}, these terms can be treated with perturbation theory:
\be\label{FeynCorr2}
\left\la\prod_{i=3}^{n}V_i\right\ra^{\mathrm{conn.,\,tree}}_{\mathrm{SDTN}}=\sum_{t=0}^{\infty}\sum_{p+q+r=t}\left\la\prod_{i=3}^{n}V_i\,\prod_{a=1}^{p}U_a^{+}\,\prod_{b=1}^{q}U_{b}^{-}\,\prod_{c=1}^{r}U_c^{0}\right\ra^{\mathrm{conn.,\,tree}}_{\mathrm{free}}\,,
\ee
where the correlator is evaluated in the free 2d CFT
\be\label{FreeTSM}
S_{\mathrm{free}}=\int_{\P^1}\frac{\D\lambda}{\la1\,\lambda\ra^2\,\la2\,\lambda\ra^2}\left[m^{\dot\alpha}\left(\epsilon_{\dot\beta\dot\alpha}\,\dbar+\frac{\partial^2 \sh}{\partial\mu^{\dot\alpha}\partial\mu^{\dot\beta}}(\sF,\lambda)\right)m^{\dot\beta}\right]\,,
\ee
and the effect of the `non-free' background terms from $S_{\mathrm{SDTN}}$ is encapsulated by insertion of the background vertex operators
\be\label{BVO+}
U_a^+=\frac{1}{24\,M}\int_{\P^1}\frac{\varpi_a}{\la1\,\lambda_a\ra^2\,\la2\,\lambda_a\ra^2}\,\sF^+(\lambda_a)\,[m\,\tilde{o}]\,[m\,\tilde{\iota}]^2\,,
\ee
\be\label{BVO-}
U_a^-=\frac{1}{24\,M}\int_{\P^1}\frac{\varpi_a}{\la1\,\lambda_a\ra^2\,\la2\,\lambda_a\ra^2}\,\sF^-(\lambda_a)\,[m\,\tilde{\iota}]\,[m\,\tilde{o}]^2\,,
\ee
and
\be\label{BVO0}
U_a^0=\frac{1}{8\,M}\int_{\P^1}\frac{\varpi_a}{\la1\,\lambda_a\ra^2\,\la2\,\lambda_a\ra^2}\,[m\,\tilde{o}]^2\,[m\,\tilde{\iota}]^2\,,
\ee
for
\be\label{spherekahler}
\varpi:=\frac{\D\lambda\wedge\D\hat{\lambda}}{\la\lambda\,\hat{\lambda}\ra^2}\,,
\ee
the K\"ahler form on $\P^1$. 

Thus, the challenge of computing the $n$-point MHV graviton amplitude on SDTN is reduced to the task of computing the connected tree-level correlation functions \eqref{FeynCorr2} in the free, classical CFT on $\P^1$ defined by \eqref{FreeTSM}. In particular, the $n$-point MHV amplitude is now given by
\be\label{MHV-preamp1}
\cM_{n}=\la1\,2\ra^6 \int_{\CM}\d^{4}x\,\sqrt{|g|}\,\phi_{1}^{(q_1)}\,\phi_{2}^{(q_2)}\,\left\la\prod_{i=3}^{n}V_i\right\ra^{\mathrm{conn.,\,tree}}_{\mathrm{SDTN}}\,,
\ee
having now converted the integrand back to Gibbons-Hawking coordinates, which produces an additional factor of $\la1\,2\ra^2$ along with the square root of the metric determinant $|g|=(1+2M/r)^2$.


\subsection{Computing the twistor sigma model correlator}

At this point, the computation of the graviton MHV amplitude on SDTN -- an apparently difficult problem in perturbative gravity on a curved manifold -- has been translated into the computation of a (set of) connected, tree-level correlation function in a free, chiral CFT on the Riemann sphere. This new problem can be approached using fairly standard methods in 2d CFT and graph theory.

To begin, let us first consider the $t=0$ term in the correlator \eqref{FeynCorr2}:
\be\label{notailCorr1}
\mathcal{C}_{n}[0,0,0]:=\left\la\prod_{i=3}^{n}V_i\right\ra^{\mathrm{conn.,\,tree}}_{\mathrm{free}}\,.
\ee
This correlator is computed by summing all connected, tree-level Feynman diagrams on the insertions of the vertex operators $\{V_i\}$ in the free CFT \eqref{FreeTSM}. The only quantum field in this theory is $m^{\dot\alpha}(\lambda)$, whose propagator is given by inverting the differential operator $\nbar|_{X}$ with appropriate weighting factors:
\be\label{mprop}
\la m^{\dot\alpha}(\lambda)\,m^{\dot\beta}(\lambda')\ra=\frac{\sH^{\dot\alpha}{}_{\dot\gamma}(\lambda)\,\sH^{\dot\beta\dot\gamma}(\lambda')}{\la\lambda\,\lambda'\ra}\,\la1\lambda\ra\,\la2\lambda\ra\,\la1\,\lambda'\ra\,\la2\,\lambda'\ra\,,
\ee
where all dependence on $x$ (the choice of twistor curve) has been suppressed and $\sH^{\dot\alpha}{}_{\dot\beta}$ is the matrix \eqref{eq:dotted_frame} defining the holomorphic frame for the twisted normal bundle of the twistor curves in the twistor space of SDTN. That \eqref{mprop} is indeed the Green's function for $\nbar|_X$ follows from \eqref{framedef}, which implies that
\be\label{mprop2}
\dbar|_{X}H^{\dot\alpha}{}_{\dot\beta}(x,\lambda)=\left.\frac{\partial^2\sh}{\partial\mu^{\dot\gamma}\partial\mu_{\dot\alpha}}\right|_{X}\,H^{\dot\gamma}{}_{\dot\beta}(x,\lambda)\,,
\ee
on any twistor curve.

Thus, all of the Feynman diagrams contributing to the correlator \eqref{notailCorr1} are spanning tree graphs on $n-2$ vertices, where the vertices are the vertex operators $\{V_i\}$ and the edges of the graph correspond to single Wick contractions, via the propagator \eqref{mprop}, between these vertex operators. Now, such a Wick contraction will contribute to any Feynman diagram in which the corresponding edge appears as
\be\label{VOWick1}
\left\la V_i\,V_j\right\ra=\int\limits_{(\P^1)^2}\frac{\D\lambda_i\,\D\lambda_j}{\la1\,\lambda_i\ra\,\la2\,\lambda_i\ra\,\la1\,\lambda_j\ra\,\la2\,\lambda_j\ra}\,\frac{\sH^{\dot\alpha}{}_{\dot\gamma}(\lambda_i)\,\sH^{\dot\beta\dot\gamma}(\lambda_j)}{\la\lambda_i\,\lambda_j\ra}\,\frac{\partial h_{i}}{\partial \mu^{\dot\alpha}}(\lambda_i)\,\frac{\partial h_{j}}{\partial\mu^{\dot\beta}}(\lambda_j)\,,
\ee
where $h_{i}\equiv h_i^{(q_i)}$ is the twistor representative \eqref{phelrep} for the quasi-momentum eigenstate wavefunction of the $i^{\mathrm{th}}$ positive helicity graviton.

It seems that the partial derivatives of the twistor wavefunctions in this Wick contraction will be a mess: the wavefunctions \eqref{phelrep} have both polynomial and exponential dependence on $\mu^{\dot\alpha}$. However, the relation \eqref{frameident2} linking $\sH^{\dot\alpha}_{\dot\beta}$ to the dressing matrix $\sG^{\dot\alpha\dot\beta}$ \eqref{dmatrix2} now comes to the rescue; adapted to the graviton wavefunctions, this gives
\be\label{frameident3}
\sH^{\dot\alpha}{}_{\dot\gamma}(\lambda_i)\,\lambda_i^{\alpha}\,\frac{\partial h_i(\lambda_i)}{\partial\mu^{\dot\alpha}}=\im\,\kappa_i^{\alpha}\,\tilde{K}_{i\,\dot\gamma}\,h_i(\lambda_i)\,,
\ee
where $\tilde{K}_{i\,\dot\alpha}\equiv\tilde{K}_{i\,\dot\alpha}^{(q_i)}$ is the dressed dotted momentum spinor \eqref{dmatrix1}.

Combining \eqref{frameident3} with the fact that, on the support of the holomorphic delta functions inside $h_i$, one can identify $s_i\,\lambda^{\alpha}_i=\kappa_i^{\alpha}$, it follows that the Wick contraction \eqref{VOWick1} actually takes a remarkably simple form:
\be\label{VOWick2}
\left\la V_i\,V_j\right\ra=-\int\limits_{(\P^1)^2}\frac{\D\lambda_i\,\D\lambda_j}{\la1\,\lambda_i\ra\,\la2\,\lambda_i\ra\,\la1\,\lambda_j\ra\,\la2\,\lambda_j\ra}\,\frac{s_i\,s_j\,[\![i\,j]\!]}{\la\lambda_i\,\lambda_j\ra}\,h_i(\lambda_i)\,h_{j}(\lambda_j)\,,
\ee
where we have adopted the notation
\be\label{dressedsqr}
[\![i\,j]\!]:=\tilde{K}_i^{\dot\alpha}\,\tilde{K}_{j\,\dot\alpha}\,,
\ee
for the contraction of dressed, dotted momentum spinors. In \eqref{VOWick2}, we have abused notation by writing explicit powers of the scaling parameters $s_i,s_j$, which are integrated over inside of $h_i,h_j$. What is meant by this is that one multiplies the measure inside $h_i$ by $s_i$ prior to integration over $\C^*$; this definition is unambiguous and saves having to rewrite all powers of $s_i$ and its measure in $h_i$ every time a Wick contraction is taken.

\begin{figure}[t]
\centering
\includegraphics[scale=.85]{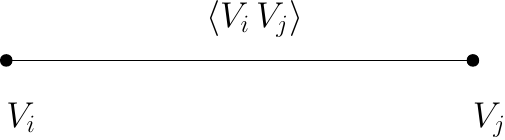}
\caption{The basic building block of the graph theory problem: vertices corresponding to external gravitons and edges given by Wick contractions between them.}
\label{Graph1}
\end{figure}

This means that we are left with a counting problem: sum over all connected, tree-level graphs on the $n-2$ vertices, weighted by the contributions of the Wick contractions corresponding to each edge -- see Figure~\ref{Graph1}. Fortunately, this weighted sum over spanning tree graphs is performed by a well-known theorem in algebraic combinatorics -- the (weighted) matrix tree theorem (cf., \cite{Stanley:1999,vanLint:2001,Stanley:2012} for textbook treatments). 

In particular, let $W$ be the weighted Laplacian matrix associated to the totally connected graph on all of the positive helicity graviton vertex operators. Using \eqref{VOWick2}, this is the $(n-2)\times(n-2)$ matrix with entries
\be\label{wLap1}
\begin{split}
W_{ij}&=-s_i\,s_j\,\frac{[\![i\,j]\!]}{\la\lambda_i\,\lambda_j\ra}\,\la1\,\lambda_i\ra\,\la2\,\lambda_i\ra\,\la1\,\lambda_j\ra\,\la2\,\lambda_j\ra\,, \quad i\neq j\,, \\
W_{ii}&=-\sum_{j\neq i}W_{ij}\,.
\end{split}
\ee
By definition, this matrix has co-rank one, and the matrix tree theorem states that the weighted sum over all spanning tree graphs on the set of graviton vertex operators is given by taking the determinant of the once-reduced minor:
\be\label{wLap2}
\left|W^{i}_{i}\right|\,,
\ee
where $i$ corresponds to one of the positive helicity graviton insertions (one of $3,\ldots,n$) and $W^{i}_{i}$ denotes the weighted Laplacian matrix with row and column $i$ removed. Remarkably, one can show that the value of the determinant \eqref{wLap2} is \emph{independent} of which $i$ is chosen to define the minor.

\medskip

Assembling all of the pieces, this leaves an expression for the correlator \eqref{notailCorr1} we initially set out to compute:
\be\label{notailCorr2}
\mathcal{C}_{n}[0,0,0]=\int\limits_{(\P^1)^{n-2}}\left|W^{i}_{i}\right|\prod_{j=3}^{n}\frac{\D\lambda_j}{\la1\,\lambda_j\ra^2\,\la2\,\lambda_j\ra^2}\,h_{j}(\lambda_i)\,.
\ee
All of the scale integrals over the $\{s_j\}$ and all of the $\P^1$ integrals over the $\{\lambda_j\}$ can now be performed against the holomorphic delta functions in the twistor wavefunctions, leaving
\be\label{notailCorr3}
\mathcal{C}_{n}[0,0,0]=\frac{|\HH^{i}_{i}|}{\la1\,i\ra^2\,\la2\,i\ra^2}\,\prod_{j=3}^{n}\phi^{(q_j)}_{j}(x)\,,
\ee
where $\HH$ is the $(n-2)\times(n-2)$ matrix with entries
\be\label{Hodges1}
\HH_{jk}=-\frac{[\![j\,k]\!]}{\la j\,k\ra}\,,
\ee
\begin{equation*}
   \HH_{jj}=\sum_{k\neq j}\frac{[\![j\,k]\!]}{\la j\,k\ra}\,\frac{\la1\,k\ra\,\la2\,k\ra}{\la1\,j\ra\,\la2\,j\ra}\,.
\end{equation*}
Note that this matrix $\HH$ is related to the weighted Laplacian by removing a factor of $\la1\,j\ra\,\la2\,j\ra$ from the $j^{\mathrm{th}}$ row and column of the Laplacian matrix, for each $j$, after performing all integrals in \eqref{notailCorr2}.

\medskip

\begin{figure}[t]
\centering
\includegraphics[scale=.85]{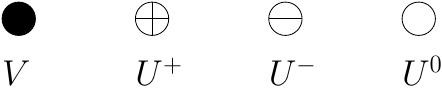}
\caption{The different kinds of vertices contributing to the Feynman graphs associated to $\mathcal{C}_n[p,q,r]$.}
\label{Vertices}
\end{figure}

Of course, this only gives the $t=0$ term in the general correlator \eqref{FeynCorr2} that we need to compute in order to determine the MHV amplitude on SDTN:
\be\label{FeynCorr3}
\begin{split}
\left\la\prod_{i=3}^{n}V_i\right\ra^{\mathrm{conn.,\,tree}}_{\mathrm{SDTN}}&=\sum_{t=0}^{\infty}\sum_{p+q+r=t}\left\la\prod_{i=3}^{n}V_i\,\prod_{a=1}^{p}U_a^{+}\,\prod_{b=1}^{q}U_{b}^{-}\,\prod_{c=1}^{r}U_c^{0}\right\ra^{\mathrm{conn.,\,tree}}_{\mathrm{free}}\\
&:=\sum_{t=0}^{\infty}\sum_{p+q+r=t}\mathcal{C}_{n}[p,q,r]\,.
\end{split}
\ee
Computing a given term in this sum can again be reduced to the problem of summing over weighted spanning tree graphs, now on an enhanced set of vertices which includes not only the $n-2$ graviton vertex operators but also $p$ of the background vertex operators $U^+$, $q$ of the $U^-$ and $r$ of the $U^0$ -- see Figure~\ref{Vertices}.

At first, it might seem that this problem is no worse than the one we solved with \eqref{notailCorr3}: one simply writes the weighted Laplacian matrix for the given set of vertices and computes the suitable reduced minor. The only difference is that this is now a $(n+t-2)\times(n+t-2)$ matrix, with new entries corresponding to Wick contractions between the graviton vertex operators and the background vertex operators as well as between the background vertex operators themselves.


However, a quick inspection reveals that the remaining $t>0$ terms involve an additional subtlety. For concreteness, let us consider one of the three $t=1$ contributions to \eqref{FeynCorr3}:
\be\label{tailCorr1*}
\mathcal{C}_{n}[1,0,0]=\left\la\prod_{i=3}^{n}V_i\right\ra^{\mathrm{conn.,\,tree}}_{\mathrm{SDTN}}=\left\la U^+\,\prod_{i=3}^{n}V_i\right\ra^{\mathrm{conn.,\,tree}}_{\mathrm{free}}\,.
\ee
Naively, we can proceed to evaluate this correlator by again applying the matrix tree theorem, now to the collection of $n-2$ external graviton vertex operators along with the background vertex operator $U^+$. The only new ingredient required at this point is the Wick contraction between a graviton vertex operator and $U^+$:
\begin{multline}\label{gravU+}
\left\la V_i\,U^+\right\ra=\frac{\im}{24\,M}\int_{(\P^1)^2}\frac{\D\lambda_i\,\varpi}{\la1\,\lambda_i\ra\,\la2\,\lambda_i\ra\,\la1\,\lambda\ra\,\la2\,\lambda\ra}\frac{s_i\,\tilde{K}_i^{\dot\alpha}\,\sH^{\dot\beta}{}_{\dot\alpha}(\lambda)}{\la\lambda_i\,\lambda\ra} \\
\times\,\left(\tilde{o}_{\dot\beta}\,[m\,\tilde{\iota}]^2+2\tilde{\iota}_{\dot\beta}\,[m\,\tilde{o}]\,[m\,\tilde{\iota}]\right)\,h_i(\lambda_i)\,\sF^{+}(\lambda)\,,
\end{multline}
where we have used \eqref{mprop} and \eqref{frameident3}, and
\be\label{U^+Kahler}
\varpi=\frac{\D\lambda\wedge\D\hat{\lambda}}{\la\lambda\,\hat{\lambda}\ra^2}\,,
\ee
is the integration measure on the copy of $\P^1$ associated with the background vertex operator. In the second line, the first term corresponds to Wick contraction between the graviton vertex operator and the (linear) dependence of $U^+$ on $[m\,\tilde{o}]=m^+$, while the second term is given by Wick contraction into the quadratic dependence of $U^+$ on $[m\,\tilde{\iota}]=m^-$.

Now it becomes clear why a naive application of the matrix tree theorem is insufficient in this case. Since $m^{\dot\alpha}$ is a quantum mechanical field with no zero mode insertions, all of its insertions in the correlator \eqref{tailCorr1*} must be Wick contracted away. From \eqref{gravU+} we see that a single Wick contraction between $U^+$ and a graviton vertex operator leaves an object which is still a quadratic function of $m^{\dot\alpha}$. To give a non-vanishing correlator, these must be removed by two additional Wick contractions; to ensure that these arise from a tree graph the additional contractions must come from two other graviton vertex operators. Such a contribution corresponds to:
\begin{multline}\label{gravU+3}
\left\la (V_i\,V_j\,V_k)\rightarrow U^+\right\ra=\frac{-\im}{12\,M}\int_{(\P^1)^4}\frac{\D\lambda_i\,\D\lambda_j\,\D\lambda_k\,\varpi\,\la1\,\lambda\ra\,\la2\,\lambda\ra}{\la1\,\lambda_i\ra\,\la2\,\lambda_i\ra\cdots\la1\,\lambda_k\ra\,\la2\,\lambda_k\ra}\,\frac{s_i\,s_j\,s_k\,\tilde{K}_i^{\dot\alpha}\,\tilde{K}_j^{\dot\beta}\,\tilde{K}_k^{\dot\gamma}}{\la\lambda_i\,\lambda\ra\,\la\lambda_j\,\lambda\ra\,\la\lambda_k\,\lambda\ra} \\
\times\,\sH^{\dot\delta}{}_{\dot\alpha}(\lambda)\,\sH^{\dot\rho}{}_{\dot\beta}(\lambda)\,\sH^{\dot\sigma}{}_{\dot\gamma}(\lambda)\left(\tilde{o}_{\dot\delta}\,\tilde{\iota}_{\dot\rho}\,\tilde{\iota}_{\dot\sigma}+\tilde{\iota}_{\dot\delta}\,\tilde{o}_{\dot\rho}\,\tilde{\iota}_{\dot\sigma}+\tilde{\iota}_{\dot\delta}\,\tilde{\iota}_{\dot\rho}\,\tilde{o}_{\dot\sigma}\right)\,h_{i}(\lambda_i)\,h_{j}(\lambda_j)\,h_{k}(\lambda_k)\,\sF^{+}(\lambda)\,.
\end{multline}
At this point, the $\P^1$ integrals corresponding to the graviton vertex operators can be performed explicitly against the holomorphic delta functions, leaving only a single $\P^1$ integral associated with the background vertex operator. To incorporate contributions of the form \eqref{gravU+3} into the sum over tree diagrams via the matrix tree theorem, the weighted Laplacian matrix must be modified to ensure that tree graphs with such contributions -- and \emph{only} these graphs -- are included.

This can be done by exploiting the multi-linearity of determinants. Consider the following $(n-1)\times(n-1)$ weighted Laplacian matrix:
\be\label{tailCorr2}
\cH[1,0,0]=\left(\begin{array}{cc}
                 \HH & \mathfrak{h} \\
                 \mathfrak{h}^{\mathrm{T}} & \mathbb{T}
                 \end{array}\right)\,,
\ee
where the $(n-2)\times(n-2)$ block $\HH$ is the same as that given by \eqref{Hodges1} for the $t=0$ correlator except for a modification of the diagonal entries:
\be\label{Hodges1*}
\HH_{jj}\to \HH_{jj}-\im\,\frac{\tilde{K}^{\dot\alpha}_{j}\,\sH^{\dot\beta}{}_{\dot\alpha}(\lambda)}{\la j\,\lambda\ra}\left(\veps^+\,\tilde{o}_{\dot\beta}+\veps^{-}\,\tilde{\iota}_{\dot\beta}\right)\frac{\la 1\,\lambda\ra\,\la2\,\lambda\ra}{\la1\,j\ra\,\la2\,j\ra}\,,
\ee
while the $(n-2)$-component vector $\mathfrak{h}$ has entries
\be\label{frakh1*}
\mathfrak{h}_j=\im\,\frac{\tilde{K}^{\dot\alpha}_{j}\,\sH^{\dot\beta}{}_{\dot\alpha}(\lambda)}{\la j\,\lambda\ra}\left(\veps^+\,\tilde{o}_{\dot\beta}+\veps^{-}\,\tilde{\iota}_{\dot\beta}\right)\la1\,\lambda\ra\,\la2\,\lambda\ra\,,
\ee
and
\be\label{TT1*}
\T=-\im\sum_{j=3}^{n} \frac{\tilde{K}^{\dot\alpha}_{j}\,\sH^{\dot\beta}{}_{\dot\alpha}(\lambda)}{\la j\,\lambda\ra}\left(\veps^+\,\tilde{o}_{\dot\beta}+\veps^{-}\,\tilde{\iota}_{\dot\beta}\right)\frac{\la1\,\lambda\ra\,\la2\,\lambda\ra}{\la1\,j\ra\,\la2\,j\ra}\,.
\ee
The role of the formal parameters $\veps^+,\,\veps^-$ will soon become apparent, but they can be assigned scaling weight $-1$ with respect to the homogeneous coordinate $\lambda_{\alpha}$, rendering the matrix entries $\mathfrak{h}_j$ and $\T$ weightless under projective rescalings of $\lambda_{\alpha}$.

From this object, consider the quantity:
\be\label{tailCorr3}
\frac{1}{24\,M}\int_{\P^1}\frac{\varpi}{\la1\,\lambda\ra^2\,\la2\,\lambda\ra^2}\,\frac{\sF^{+}(\lambda)}{\veps^+\,(\veps^-)^2}\,\frac{\left|\cH[1,0,0]^{i}_{i}\right|}{\la1\,i\ra^2\,\la2\,i\ra^2}\,\prod_{j=3}^{n}\phi_{j}^{(q_j)}(x)\,,
\ee
where $|\cH[1,0,0]^{i}_{i}|$ is the once-reduced determinant of the weighted Laplacian matrix \eqref{tailCorr2}. It can easily be checked that this formula is mathematically well-defined, in the sense that the integral over $\P^1$ makes sense, with the integrand having projective scaling weight zero. Now, basic properties of determinants ensure that $|\cH[1,0,0]^{i}_{i}|$ is a polynomial in the formal parameters $\veps^{\pm}$, whose highest-order terms are of the form $(\veps^+)^a\,(\veps^-)^b$ for $a+b=n-2$.

By the matrix tree theorem, this determinant gives -- up to factors which are compensated for by the remaining ingredients in \eqref{tailCorr3} -- the sum over all tree diagrams involving the $n-2$ graviton vertex operators and a `fictional' version of $U^+$ which can absorb arbitrarily many Wick contractions. The placement of the parameters $\veps^{\pm}$ in the matrix $\cH[1,0,0]$ ensures that the order $(\veps^+)^a\,(\veps^-)^b$ term in the reduced determinant corresponds to diagrams with $a+b$ graviton vertex operators Wick contracting into $U^+$, $a$ of which contract into an insertion of $m^+$ in the background vertex operator, with the remaining $b$ contracting into an insertion of $m^-$.

\begin{figure}[t]
\centering
\includegraphics[scale=.85]{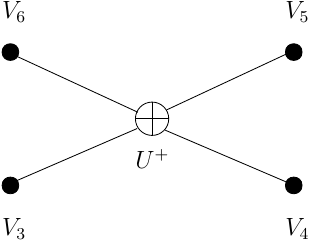}
\caption{An example of a graph whose contribution to $\mathcal{C}_6[1,0,0]$ vanishes, but which would be counted by na\"ive application of the matrix tree theorem.}
\label{Graph2}
\end{figure}

It is then clear that $|\cH[1,0,0]^{i}_{i}|$ over-counts the correct number of tree diagrams contributing to the correlator $\mathcal{C}_n[1,0,0]$. As we already observed, $U^+$ must absorb \emph{exactly} one Wick contraction into $m^+$ and two Wick contractions into $m^{-}$ -- any other number results in a vanishing contribution to the correlator. Consequently, only the terms proportional to $\veps^+\,(\veps^-)^2$ in $|\cH[1,0,0]^{i}_{i}|$ correspond to admissible contributions to the correlator, whereas the full object \eqref{tailCorr3} includes many contributions of higher (as well as lower) order which should not contribute -- see Figure~\ref{Graph2} for an example. 

Indeed, upon collecting all the terms of order $\veps^+\,(\veps^-)^2$ from $|\cH[1,0,0]^{i}_{i}|$, one finds that their contribution to \eqref{tailCorr3} is
\begin{multline}\label{tailCorr4}
\frac{-\im}{12\,M}\int_{\P^1}\varpi\,\la1\,\lambda\ra\,\la2\,\lambda\ra\,\sF^{+}(\lambda)\,\left.\frac{\left|\HH^{i}_{i}\right|}{\la1\,i\ra^2\,\la2\,i\ra^2}\right|_{\veps^+=\veps^-=0}\,\,\prod_{m=3}^{n}\phi_{m}^{(q_m)}(x) \\
\times\,\sum_{j\neq k\neq l}\frac{\tilde{K}^{\dot\alpha}_j\,\tilde{K}^{\dot\beta}_{k}\,\tilde{K}_{l}^{\dot\gamma}}{\la j\,\lambda\ra\,\la k\,\lambda\ra\,\la l\,\lambda\ra}\,\sH^{\dot\delta}{}_{\dot\alpha}(\lambda)\,\sH^{\dot\rho}{}_{\dot\beta}(\lambda)\,\sH^{\dot\sigma}{}_{\dot\gamma}(\lambda)\left(\tilde{o}_{\dot\delta}\,\tilde{\iota}_{\dot\rho}\,\tilde{\iota}_{\dot\sigma}+\tilde{\iota}_{\dot\delta}\,\tilde{o}_{\dot\rho}\,\tilde{\iota}_{\dot\sigma}+\tilde{\iota}_{\dot\delta}\,\tilde{\iota}_{\dot\rho}\,\tilde{o}_{\dot\sigma}\right)\,,
\end{multline}
corresponding precisely to the desired diagrams of the form \eqref{gravU+3}. In other words, the $O(1)$ terms in \eqref{tailCorr3} give the correct answer for $\mathcal{C}_{n}[1,0,0]$, with all other terms corresponding to tree diagrams which should actually give a vanishing contribution to the correlator (due to over- or under-saturating the background vertex operator with Wick contractions).

This correct contribution can be isolated directly by simply taking an appropriate number of derivatives of the determinant $|\cH[1,0,0]^{i}_{i}|$ with respect to $\veps^{\pm}$ and then setting these parameters to zero. In particular, it follows that
\be\label{tailCorr5}
\mathcal{C}_n[1,0,0]=\int_{\P^1}\frac{\varpi}{\la1\,\lambda\ra^2\,\la2\,\lambda\ra^2}\,\frac{\sF^{+}(\lambda)}{24\,M}\,\frac{\partial}{\partial\veps^+}\,\frac{\partial^2}{\partial\veps^{-\,2}}\left.\frac{\left|\cH[1,0,0]^{i}_{i}\right|}{\la1\,i\ra^2\,\la2\,i\ra^2}\right|_{\veps^{\pm}=0}\,\prod_{j=3}^{n}\phi_{j}^{(q_j)}(x)\,,
\ee
with the derivatives with respect to $\veps^{\pm}$ now ensuring that the correct contribution, corresponding to a single insertion of $U^+$, is extracted.

\medskip

This strategy can now be applied to evaluate all other correlators $\mathcal{C}_n[p,q,r]$ contributing to the MHV amplitude. A general $(n+t-2)\times (n+t-2)$ weighted Laplacian matrix $\cH[t]$ can be defined for any $t\geq0$: 
\be\label{genHmat1}
\cH[t]=\left(\begin{array}{cc}
                      \HH & \mathfrak{h} \\
                      \mathfrak{h}^{\mathrm{T}} & \T
                      \end{array}\right)\,,
\ee
with $\HH$, $\mathfrak{h}$ and $\T$ being $(n-2)\times(n-2)$, $(n-2)\times t$ and $t\times t$ matrices, respectively. Their entries are given by
\be\label{HH1}
\HH_{jk}=-\frac{[\![j\,k]\!]}{\la j\,k\ra}\,,
\ee
\begin{equation*}
   \HH_{jj}=\sum_{k\neq j}\frac{[\![j\,k]\!]}{\la j\,k\ra}\,\frac{\la1\,k\ra\,\la2\,k\ra}{\la1\,j\ra\,\la2\,j\ra}-\im\sum_{\m=1}^{t}\frac{\tilde{K}^{\dot\alpha}_{j}\,\sH^{\dot\beta}{}_{\dot\alpha}(\lambda_{\m})}{\la j\,\lambda_\m\ra}\left(\veps_{\m}^+\,\tilde{o}_{\dot\beta}+\veps_\m^{-}\,\tilde{\iota}_{\dot\beta}\right)\frac{\la 1\,\lambda_\m\ra\,\la2\,\lambda_\m\ra}{\la1\,j\ra\,\la2\,j\ra}\,,
\end{equation*}
for $j,k=3,\ldots,n$,
\be\label{h1}
\mathfrak{h}_{j\m}=\im\,\frac{\tilde{K}^{\dot\alpha}_{j}\,\sH^{\dot\beta}{}_{\dot\alpha}(\lambda_{\m})}{\la j\,\lambda_\m\ra}\left(\veps_{\m}^+\,\tilde{o}_{\dot\beta}+\veps_\m^{-}\,\tilde{\iota}_{\dot\beta}\right)\la1\,\lambda_\m\ra\,\la2\,\lambda_m\ra\,,
\ee
and
\be\label{TT1}
\T_{\mathrm{mn}}=\frac{\sH^{\dot\alpha\dot\gamma}(\lambda_\m)\,\sH^{\dot\beta}{}_{\dot\gamma}(\lambda_{\mathrm{n}})}{\la\lambda_\m\,\lambda_{\mathrm{n}}\ra} \left(\veps_{\m}^+\,\tilde{o}_{\dot\alpha}+\veps_\m^{-}\,\tilde{\iota}_{\dot\alpha}\right)\left(\veps_{\mathrm{n}}^+\,\tilde{o}_{\dot\beta}+\veps_{\mathrm{n}}^{-}\,\tilde{\iota}_{\dot\beta}\right)\,\la1\,\lambda_\m\ra\,\la2\,\lambda_\m\ra\,\la1\,\lambda_{\mathrm{n}}\ra\,\la2\,\lambda_{\mathrm{n}}\ra
\ee
\begin{equation*}
\T_{\m\m}=-\sum_{\mathrm{n}\neq\m}\T_{\mathrm{mn}}-\im\sum_{j=3}^{n} \frac{\tilde{K}^{\dot\alpha}_{j}\,\sH^{\dot\beta}{}_{\dot\alpha}(\lambda_{\m})}{\la j\,\lambda_\m\ra}\left(\veps_{\m}^+\,\tilde{o}_{\dot\beta}+\veps_\m^{-}\,\tilde{\iota}_{\dot\beta}\right)\la1\,\lambda_\m\ra\,\la2\,\lambda_m\ra\,\la1\,j\ra\,\la2\,j\ra\,,
\end{equation*}
for $\mathrm{m,n}=1,\ldots,t$. 

The role of the formal parameters $\{\veps^{\pm}_{1},\ldots,\veps^{\pm}_{t}\}$ is the same as in the example of $\mathcal{C}_n[1,0,0]$ studied above: $\veps_{\m}^+$ accompanies a Wick contraction into an insertion of $m^+$ background vertex operator $\m$, while $\veps_\m^-$ accompanies a Wick contraction into an insertion of $m^-$. When $t\geq 2$ these Wick contractions can occur between two background vertex operators, as well as between graviton vertex operators and background vertex operators; the new contractions are encoded in the $\T$ block of $\cH[t]$. By the matrix tree theorem, to obtain the correct contribution to $\mathcal{C}_n[p,q,r]$, one must extract the portion of $|\cH[t]^i_i|$ which is proportional to
\be\label{tailCorr6}
\prod_{a=1}^{p}\veps^+_a\,(\veps^{-}_a)^2\,\prod_{b=1}^{q}(\veps_b^+)^2\,\veps_b^-\,\prod_{c=1}^{r}(\veps_c^+)^2\,(\veps_c^-)^2\,.
\ee
This is done by acting with an appropriate set of differential operators in the formal parameters and then evaluating the remaining object with $\veps_1^{\pm}=\cdots\veps_t^{\pm}=0$.

Consequently, one arrives at the expression for a generic term in the sum of correlators \eqref{FeynCorr3}:
\begin{multline}\label{tailCorr1}
\mathcal{C}_n[p,q,r]=\int_{(\P^1)^t}\prod_{\m=1}^{t}\frac{\varpi_\m}{\la1\,\lambda_\m\ra^2\,\la2\,\lambda_\m\ra^2}\prod_{a=1}^{p}\frac{\sF^+(\lambda_a)}{24\,M}\,\frac{\partial}{\partial\veps_a^+}\,\frac{\partial^2}{\partial\veps_a^{-\,2}} \prod_{b=1}^{q}\frac{\sF^-(\lambda_b)}{24\,M}\,\frac{\partial^2}{\partial\veps_b^{+\,2}}\,\frac{\partial}{\partial\veps_b^{-}}  \\
\times\left.\,\prod_{c=1}^{r}\frac{1}{8\,M}\frac{\partial^2}{\partial\veps_c^{+\,2}}\,\frac{\partial^2}{\partial\veps_c^{-\,2}}\,\frac{\left|\cH[t]^{i}_{i}\right|}{\la1\,i\ra^2\,\la2\,i\ra^2}\right|_{\boldsymbol{\veps}=0}\,\prod_{j=3}^{n}\phi_j^{(q_j)}(x)\,.
\end{multline}
A few comments about the structure of the result \eqref{tailCorr1} are in order. Firstly, there remain $t$ integrals over the Riemann sphere -- one corresponding to each background vertex operator -- which have not been performed analytically due to the complicated dependence on the vertex operator insertion points in the reduced determinant $|\cH[t]^{i}_{i}|$. Unlike the graviton vertex operator insertions, these are \emph{not} localised against delta functions.


One can also confirm that these integrals are projectively well-defined on each copy of $\P^1$: effectively, each formal parameter $\veps^{\pm}_{\m}$ carries scaling weight $-1$ with respect to the homogeneous coordinate $\lambda_\m$, so that the entries of the matrix $\cH[t]$ are weightless in $\lambda_\m$. The integral measure
\be\label{tailmeasure}
\frac{\varpi_\m}{\la1\,\lambda_\m\ra^2\,\la2\,\lambda_\m\ra^2}\,,
\ee
is weight $-4$, which is then balanced by the weight $+4$ differential operator in the formal parameters which extracts the appropriate terms from the reduced determinant. Equivalently, it is easy to see that once the formal parameters have been removed, the remaining terms from the determinant are weight $+4$ in each $\lambda_\m$.

\medskip

The correlator \eqref{FeynCorr3} that underpins the gravitational MHV amplitude on SDTN is given by a sum -- in principle, an infinite sum -- over these building blocks, graded by the number of background vertex operator insertions. However, this sum is actually finite for any given $n$. Each background vertex operator must absorb a minimum of three Wick contractions, with the resulting Feynman graph restricted to be a spanning tree on the set of all vertices. An inductive argument easily shows that the maximum number of background insertions $t$ for which this is possible for fixed $n$ is given by $n-4$.

So we have finally established an explicit formula for the full correlation function of interest:
\begin{multline}\label{FeynCorr4}
\left\la\prod_{i=3}^{n}V_i\right\ra^{\mathrm{conn.,\,tree}}_{\mathrm{SDTN}}=\sum_{t=0}^{n-4}\sum_{p+q+r=t}\int_{(\P^1)^t}\prod_{\m=1}^{t}\frac{\varpi_\m}{\la1\,\lambda_\m\ra^2\,\la2\,\lambda_\m\ra^2}\prod_{a=1}^{p}\frac{\sF^+(\lambda_a)}{24\,M}\,\frac{\partial}{\partial\veps_a^+}\,\frac{\partial^2}{\partial\veps_a^{-\,2}}  \\
\times\left.\prod_{b=1}^{q}\frac{\sF^-(\lambda_b)}{24\,M}\,\frac{\partial^2}{\partial\veps_b^{+\,2}}\,\frac{\partial}{\partial\veps_b^{-}} \,\prod_{c=1}^{r}\frac{1}{8\,M}\frac{\partial^2}{\partial\veps_c^{+\,2}}\,\frac{\partial^2}{\partial\veps_c^{-\,2}}\,\frac{\left|\cH[t]^{i}_{i}\right|}{\la1\,i\ra^2\,\la2\,i\ra^2}\right|_{\boldsymbol{\veps}=0}\,\prod_{j=3}^{n}\phi_j^{(q_j)}(x)\,,
\end{multline}
which encapsulates the perturbative expansion of the MHV generating functional on the SDTN metric via \eqref{MHV-preamp1}.


\subsection{The MHV amplitude}

At this point, the result \eqref{FeynCorr4} for the twistor sigma model correlator can be fed back into \eqref{MHV-preamp1} to obtain a final expression for the graviton MHV amplitude on SDTN. After re-writing the generating functional in the Gibbons-Hawking coordinates, we are left with:
\begin{multline}\label{MHV1}
\cM_{n}=\frac{\la1\,2\ra^{6}}{\la1\,i\ra^2\,\la2\,i\ra^2}\int_{\CM}\d^{4}x\,\sqrt{|g|}\,\sum_{t=0}^{n-4}\sum_{p+q+r=t}\int_{(\P^1)^t}\prod_{\m=1}^{t}\frac{\varpi_\m}{\la1\,\lambda_\m\ra^2\,\la2\,\lambda_\m\ra^2} \\
\times\,\prod_{a=1}^{p}\frac{\sF^+(\lambda_a)}{24\,M}\,\frac{\partial}{\partial\veps_a^+}\,\frac{\partial^2}{\partial\veps_a^{-\,2}} \prod_{b=1}^{q}\frac{\sF^-(\lambda_b)}{24\,M}\,\frac{\partial^2}{\partial\veps_b^{+\,2}}\,\frac{\partial}{\partial\veps_b^{-}} \,\prod_{c=1}^{r}\frac{1}{8\,M}\frac{\partial^2}{\partial\veps_c^{+\,2}}\,\frac{\partial^2}{\partial\veps_c^{-\,2}} \\
\times\,\left|\cH[\mathbf{t}]^{i}_{i}\right|\Big|_{\boldsymbol{\veps}=0}\,\prod_{j=1}^{n}\phi_j^{(q_j)}(x)\,.
\end{multline}
where $|g|$ is the determinant of the SDTN metric and the gravitons 1 and 2 have negative helicity, while $3,\ldots,n$ have positive helicity. Making use of the explicit form of the scalar quasi-momentum eigenstates \eqref{scalarqme} and the fact that $\sqrt{|g|}=V=1+2M/r$ in the Gibbons-Hawking coordinate system, this formula can be made more explicit as
\begin{multline}\label{MHV2}
\cM_{n}=2\pi\,\kappa^{n-2}\,\delta(\omega)\,\frac{\la1\,2\ra^{6}}{\la1\,i\ra^2\,\la2\,i\ra^2}\,\sum_{t=0}^{n-4}\sum_{p+q+r=t}\,\int\limits_{\R^3\times(\P^1)^t}\d^{3}\vec{x}\,\e^{\im\,\vec{k}\cdot\vec{x}}\left(1+\frac{2\,M}{r}\right) \\
\times\,\prod_{\m=1}^{t}\frac{\varpi_\m}{\la1\,\lambda_\m\ra^2\,\la2\,\lambda_\m\ra^2}\,\prod_{j=1}^{n}\left(\frac{r}{1+\zeta\,\bar{\zeta}}\right)^{q_j}\,(\zeta-z_j)^{q_j-2M\omega_j}\,(\bar{\zeta}\,z_j+1)^{q_j+2M\omega_j}\\
\times\,\prod_{a=1}^{p}\frac{\sF^+(\lambda_a)}{24\,M}\,\frac{\partial}{\partial\veps_a^+}\,\frac{\partial^2}{\partial\veps_a^{-\,2}} \prod_{b=1}^{q}\frac{\sF^-(\lambda_b)}{24\,M}\,\frac{\partial^2}{\partial\veps_b^{+\,2}}\,\frac{\partial}{\partial\veps_b^{-}} \,\prod_{c=1}^{r}\frac{1}{8\,M}\frac{\partial^2}{\partial\veps_c^{+\,2}}\,\frac{\partial^2}{\partial\veps_c^{-\,2}}\,\left|\cH[t]^{i}_{i}\right|\Big|_{\boldsymbol{\veps}=0}\,,
\end{multline}
where we have reinstated the appropriate powers of the gravitational coupling constant $\kappa$ and abbreviated
\be\label{totalmom}
\omega:=\sum_{j=1}^{n}\omega_j\,, \qquad \vec{k}:=\sum_{j=1}^{n}\vec{k}_j\,.
\ee
The integral over $\R^3$ is understood to be $\R_+\times S^2$, with the coordinates $r\in(0,\infty)$ and $(\zeta,\bar{\zeta})\in S^2$.

As the sum over $t$ in this formula corresponds to the number of background vertex operator insertions in the twistor sigma model, it is clear that these contributions to the amplitude encode the external gravitons scattering off the non-trivial geometry of the SDTN metric itself. This is a ubiquitous feature of gravitational scattering in curved spacetimes, known as \emph{tails}, resulting from the violation of Huygens' principle. Indeed, for scalar scattering it was established long ago that the only metrics which do not lead to tails are Minkowski space and vacuum plane waves~\cite{Friedlander:2010eqa}, while for gravitons only flat space scattering does not produce tails~\cite{Noonan:1989,Wunsch:1990,Harte:2013dba,Adamo:2017nia}. However, it should be emphasized that while the $t\geq1$ terms in the amplitude are unambiguously tail effects, tails are present even in the $t=0$ contributions, through the dressed momentum spinors and quasi-momentum eigenstate wavefunctions.

\medskip

Using our earlier observations from Section~\ref{sec:spin}, it is straightforward to extend this formula to one for MHV graviton scattering on the self-dual Kerr-Taub-NUT metric. Indeed, this is accomplished by the simple Newman-Janis shift:
\be\label{KTN-MHV}
\cM^{\vec{a}}_{n}=\e^{-\vec{k}\cdot\vec{a}}\,\cM_{n}\,,
\ee
where $\vec{a}$ is the spin vector of the metric. Recall, this follows because SDTN is sent to self-dual Kerr-Taub-NUT by the simple shift $\vec{x}\to\vec{x}-\im\vec{a}$.


\subsection{Properties of the amplitude}

We can comment on some general features of the MHV amplitude formula \eqref{MHV2}, including its explicit expansion for small $n$ and flat space limit.

\medskip

\paragraph{Low-point examples:} For the number of external gravitons $n\leq4$, the sum over explicit tail terms is absent, and the formula for the MHV amplitude simplifies to
\begin{multline}\label{no-tail1}
\cM_{n\leq4}=2\pi\,\kappa^{n-2}\,\delta(\omega)\,\frac{\la1\,2\ra^{6}}{\la1\,i\ra^2\,\la2\,i\ra^2}\,\int_{\R^3}\d^{3}\vec{x}\,\e^{\im\,\vec{k}\cdot\vec{x}}\left(1+\frac{2\,M}{r}\right) \\
\times\,\left|\HH^{i}_{i}\right|\,\prod_{j=1}^{n}\left(\frac{r}{1+\zeta\,\bar{\zeta}}\right)^{q_j}\,(\zeta-z_j)^{q_j-2M\omega_j}\,(\bar{\zeta}\,z_j+1)^{q_j+2M\omega_j}\,,
\end{multline}
where $\HH$ is the $(n-2)\times(n-2)$ matrix with entries \eqref{Hodges1}.

To further simplify matters, we can restrict our attention to the scattering of \emph{minimal} quasi-momentum eigenstates, for which the topological charges obey $q_i=2M|\omega_i|$; by \eqref{min2qmom}, it follows that the amplitudes for more general configurations will follow by acting with differential operators in momentum space. In this case, the set of external gravitons can be partitioned into those with positive/negative frequency as $\{1,\ldots,n\}=\mathfrak{n}_+\sqcup\mathfrak{n}_-$, and the amplitude becomes
\begin{multline}\label{no-tail2}
\cM_{n\leq4}=2\pi\,\kappa^{n-2}\,\delta(\omega)\,\frac{\la1\,2\ra^{6}}{\la1\,i\ra^2\,\la2\,i\ra^2}\,\int_{\R^3}\d^{3}\vec{x}\,\e^{\im\,\vec{k}\cdot\vec{x}}\left(1+\frac{2\,M}{r}\right) \\
\times\,\left|\HH^{i}_{i}\right|\,\prod_{j\in\mathfrak{n_+}}\left(\frac{r}{1+\zeta\,\bar{\zeta}}\right)^{2M\omega_j}\,(\bar{\zeta}\,z_j+1)^{4M\omega_j}\,\prod_{k\in\mathfrak{n}_-}\left(\frac{r}{1+\zeta\,\bar{\zeta}}\right)^{-2M\omega_k}\,(\zeta-z_k)^{-4M\omega_k}\,\,.
\end{multline}
Clearly, overall energy conservation implies that there must be at least one external graviton of both positive and negative frequency in order to obtain a non-vanishing amplitude.

For $n=3$, one can assume without loss of generality that graviton 1 has negative frequency, so that
\begin{multline}\label{n=3.1}
\cM_{3}=2\pi\,\kappa\,\delta(\omega)\,\frac{\la1\,2\ra^{6}}{\la1\,3\ra^2\,\la2\,3\ra^2}\,\int_{\R^3}\d^{3}\vec{x}\,\e^{\im\,\vec{k}\cdot\vec{x}}\left(1+\frac{2\,M}{r}\right) \\
\times \left[\frac{r\,(\zeta-z_1)\,(\bar{\zeta}\,z_2+1)}{1+\zeta\,\bar{\zeta}}\right]^{4M\omega_2}\,\left[\frac{r\,(\zeta-z_1)\,(\bar{\zeta}\,z_3+1)}{1+\zeta\,\bar{\zeta}}\right]^{4M\omega_3}\,,
\end{multline}
using the fact that $-\omega_1=\omega_2+\omega_3$ on the support of conservation of energy. This enables the amplitude to be written in terms of two basic integrals:
\be\label{n=3.2}
\cM_{3}=4\pi\,\im\,\kappa\,\delta(\omega)\,\frac{\la1\,2\ra^{6}}{\la1\,3\ra^2\,\la2\,3\ra^2}\,\left[\mathcal{I}_3+2M\,\mathcal{J}_3\right]\,,
\ee
where
\begin{multline}\label{I_3}
\mathcal{I}_3=\int_{0}^{\infty}\d r\,r^2 \int_{\C}\d\zeta\,\d\bar{\zeta}\,(1+|\zeta|^2)\,\left(r\,(\zeta-z_1)\,(\bar{\zeta}\,z_2+1)\right)^{4M\omega_2} \left(r\,(\zeta-z_1)\,(\bar{\zeta}\,z_3+1)\right)^{4M\omega_3} \\
\times\,\exp\!\left[\im\,r\,k^3\left(1-|\zeta|^2+\bar{\zeta}\,w+\zeta\,\bar{w}\right)\right]\,,
\end{multline}
\begin{multline}\label{J_3}
\mathcal{J}_3=\int_{0}^{\infty}\d r\,r \int_{\C}\d\zeta\,\d\bar{\zeta}\,\left(r\,(\zeta-z_1)\,(\bar{\zeta}\,z_2+1)\right)^{4M\omega_2} \left(r\,(\zeta-z_1)\,(\bar{\zeta}\,z_3+1)\right)^{4M\omega_3} \\
\times\,\exp\!\left[\im\,r\,k^3\left(1-|\zeta|^2+\bar{\zeta}\,w+\zeta\,\bar{w}\right)\right]\,,
\end{multline}
for
\be\label{wdef}
w:=\frac{k^{1}+\im\,k^2}{k^3}\,,
\ee
built from the components of the total spatial momentum $\vec{k}=\vec{k}_1+\vec{k}_2+\vec{k}_3$.

The integrals $\mathcal{I}_3$ and $\mathcal{J}_3$ can be evaluated as Gaussian integrals on the complex plane, followed by a Mellin integration over $r\in\R_+$ (cf., Appendix A of~\cite{Adamo:2024xpc}). This leads to a final expression
\begin{multline}\label{n=3.3}
\cM_3=4\pi^2\,\im\,\kappa\,\delta(\omega)\,\frac{\la1\,2\ra^{6}}{\la1\,3\ra^2\,\la2\,3\ra^2}\,\frac{(-1)^{4M\omega_1}\,(\im\,\alpha_{21;3})^{4M\omega_2}\,(\im\,\alpha_{31;3})^{4M\omega_3}}{|\vec{k}|^{2+8M|\omega_1|}} \\
\times\left[2M+(4M|\omega_1|)!\left(\frac{4M\omega_2\,z_{21}}{\alpha_{21;3}}+\frac{4M\omega_3\,z_{31}}{\alpha_{31;3}}\right)\right]\,,
\end{multline}
in terms of the quantities
\be\label{alphadef}
\alpha_{ij;3}:=k^3\,\left(w-z_i-z_j-\bar{w}\,z_i\,z_j\right)\,.
\ee
Here, we see that although the same spinor ratio appears as for the flat-space 3-point MHV amplitude, this is dressed by factors which depend explicitly on the background.

\medskip

Already at $n=4$, the structure of the amplitude becomes more complicated; in some sense, this is inherited from flat space: unlike its gauge theory cousin, the MHV graviton amplitude depends on both angle and square bracket kinematic invariants beyond 3-points. However, in SDTN the square brackets are themselves dressed by the background, leading to more terms which contribute to the integral over $\R^3$.

For the 4-point scattering of minimal quasi-momentum eigenstates and graviton 1 the only negative-frequency state, the MHV amplitude is given by
\begin{multline}\label{n=4.1}
\cM_4=2\pi\,\kappa^2\,\delta(\omega)\,\frac{\la1\,2\ra^6}{\la1\,3\ra\,\la1\,4\ra\,\la2\,3\ra\,\la2\,4\ra\,\la4\,3\ra} \\
\times\int\d^{3}\vec{x}\,\e^{\im\vec{k}\cdot\vec{x}}\,\left(1+\frac{2M}{r}\right)\, [\![4\,3]\!]\,\prod_{i=2}^{4}\left(\frac{r\,(\zeta-z_1)\,(\bar{\zeta}\,z_i+1)}{1+|\zeta|^2}\right)^{4M\omega_i}\,.
\end{multline}
The dressed square bracket can be evaluated explicitly to give
\begin{multline}\label{n=4.2}
[\![4\,3]\!]=\left(1+\frac{2M}{r}\right)^{-1}\left[[4\,3]-\frac{4M\,\omega_3\,\omega_4}{r\,(\bar{\zeta}\,z_3+1)\,(\bar{\zeta}\,z_4+1)}\left(\frac{(\bar{\zeta}\,z_3+1)\,(\bar{\zeta}-\tilde{z}_4)}{1+|z_4|^2}\right.\right. \\
-\left.\left.\frac{(\bar{\zeta}\,z_4+1)\,(\bar{\zeta}-\tilde{z}_3)}{1+|z_3|^2}\right)\right]\,,
\end{multline}
where $|z_i|^2=z_i\,\tilde{z}_i$ is not a real-valued quantity, since $\tilde{z}_i\neq\bar{z}_i$ for complex graviton momenta. This can be used to express the integral in \eqref{n=4.1} as
\begin{multline}\label{n=4.3}
[4\,3]\int_{0}^{\infty}\d r\,r^2\int_{\C}\d\zeta\,\d\bar{\zeta}\,(1+|\zeta|^2)\,\e^{\im\vec{k}\cdot\vec{x}}\,\prod_{i=2}^{4}\left[r\,(\zeta-z_1)\,(\bar{\zeta}\,z_i+1)\right]^{4M\omega_i} \\
-4M\,\omega_3\,\omega_4\int_{0}^{\infty}\d r\,r\int_{\C}\d\zeta\,\d\bar{\zeta}\,\e^{\im\vec{k}\cdot\vec{x}}\left[\frac{(\bar{\zeta}-\tilde{z}_4)}{(\bar{\zeta}\,z_4+1)\,(1+|z_4|^2)}-\frac{(\bar{\zeta}-\tilde{z}_3)}{(\bar{\zeta}\,z_3+1)\,(1+|z_3|^2)}\right] \\
\times\,\prod_{i=2}^{4}\left[r\,(\zeta-z_1)\,(\bar{\zeta}\,z_i+1)\right]^{4M\omega_i}\,,
\end{multline}
with 
\be\label{n=4.4}
\vec{k}\cdot\vec{x}=r\,k^3\left(1-|\zeta|^2+\bar{\zeta}\,w+\zeta\,\bar{w}\right)\,,
\ee
defined in the same way as above, but now for the total 4-point spatial momentum. These integrals can be evaluated using the same methods as at 3-points, although the resulting expressions are not particularly enlightening.

\medskip

It is also illustrative to consider the case $n=5$, where the first explicit tail contributions appear in the amplitude thanks to the insertion of background vertex operators in the twistor sigma model. In this case, only the $(p,q,r)=(1,0,0)$ and $(p,q,r)=(0,1,0)$ terms contribute to the amplitude beyond $t=0$. For instance, the former is given by 
\begin{multline}\label{n=5.1}
\frac{\pi\,\kappa^3}{12\,M}\,\delta(\omega)\,\frac{\la1\,2\ra^6}{\la1\,3\ra^2\,\la2\,3\ra^2}\int\limits_{\R^3\times S^2}\frac{\d^{3}\vec{x}\,\D\lambda\wedge\D\hat{\lambda}}{\la\lambda\,\hat{\lambda}\ra^2\,\la1\,\lambda\ra^2\,\la2\,\lambda\ra^2} \\
\times\,\left.\left(\sF^+(\lambda)\,\frac{\partial}{\partial\veps^+}\,\frac{\partial^2}{\partial\veps^{-\,2}}+\sF^-(\lambda)\,\frac{\partial^2}{\partial\veps^{+\,2}}\,\frac{\partial}{\partial\veps^{-}}\right)\left|\cH^{3}_{3}[1]\right|\right|_{\veps^{\pm}=0}\,,
\end{multline}
with the minor given by
\be\label{n=5.2}
\left|\cH^{3}_{3}[1]\right|=\HH_{44}\left(\mathbb{T}\,\HH_{55}-\mathfrak{h}_5^2\right)-\HH_{45}\left(\HH_{45}\,\mathbb{T}-\mathfrak{h}_4\,\mathfrak{h}_5\right)+\mathfrak{h}_4\left(\HH_{45}\,\mathfrak{h}_5-\HH_{55}\,\mathfrak{h}_4\right)\,,
\ee
the tail index on matrix entries being suppressed as it is irrelevant in this case. 

The matrix entries $\HH_{45}$ contain no powers of the formal parameters, while all other entries appearing in \eqref{n=5.2} are at most linear in $\veps^{\pm}$, making it clear that only
\be\label{n=5.3}
\HH_{44}\,\HH_{55}\,\mathbb{T}-\HH_{55}\,\mathfrak{h}_{4}^{2}-\HH_{44}\,\mathfrak{h}_{5}^2\,,
\ee
can give non-vanishing contributions to \eqref{n=5.1}. Introducing the notation
\be\label{n=5.4}
[\![i\,\sH^{\pm}]:=\tilde{K}_{i}^{\dot\alpha}\,\sH^{\pm}{}_{\dot\alpha}(\lambda)\,, \qquad \sH^{+}{}_{\dot\alpha}(\lambda)=\sH^{\dot\beta}{}_{\dot\alpha}(\lambda)\,\tilde{o}_{\dot\beta}\,, \quad \sH^{-}{}_{\dot\alpha}(\lambda)=\sH^{\dot\beta}{}_{\dot\alpha}(\lambda)\,\tilde{\iota}_{\dot\beta}\,,
\ee
the coefficient of $\veps^+\,(\veps^-)^2$ can be extracted from \eqref{n=5.3} to give
\begin{multline}\label{n=5.5}
\left|\cH^{3}_{3}[1]\right|\Big|_{\veps^+(\veps^-)^2}=\frac{\la1\,\lambda\ra^3\,\la2\,\lambda\ra^3}{\la3\,\lambda\ra\,\la4\,\lambda\ra\,\la5\,\lambda\ra\,\la1\,3\ra\,\la2\,3\ra\,\la1\,4\ra\,\la2\,4\ra\,\la1\,5\ra\,\la2\,5\ra} \\
\times\left([\![3\,\sH^+]\,[\![4\,\sH^-]\,[\![5\,\sH^-]+[\![3\,\sH^-]\,[\![4\,\sH^+]\,[\![5\,\sH^-]+[\![3\,\sH^-]\,[\![4\,\sH^-]\,[\![5\,\sH^+]\right)\,.
\end{multline}
As expected, this corresponds precisely to the sum of tree diagrams in the twistor sigma model including three external graviton vertex operators and a single insertion of the background vertex operator $U^+$. Extracting the coefficient of $(\veps^+)^2\veps^-$ from \eqref{n=5.3} gives a similar result with a single insertion of $U^-$.

\medskip

\paragraph{Flat limit:} A basic consistency check on the amplitude \eqref{MHV2} is that it should be equal to the MHV graviton scattering amplitude on flat space when the curvature of the background metric is turned off. Naively, one might assume that the flat limit of SDTN corresponds to $M\to 0$; indeed, the metric \eqref{eq:metric} certainly becomes flat in this limit. However, the resulting flat manifold has topology $S^1\times\R^3$, rather than $\R^4$: this is because the Euclidean time coordinate is compactified (recall that $t\sim t+8\pi M)$ to a circle with radius $4M$, so in the $M\to 0$ limit this circle becomes infinitesimally small.

To compare with flat space scattering amplitudes, one requires a flat limit with trivial topology -- that is, resulting in $\R^4$. This requires de-compactifying the Euclidean time coordinate, which (somewhat non-intuitively) actually corresponds to taking the $M\to\infty$ limit of SDTN. From the perspective its twistor description, this is clearly the correct limit, as $\sh\to 0$ when $M\to\infty$ and thus the complex structure reduces to that of the flat twistor space $\PT$. To see that this is indeed the correct limit from the metric perspective, consider the rescaling
\be\label{flatlim1}
t\to M\,t\,, \qquad r\to \frac{r}{M}\,,
\ee
under which the SDTN metric goes to
\be\label{flatlim2}
\d s^2=\left(\frac{1}{M^2}+\frac{2}{r}\right)^{-1}\left(\d t-2(1-\cos\theta)\d\phi\right)^2+\left(\frac{1}{M^2}+\frac{2}{r}\right)\left(\d r^2+r^2\,\d\Omega_2^2\right)\,.
\ee
In these rescaled coordinates, it follows that
\be\label{flatlim3}
\lim_{M\to\infty}\d s^2=\frac{r}{2}\left(\d t-2(1-\cos\theta)\d\phi\right)^2+\frac{2}{r}\left(\d r^2+r^2\,\d\Omega_2^2\right)\,,
\ee
which is precisely the flat hyperk\"ahler metric on $\R^4$ in Gibbons-Hawking coordinates~\cite{Hawking:1976jb,Gibbons:1978tef}. This highlights the fact that although $M$ resembles the ADM mass parameter of a black hole metric, at the self-dual point it is also conflated with the topological NUT charge, making the flat space limit somewhat non-intuitive.

This flat limit can now be implemented at the level of the MHV amplitude \eqref{MHV2}. Under the scaling \eqref{flatlim1}, it follows that
\be\label{flatlim4}
\omega\to\frac{\omega}{M}\,, \qquad \vec{k}\to\vec{k}\,M\,,
\ee
to ensure that the quantity $k\cdot x$ remains finite. Under this, the important scalings of the MHV amplitude are captured by the collection:
\begin{multline}\label{flatlim5}
\delta(\omega)\,\d^{3}\vec{x}\,\left(1+\frac{2M}{r}\right)\prod_{j=1}^{n}\left(\frac{r}{1+\zeta\,\bar{\zeta}}\right)^{q_j} (\zeta-z_j)^{q_j-2M\omega_j}\,(\bar{\zeta}\,z_j+1)^{q_j+2M\omega_j} \\
\to \delta(\omega)\,\d^{3}\vec{x}\,\left(\frac{1}{M^2}+\frac{2}{r}\right)\prod_{j=1}^{n}M^{-q_j}\,\left(\frac{r}{1+\zeta\,\bar{\zeta}}\right)^{q_j} (\zeta-z_j)^{q_j-2\omega_j}\,(\bar{\zeta}\,z_j+1)^{q_j+2\omega_j}\,.
\end{multline}
Now, in the $M\to\infty$ limit the topology of the metric becomes trivial, so all topological charges vanish, meaning that (after the rescaling) $q_j\pm2\omega_j\to 0$ and $\sum_j q_j\to\pm2\sum_j\omega_j=0$ in the flat limit. 

Thus, one finds that
\be\label{flatlim6}
\lim_{M\to\infty}\cM_{n}=4\pi\,\kappa^{n-2}\,\delta(\omega)\,\frac{\la1\,2\ra^6}{\la1\,i\ra^2\,\la2\,i\ra^2}\,\int_{\R^3}\frac{\d^3\vec{x}}{r}\,\e^{\im\,\vec{k}\cdot\vec{x}}\,|\HH^{i}_{i}|\,,
\ee
with all $t>0$ terms vanishing as powers of $1/M$. After performing the diffeomorphism from the flat Gibbons-Hawking coordinates to standard spherical polar coordinates on $\R^3$, the remaining integrals can be performed trivially to give
\be\label{flatlim7}
\lim_{M\to\infty}\cM_{n}=(2\pi)^4\,\kappa^{n-2}\,\delta(\omega)\,\delta^{3}(\vec{k})\,\frac{\la1\,2\ra^6}{\la1\,i\ra^2\,\la2\,i\ra^2}\,|\HH^{i}_{i}|\,,
\ee
which is precisely Hodges' formula for $n$-graviton MHV scattering in Minkowski spacetime~\cite{Hodges:2012ym}. One might worry that we have seemingly obtained a formula for scattering in Lorentzian Minkowski space from scattering on Euclidean $\R^4$. In flat space there is, in fact, no distinction, as the amplitude is a rational function whose analytic continuation to the complex momenta of the Euclidean setting is trivial.


\section{Holomorphic collinear limits on self-dual Taub-NUT}\label{sec:celestial_ope}

Armed with the explicit formula \eqref{MHV2} for $n$-point MHV graviton scattering on SDTN, we can explore its behaviour in the limit where two of the external graviton momenta become \emph{holomorphically collinear}. On a flat background, this leads to the celestial OPE for the so-called $\mathcal{L}w^{\wedge}_{1+\infty}$ or $\cL\mathfrak{ham}(\C^2)$ algebra~\cite{Guevara:2021abz,Strominger:2021lvk} and one can ask whether this is deformed on our SDTN background.  

To investigate,let gravitons $i,j$ be positive helicity with initial, un-dressed momenta $k_i^{\alpha\dot\alpha}=\kappa_i^{\alpha}\,\tilde{\kappa}_i^{\dot\alpha}$ and $k_j^{\alpha\dot\alpha}=\kappa_j^{\alpha}\,\tilde{\kappa}_j^{\dot\alpha}$, respectively. The holomorphic collinear limit is then parametrized as
\be\label{hcl1}
k_i+k_j=P+\epsilon^2\,q\,,
\ee
where $P^{\alpha\dot\alpha}=\kappa_P^{\alpha}\,\tilde{\kappa}_P^{\dot\alpha}$ is the collinear null momentum, $q^{\alpha\dot\alpha}=\xi^{\alpha}\,\tilde{\xi}^{\dot\alpha}$ is an arbitrarily-chosen reference null momentum and the small parameter $\epsilon$ controls the collinear limit, as $k_i\cdot k_j=\epsilon^2\,P\cdot q$. The fact that this is a \emph{holomorphic} collinear limit is simply the statement that $\la i\,j\ra\sim\epsilon$ in the $\epsilon\to 0$ limit, while $[i\,j]$ does not scale with $\epsilon$ and remains finite.

To analyze the holomorphic collinear limit of $\cM_n$ we extract the leading singularity in $\la i\,j\ra$ and evaluate the coefficient of this singularity on
\be\label{hcl2}
\kappa_i^{\alpha}=\frac{\la\xi\,i\ra}{\la\xi\,P\ra}\,\kappa_P^{\alpha}+O(\epsilon)\,, \qquad \kappa_j^{\alpha}=\frac{\la\xi\,j\ra}{\la\xi\,P\ra}\,\kappa_P^{\alpha}+O(\epsilon)\,.
\ee
Although the analysis of the holomorphic collinear limit of the MHV amplitude on SDTN follows that for generic self-dual radiative metrics~\cite{Adamo:2023zeh}, the non-trivial topology of Euclidean time in SDTN leads to some new features. In particular, the topological charges of the two collinear gravitons must also respect the collinear limit; the natural way to parametrize this is in terms of the longitudinal momentum fraction $s\in[0,1]$:
\be\label{lmomfrac}
\omega_i=s\,\omega_{P}\,, \qquad \omega_j=(1-s)\,\omega_P\,,
\ee
for $\omega_P$ the frequency of the collinear momentum $P^{\alpha\dot\alpha}$. In order to preserve the topological quantization condition $q_P\pm 2M\omega_P$, it follows that the topological charges must follow the same parametrization
\be\label{topfrac}
q_i=s\,q_P\,, \qquad q_j=(1-s)\,q_P\,,
\ee
in the collinear limit where $q_i+q_j=q_P$.

The central ingredient in each term of \eqref{MHV2} is the once-reduced determinant of the matrix $\cH[t]$, with the row and column removed to create the minor being arbitrarily chosen. As such, we can freely choose to remove the row and column corresponding to one of the two holomorphically collinear gravitons, say $j$. By expanding the resulting minor along the $i^{\mathrm{th}}$ row and exploiting the properties of the holomorphic collinear limit, it follows that 
\be\label{hcl3}
\left|\cH^{j}_{j}[t]\right|=\frac{[\![i\,j]\!]}{\la i\,j\ra}\,\left|\cH^{ij}_{ij}[t]\right|+O(\epsilon^0)\,,
\ee
exposes the leading holomorphic collinear singularity of the amplitude; all other ingredients of $\cM_{n}$ are regular as $\la i\,j\ra\to 0$. Thus, the remaining parts of the amplitude can be evaluated to leading order in the collinear limit using \eqref{hcl1} -- \eqref{hcl2}.

Now, recall that
\be\label{dressedsqr2}
[\![i\,j]\!]=\tilde{\kappa}_{i\,\dot\beta}\,\tilde{\kappa}_{j\,\dot\gamma}\,\mathsf{G}^{\dot\beta\dot\alpha}(x;k_i,q_i)\,\mathsf{G}^{\dot\gamma}{}_{\dot\alpha}(x;k_j,q_j)\,,
\ee
for the dressing matrix $\mathsf{G}^{\dot\alpha}{}_{\dot\beta}$ given by \eqref{dmatrix2}. The dressing matrix is a homogeneous function of the un-dotted momentum spinor, so in the holomorphic collinear limit \eqref{hcl2}, it follows that 
\be\label{hclMat1}
\mathsf{G}^{\dot\beta\dot\alpha}(x;k_i,q_i)=\mathsf{G}^{\dot\beta\dot\alpha}(x;P,q_i)+O(\epsilon)\,,
\ee
and similarly for $\mathsf{G}^{\dot\beta\dot\alpha}(x;k_j,q_j)$. A computation then reveals that
\be\label{hcl4}
\mathsf{G}^{\dot\beta\dot\alpha}(x;P,q_i)\,\mathsf{G}^{\dot\gamma}{}_{\dot\alpha}(x;P,q_j)=\epsilon^{\dot\gamma\dot\beta}\,,
\ee
and hence that 
\be\label{hcl5}
\left|\cH^{j}_{j}[t]\right|=\frac{[i\,j]}{\la i\,j\ra}\,\left|\cH^{ij}_{ij}[t]\right|+O(\epsilon^0)\,,
\ee
in the holomorphic collinear limit.

In the minor $|\cH^{ij}_{ij}[t]|$, the only remaining dependence on the collinear momenta is through the diagonal entries of the matrix $\cH$, since the rows and columns corresponding to the collinear momenta have been removed. This dependence is controlled in the diagonal entries $\HH_{kk}$ through linear combinations of the form
\begin{multline}\label{hcldiag1}
\frac{[\![k\,i]\!]\,\la1\,i\ra\,\la2\,i\ra}{\la k\,i\ra\,\la1\,k\ra\,\la2\,k\ra}+\frac{[\![k\,j]\!]\,\la1\,j\ra\,\la2\,j\ra}{\la k\,j\ra\,\la1\,k\ra\,\la2\,k\ra} \\
=\frac{\la1\,P\ra\,\la2\,P\ra}{\la k\,P\ra\,\la1\,k\ra\,\la2\,k\ra\,\la\xi\,P\ra}\,\big([\![k\,i]\!]\,\la\xi\,i\ra+[\![k\,j]\!]\,\la\xi\,j\ra\big)+O(\epsilon)\,,
\end{multline}
with a similar formula controlling the dependence of diagonal entries of the block $\mathbb{T}$. 

In order to further simplify such expressions, one must account for the fact that the dressed momentum spinors $\tilde{K}_{i,j}^{\dot\alpha}$ depend on both the frequencies and topological charges of the dressed momenta. In particular, 
\begin{multline}\label{hcldiag2}
[\![k\,i]\!]\,\la\xi\,i\ra+[\![k\,j]\!]\,\la\xi\,j\ra=\frac{\tilde{K}_{k}^{\dot\alpha}}{\sqrt{V}}\left[\la\xi\,i\ra\,\tilde{\kappa}_{i\,\dot\alpha}+\la\xi\,j\ra\,\tilde{\kappa}_{j\,\dot\alpha}+(q_i+2M\omega_i)\,\frac{\la P|T|_{\dot\alpha}\,\la\chi_+|T|i]\,\la\xi\,i\ra}{\omega_i\,r\,\la P\,\chi_+\ra} \right. \\
+(q_j+2M\omega_j)\,\frac{\la P|T|_{\dot\alpha}\,\la\chi_+|T|j]\,\la\xi\,j\ra}{\omega_j\,r\,\la P\,\chi_+\ra}-(q_i-2M\omega_i)\,\frac{\la P|T|_{\dot\alpha}\,\la\chi_-|T|i]\,\la\xi\,i\ra}{\omega_i\,r\,\la P\,\chi_-\ra} \\
\left.-(q_j-2M\omega_j)\,\frac{\la P|T|_{\dot\alpha}\,\la\chi_-|T|j]\,\la\xi\,j\ra}{\omega_j\,r\,\la P\,\chi_-\ra}\right]\,,
\end{multline}
having abbreviated $\la P|T|_{\dot\alpha}=\kappa_P^{\alpha}\,T_{\alpha\dot\alpha}$ and $\la\chi_\pm|T|i]=\chi_{\pm}^{\alpha}\,T_{\alpha}{}^{\dot\alpha}\,\tilde{\kappa}_{i\,\dot\alpha}$, etc. Using the collinear parametrizations \eqref{lmomfrac}, \eqref{topfrac}, this simplifies to
\begin{multline}\label{hcldiag3}
\frac{\tilde{K}_{k}^{\dot\alpha}}{\sqrt{V}}\left[\la\xi\,i\ra\,\tilde{\kappa}_{i\,\dot\alpha}+\la\xi\,j\ra\,\tilde{\kappa}_{j\,\dot\alpha}+(q_P+2M\omega_P)\,\frac{\la P|T|_{\dot\alpha}\,(\la\chi_+|T|i]\,\la\xi\,i\ra+\la\chi_+|T|j]\,\la\xi\,j\ra)}{\omega_P\,r\,\la P\,\chi_+\ra}\right. \\
\left. -(q_P-2M\omega_P)\,\frac{\la P|T|_{\dot\alpha}\,(\la\chi_-|T|i]\,\la\xi\,i\ra+\la\chi_-|T|j]\,\la\xi\,j\ra)}{\omega_P\,r\,\la P\,\chi_-\ra}\right]+O(\epsilon)\,.
\end{multline}
Observing that in the collinear limit \eqref{hcl1} -- \eqref{hcl2},
\be\label{hcldiag4}
\la\xi\,i\ra\,\tilde{\kappa}_{i}^{\dot\alpha}+\la\xi\,j\ra\,\tilde{\kappa}_j^{\dot\alpha}=\la\xi\,P\ra\,\tilde{\kappa}_P^{\dot\alpha}+O(\epsilon)\,,
\ee
it then follows that
\be\label{hcldiag5}
[\![k\,i]\!]\,\la\xi\,i\ra+[\![k\,j]\!]\,\la\xi\,j\ra=[\![k\,P]\!]\,\la\xi\,P\ra+O(\epsilon)\,,
\ee
so that \eqref{hcldiag1} simplifies to
\be\label{hcldiag6}
\frac{[\![k\,i]\!]\,\la1\,i\ra\,\la2\,i\ra}{\la k\,i\ra\,\la1\,k\ra\,\la2\,k\ra}+\frac{[\![k\,j]\!]\,\la1\,j\ra\,\la2\,j\ra}{\la k\,j\ra\,\la1\,k\ra\,\la2\,k\ra} \\
=\frac{[\![k\,P]\!]\,\la1\,P\ra\,\la2\,P\ra}{\la k\,P\ra\,\la1\,k\ra\,\la2\,k\ra}+O(\epsilon)\,,
\ee
to leading order in the holomorphic collinear limit. A similar simplification occurs in the diagonal entries of $\mathbb{T}$.

Consequently, in the holomorphic collinear limit it follows that the two collinear gravitons are effectively replaced by the single collinear graviton:
\be\label{hcl6}
\left|\cH^{j}_{j}[t]\right|=\frac{[i\,j]}{\la i\,j\ra}\,\left|\widehat{\cH}^{P}_{P}[t]\right|+O(\epsilon^0)\,,
\ee
where $\widehat{\cH}[t]$ is the matrix defined for the set of external positive helicity gravitons in which $i,j$ have been removed and replaced by $P$ with its associated collinear quantum numbers. The other ingredients in the $n$-point MHV amplitude \eqref{MHV2} can likewise be evaluated in the strict collinear limit, where
\be\label{hcl7}
\sum_{k=1}^{n}\omega_k\to \sum_{k\neq i,j}\omega_k+\omega_P\,, \qquad \sum_{l=1}^{n}\vec{k}_{l}\to\sum_{l\neq i,j}\vec{k}_l+\vec{k}_{P}\,,
\ee
\be\label{hcl8}
\frac{\la1\,2\ra^{6}}{\la1\,j\ra^2\,\la2\,j\ra^2}\to\frac{\la1\,2\ra^{6}}{\la1\,P\ra^2\,\la2\,P\ra^2}\,\frac{\la\xi\,P\ra^4}{\la\xi\,i\ra^2\,\la\xi\,j\ra^2}\,,
\ee
and
\begin{multline}\label{hcl9}
\prod_{k=i,j}\left(\frac{r}{1+\zeta\,\bar{\zeta}}\right)^{q_k}\,(\zeta-z_k)^{q_k-2M\omega_k}\,(\bar{\zeta}\,z_k+1)^{q_k+2M\omega_k} \\
\to\left(\frac{r}{1+\zeta\,\bar{\zeta}}\right)^{q_P}\,(\zeta-z_P)^{q_P-2M\omega_P}\,(\bar{\zeta}\,z_P+1)^{q_P+2M\omega_P}\,,
\end{multline}
as $z_{i},z_j\to z_P$ in the holomorphic collinear limit.

Finally, it can be seen (following an argument identical to the self-dual radiative case -- see~\cite{Adamo:2023zeh}) that the $t=n-4$ terms in the MHV amplitude cannot contain the holomorphic collinear singularity \eqref{hcl6}. In other words, for $t=n-4$, the process of extracting polynomials of the correct order in the formal parameters $\veps_{\mathrm{m}}$ annihilates the holomorphic collinear singularity. Thus, only terms with $t$ running from zero to $n-5$ actually contribute to the leading holomorphic collinear limit.

\medskip

Pulling all of these pieces together, one obtains
\be\label{hclFinal}
\lim_{\epsilon\to0}\cM_{n}\to \kappa\,\frac{[i\,j]}{\la i\,j\ra}\,\frac{\la\xi\,P\ra^4}{\la\xi\,i\ra^2\,\la\xi\,j\ra^2}\,\cM_{n-1}+O(\epsilon^0)\,,
\ee
where $\cM_{n-1}$ is the MHV amplitude on SDTN with the positive helicity external gravitons $i$ and $j$ removed and replaced with the single positive helicity collinear graviton $P$. Remarkably, the coefficient
\be\label{hclSplitting}
\kappa\,\frac{[i\,j]}{\la i\,j\ra}\,\frac{\la\xi\,P\ra^4}{\la\xi\,i\ra^2\,\la\xi\,j\ra^2}=\mathrm{Split}(i^+,\,j^+\to P^+)\,,
\ee
matches the tree-level holomorphic collinear splitting function of positive helicity gravitons in \emph{flat} space~\cite{Bern:1998sv,White:2011yy,Akhoury:2011kq}! 

It is far from obvious why this should be so from background field theory in the SDTN metric, but the twistor theory of SDTN in fact hints that this should be the case. In~\cite{Adamo:2023zeh}, it was shown that holomorphic collinear splitting functions on any self-dual radiative metric are un-deformed from flat space. While the interpretation of SDTN as a self-dual black hole makes it apparently non-radiative, the complex structure defined by the Hamiltonian \eqref{eq:hamiltonian} takes the same functional form as that of a self-dual radiative metric. 

Furthermore, it is known that holomorphic splitting functions of positive helicity gravitons are intimately connected with \emph{celestial chiral algebras}, the algebras of positive helicity gravitons on any self-dual metric formed via holomorphic collinear limits~\cite{Guevara:2021abz,Strominger:2021lvk,Himwich:2021dau,Jiang:2021ovh}. These algebras are easily identified in twistor space~\cite{Adamo:2021lrv,Adamo:2021zpw,Adamo:2025mqp,Heuveline:2025nmb}, and recently strong evidence was given for the celestial chiral algebra of SDTN to be $\cL\mathfrak{ham}(\C^2)$ (the loop algebra of the Lie algebra of Hamiltonian vector fields on $\C^2$)~\cite{Bogna:2024gnt}, which is the same as the celestial chiral algebra of Minkowski space.


\section{Conclusion}\label{sec:concl}

In this paper, we showed that the self-dual Taub-NUT metric serves as a toy model for a black hole where it is possible to compute graviton scattering amplitudes exactly in the background and to high multiplicity. In particular, we obtained an explicit formula for the tree-level graviton MHV amplitude on SDTN, and studied some of its basic properties. There are several interesting directions for future study based on this work, and we conclude with a brief overview of some of them.

Our initial motivation for studying graviton scattering on SDTN was to provide a toy model for gravitational scattering on an astrophysical black hole. Now that we have an explicit, all-multiplicity formula on the SDTN background, it remains to be seen how much information can be gleaned from this about scattering on a \emph{real} black hole. Perhaps the most obvious place to test this is already at the level of the 2-point MHV amplitude. These semi-classical 2-point amplitudes are governed by the \emph{radial action} of the background~\cite{Kol:2021jjc}, which encodes the radial geodesic motion of a probe in the background geometry. Known formulations of the classical geometry and geodesics of SDTN (cf., \cite{Gibbons:1986hz,Dunajski:2019gwz}) should enable an explicit comparison between physical observables on SDTN and Kerr; while the former will obviously be complex, it may be that real sections encode some amount of information in the Kerr result.

A related, but distinct, question is to what extent the twistor methods underlying the work in this paper can be extended away from the self-duality requirement on the background. While astrophysical black holes are definitely not self-dual, they do possess a \emph{hermitian} structure which is a Lorentzian version of the conformal K\"ahler condition~\cite{Flaherty:1974,Flaherty:1976}. This endows these black hole metrics with a 2-dimensional twistor space~\cite{Bailey:1991a,Bailey:1991b} (rather than the 3-dimensional twistor space of the self-dual setting) and enables the definition of a Penrose transform for the Teukolsky system~\cite{Araneda:2018ezs,Araneda:2019uwy} as well as explicit descriptions of the metric in terms of scalar potentials~\cite{Aksteiner:2022bwr}. Furthermore, all astrophysical black holes can be constructed from certain quadrics in (flat) twistor space~\cite{Araneda:2022xii}. More generally, the half type-D condition leads to integrability via reduction to the $\SU(\infty)$ Toda equation \cite{Przanowski:1984qq,Tod:2020ual}, a well-known integrable system. Apart from an initial attempt at solving the wave equation on Kerr perturbatively around the self-dual point~\cite{Guevara:2023wlr}, it appears that none of these facts have been exploited for amplitudes-based approaches to black hole physics, and it would be fascinating to do so in the future.

Even remaining with the self-dual background, one fairly obvious challenge is to extend the formula \eqref{MHV2} beyond the MHV helicity sector to the full tree-level S-matrix on the SDTN metric. Each extra MHV degree gives an extra perturbation away from the MHV sector. These could be taken to be scattering states (i.e., momentum eigenstates).  Alternatively, they could be taken to provide a perturbative expansion towards the full Kerr metric from the self-dual sector.  Starting at the next-to-maximal helicity violating (NMHV) configuration, the methodology used in this paper breaks down, and we lack a first principles way to derive scattering amplitudes from general relativity on spacetime. It is possible to guess an extension of the formula to N$^k$MHV by replacing the integral over points in SDTN with an integral over the moduli space of degree $k+1$ holomorphic rational maps from the Riemann sphere to $\CPT$~\cite{Bogna:2025}. Indeed, a similar strategy allowed us to conjecture formulae for the tree-level S-matrix on any self-dual radiative space~\cite{Adamo:2022mev}. This follows the known pattern for the tree-level graviton S-matrix in Minkowski space~\cite{Cachazo:2012kg}, but there is currently no method to verify these conjectures on curved backgrounds besides explicit comparison with background field calculations (which are themselves prohibitively difficult).

\medskip

However, there are several other interesting questions which could be explored within the MHV sector itself. For gluon scattering on the gauge-theoretic analogue of SDTN -- namely, the self-dual dyon -- it has been shown that MHV amplitudes~\cite{Adamo:2024xpc} can be bootstrapped directly from the chiral vertex operator algebra associated with self-dual gauge theory on the background~\cite{Garner:2024tis}. This confirms an important underlying assumption of the celestial holography program: that a 2d chiral algebra should encode S-matrix quantities on self-dual backgrounds. It would then be very interesting to see whether an extension of such a chiral algebra bootstrap can reproduce the MHV amplitude presented in this paper.

It would also be useful to further explore the extent to which `standard' amplitude structures from Minkowski space persist or are modified on the SDTN background. For example, on self-dual radiative spaces, there is a background-coupled version of the KLT kernel which relates the graviton scattering amplitudes to the `double copy' of gluon amplitudes on a self-dual radiative Yang-Mills background~\cite{Adamo:2024hme}. While the interpretation of the curved background kernel is not clear (e.g., its inverse does not have an obvious interpretation), it would still be enlightening to know if a similar KLT-like kernel exists relating graviton scattering on SDTN to gluon scattering on the self-dual dyon. Finally, while we considered holomorphic collinear limits in this paper, it would also be interesting to explore other IR features of amplitudes on SDTN, such as soft theorems or factorization, which have only been studied on plane wave backgrounds thus far~\cite{Ilderton:2012qe,Ilderton:2020rgk,Klisch:2025mxn}.


\acknowledgments
	We thank Alfredo Guevara, Simon Heuveline, Uri Kol, Martin Ro\v{c}ek, Chiara Toldo, David Skinner and Iustin Surubaru for helpful conversations. TA is supported by a Royal Society University Research Fellowship URF\textbackslash R\textbackslash241024, the Simons Collaboration on Celestial Holography CH-00001550-11, the ERC Consolidator/UKRI Frontier grant TwistorQFT EP/Z000157/1 and the STFC consolidated grant ST/X000494/1. GB is supported by a joint Clarendon Fund, Mathematical Institute and Merton College Mathematics Scholarship. LJM would like to thank the Simons Collaboration on Celestial Holography CH-00001550-11 and STFC for financial support from grant number ST/T000864/1. AS is supported by a Black Hole Initiative fellowship, funded by the Gordon and Betty Moore Foundation and the John Templeton Foundation. 


\appendix

\section{Alternative descriptions of the SDTN twistor space}\label{subsec:other_twistor_spaces}

The SDTN metric is one of the oldest explicit examples of the non-linear graviton construction, and as such it has been treated many times in the twistor literature. While all descriptions must be equivalent up to diffeomorphisms on the SDTN manifold or, equivalently, gauge transformations of the complex structure on $\CPT$, they are not all equally easy to work with from a computational point of view. In this appendix, we review alternative descriptions of the SDTN twistor space which have appeared in the literature and relate them to the description given in Section~\ref{sec:twistor_space} and used throughout the paper.

\medskip

\paragraph{Hitchin's constrained twistor space:} This description, due to Hitchin~\cite{Hitchin:1979rts} (see also Chapter 13 of~\cite{Besse:1987pua}), makes use of the fact that the SDTN metric \eqref{eq:metric} is a complete hyperk\"ahler metric on $\R^4\cong\C^2$ of Gibbons-Hawking form. Since such a metric is specified by a solution of the Bogomolny monopole equations on $\R^3$, it follows that the associated twistor space $\CPT$ must be a bundle over the \emph{minitwistor space} of $\R^3$. This minitwistor space, $\MT$, is the space of oriented geodesics in $\R^3$, which can be parametrized as the total space of $\cO(2)\to\P^1$~\cite{Hitchin:1982gh}. The twistor description used throughout the text clearly fits into this framework, with $\eta=\mu^+\,\mu^-$ acting as a coordinate on the fibres of $\MT$.

Hitchin's construction instead takes a different coordinate on the $\cO(2)$ fibres:
\be\label{Hitchcoord1}
Q:=T^{\alpha}{}_{\dot\alpha}\,\mu^{\dot\alpha}\,\lambda_{\alpha}\,.
\ee
One can then consider a line bundle $L\to\MT$ with partial connection
 \begin{equation}
     \bar{D}_L=\dbar_{\MT}+\frac{Q}{2M}\,\bar e^0\,,
 \end{equation}
 and the corresponding patching function on $U_1\cap U_0$ is easily seen to be
 \begin{equation}
     \phi_{10}=\exp\left(\frac{Q}{2M\,\lambda_0\,\lambda_1}\right)\,,
 \end{equation}
 in agreement with \eqref{eq:transition_function}. The twistor space is the sub-bundle of $(L\oplus L^{-1})\otimes\mathcal{O}(1)$ defined by
 \begin{equation}
     \mu^+\,\mu^-=Q\,,\label{eq:constraint}
 \end{equation}
 where $\mu^\pm$ are elements of the fibres of $L^{\pm1}\otimes\mathcal{O}(1)$, respectively. The Euclidean structure in this presentation of twistor space corresponds to the anti-holomorphic involution
 \begin{equation}
     (\mu^+,\,\mu^-,\,Q,\,\lambda_\alpha)\mapsto\left(\overline{\mu^-},\,-\overline{\mu^+},\,-\overline{Q},\,\hat\lambda_\alpha\right)\,.
 \end{equation}
This is a global, albeit constrained thanks to \eqref{eq:constraint}, description of the twistor space.

The fibration $\CPT\to\P^1$ is found by composing $\CPT\to\MT$ with $\MT\to\P^1$. To write a holomorphic symplectic form on the fibers of $\CPT\to\P^1$, one introduces local trivializations $\rho^+,\rho^-$ of $L,L^{-1}$ on the patch $U_0$, and $\tilde{\rho}^+,\tilde{\rho}^-$ on the patch $U_1$:
\be
\begin{split}
    \mu^+ &= \begin{cases}
    \rho^+\,\exp\!\left(\frac{-Q\,\hat{\lambda}_0}{2M\,\lambda_0\,\la\lambda\,\hat{\lambda}\ra}\right)\qquad\lambda_0\neq0\\
    \tilde{\rho}^+\,\exp\!\left(\frac{-Q\,\hat{\lambda}_1}{2M\,\lambda_1\,\la\lambda\,\hat{\lambda}\ra}\right)\qquad\lambda_1\neq0\end{cases}\,,\\
    \mu^- &= \begin{cases}
    \rho^-\,\exp\!\left(\frac{Q\,\hat{\lambda}_0}{2M\,\lambda_0\,\la\lambda\,\hat{\lambda}\ra}\right)\qquad\lambda_0\neq0\\
    \tilde{\rho}^-\,\exp\!\left(\frac{Q\,\hat{\lambda}_1}{2M\,\lambda_1\,\la\lambda\,\hat{\lambda}\ra}\right)\qquad\lambda_1\neq0\end{cases}\,,
\end{split}
\ee
where $\rho^{\pm},\tilde{\rho}^{\pm}$ are $\cO(1)$-valued. These coordinates are related by the transition functions
\be\label{hittransf1}
\tilde{\rho}^+=\rho^+\,\exp\!\left(\frac{Q}{2M\,\lambda_0\,\lambda_1}\right)\,, \qquad \tilde{\rho}^-=\rho^-\,\exp\!\left(\frac{-Q}{2M\,\lambda_0\,\lambda_1}\right)\,,
\ee
on the overlap $U_0\cap U_1$.

The holomorphic symplectic form on $U_0$ is given by
\be
\Sigma = \frac12\,\d\log\left(\frac{\rho^+}{\rho^-}\right)\wedge\d Q = \frac{1}{2}\,\frac{\rho^-\,\d \rho^+-\rho^+\,\d \rho^-}{\rho^+\,\rho^-}\wedge\d Q\,,
\ee
which appears to be non-singular only when $\rho^+\,\rho^-\neq 0$. However, upon using the constraint $Q=\mu^+\,\mu^-=\rho^+\,\rho^-$, it follows that
\be
\Sigma = \d \rho^+\wedge\d \rho^-\,,
\ee
which extends across $\rho^+\,\rho^-=0$. Using the transition functions \eqref{hittransf1} along with the relation $\rho^+\,\rho^-=Q$, it is easily checked that this takes the same form on the other patch:
\be
\Sigma=\d\rho^{+}\wedge\d\rho^{-}=\d\tilde{\rho}^{+}\wedge\d\tilde{\rho}^{-}\:\mod \d\lambda_{\alpha}\,.
\ee
Thus, $\Sigma$ is global on the fibres of $\CPT\to\P^1$ and $\rho^{\pm},\tilde{\rho}^{\pm}$ both provide Darboux coordinates on their respective domains of definition. To reconstruct the metric, one follows the usual method~\cite{Gindikin:1986} of solving for the holomorphic rational curves in twistor space, then pulling $\Sigma$ back to $\CM\times\P^1$ and reading off a basis of ASD 2-forms.

\medskip

\paragraph{Sparling's non-linear graviton:} In fact, the earliest description of the SDTN metric was by Sparling~\cite{Sparling:1976} (see also \S3.4 of~\cite{Hughston:1979tq}), in terms of a \v{C}ech description of twistor space built on a four-set open cover. This is closely related to Hitchin's construction. For $i=0,1$, let $\rho_i^\pm$ be local coordinates for $\rho^\pm$, defined by setting
\begin{align}
\rho^+&=\begin{cases}
    \rho^+_0\exp\left(-\dfrac{Q\hat\lambda_0}{2M\la\lambda\,\hat\lambda\ra\lambda_0}\right)\,,\qquad\lambda_\alpha\in U_0\\
    \rho^+_1\exp\left(-\dfrac{Q\hat\lambda_1}{2M\la\lambda\,\hat\lambda\ra\lambda_1}\right)\,,\qquad \lambda_\alpha\in U_1
\end{cases}\,,\\
\rho^-&=\begin{cases} \rho^-_0\exp\left(\dfrac{Q\hat\lambda_0}{2M\la\lambda\,\hat\lambda\ra\lambda_0}\right)\,,\qquad\lambda_\alpha\in U_0\\
    \rho^-_1\exp\left(\dfrac{Q\hat\lambda_1}{2M\la\lambda\,\hat\lambda\ra\lambda_1}\right)\,,\qquad \lambda_\alpha\in U_1
\end{cases}\,.
\end{align}
Now define the four sets
\begin{equation}
    U_{0\pm}=\{\lambda_0\neq 0,\rho_0^\pm\neq 0\}\,,\qquad U_{1\pm}=\{\lambda_1\neq 0,\rho_1^\pm\neq 0\}\,,
\end{equation}
so that $U_{i+}\cup U_{i-}=U_i$ for $i=0,1$. On each open set, one of the $\rho^{\pm}_{i}$ coordinates can be eliminated using the constraint $\rho_i^+\,\rho_i^{-}=Q$. For instance, local holomorphic coordinates on $U_{0+}$ are given by $(\rho_0^+,Q,\lambda_\alpha)$ with $\rho_0^-=Q/\rho_0^+$ fixed by the constraint. 

Now, defined the local coordinates
\begin{align}
\left.\mu^{\dot\alpha}\right|_{U_{0+}}&=2T^{1\dot\alpha}\,\frac{Q}{\lambda_0}-4M\,T^{\alpha\dot\alpha}\,\lambda_\alpha\,\log\frac{\rho^+_0}{\lambda_0}\,, \label{spartrans1}\\
    \left.\mu^{\dot\alpha}\right|_{U_{0-}}&=2T^{1\dot\alpha}\,\frac{Q}{\lambda_0}+4M\,T^{\alpha\dot\alpha}\,\lambda_\alpha\,\log\frac{\rho^-_0}{\lambda_0}\,,\\
    \left.\mu^{\dot\alpha}\right|_{U_{1+}}&=-2T^{0\dot\alpha}\,\frac{Q}{\lambda_1}-4M\,T^{\alpha\dot\alpha}\,\lambda_\alpha\,\log\frac{\rho^+_1}{\lambda_1}\,,\\
    \left.\mu^{\dot\alpha}\right|_{U_{1-}}&=-2T^{0\dot\alpha}\,\frac{Q}{\lambda_1}+4M\,T^{\alpha\dot\alpha}\,\lambda_\alpha\,\log\frac{\rho_1^-}{\lambda_1}\,. \label{spartrans4}
\end{align}
It can be check that $\mu^{\dot\alpha}T^\alpha{}_{\dot\alpha}\lambda_\alpha=Q$ holds globally. Moreover, the six transition functions are
\begin{align}
    U_{0+}\cap U_{0-}\qquad&\colon\qquad\left.\mu^{\dot\alpha}\right|_{U_{0+}}-\left.\mu^{\dot\alpha}\right|_{U_{0-}}=-4MT^{\alpha\dot\alpha}\lambda_\alpha\log\frac{Q}{\lambda_0^2}\,,\\
    U_{0+}\cap U_{1+}\qquad&\colon\qquad\left.\mu^{\dot\alpha}\right|_{U_{0+}}-\left.\mu^{\dot\alpha}\right|_{U_{1+}}=-4MT^{\alpha\dot\alpha}\lambda_\alpha\log\frac{\lambda_1}{\lambda_0}\,,\\
    U_{0+}\cap U_{1-}\qquad&\colon\qquad\left.\mu^{\dot\alpha}\right|_{U_{0+}}-\left.\mu^{\dot\alpha}\right|_{U_{1-}}=-4MT^{\alpha\dot\alpha}\lambda_\alpha\log\frac{Q}{\lambda_0\lambda_1}\,,\\
    U_{0-}\cap U_{1+}\qquad&\colon\qquad\left.\mu^{\dot\alpha}\right|_{U_{0-}}-\left.\mu^{\dot\alpha}\right|_{U_{1+}}=-4MT^{\alpha\dot\alpha}\lambda_\alpha\log\frac{\lambda_0\lambda_1}{Q}\,,\\
    U_{0-}\cap U_{1-}\qquad&\colon\qquad\left.\mu^{\dot\alpha}\right|_{U_{0-}}-\left.\mu^{\dot\alpha}\right|_{U_{1-}}=-4MT^{\alpha\dot\alpha}\lambda_\alpha\log\frac{\lambda_0}{\lambda_1}\,,\\
    U_{1+}\cap U_{1-}\qquad&\colon\qquad\left.\mu^{\dot\alpha}\right|_{U_{1+}}-\left.\mu^{\dot\alpha}\right|_{U_{1-}}=-4MT^{\alpha\dot\alpha}\lambda_\alpha\log\frac{Q}{\lambda_1^2}\,.
\end{align}
These are precisely the patching functions prescribed by Sparling~\cite{Sparling:1976}. This construction should be viewed as a gravitational analogue of the `twistor quadrille,' which gives the twistor description of a self-dual dyon~\cite{Sparling:1979}.

It is worth mentioning that graviton wavefunctions in this presentation of the SDTN twistor space are simply the same as for flat space; for instance
\be\label{flatrep}
\Phi_{\mathrm{flat}}=\int_{\C^*}\d s\,s\,\bar{\delta}^{2}(s\,\lambda-\kappa)\,\e^{\im\,s\,[\mu\,\tilde{\kappa}]}\,.
\ee
The quasi-momentum eigenstate representatives \eqref{eq:scalar_twistor_rep} for the presentation used throughout the text are then obtained by applying the coordinate transformations \eqref{spartrans1} -- \eqref{spartrans4} into \eqref{flatrep}. In fact, this is how we obtained the twistor representatives for the quasi-momentum eigenstates in the first instance!


\section{Twistor theory for general Gibbons-Hawking metrics}\label{app:GHmets}

In this appendix, we demonstrate how the presentation of SDTN twistor space first introduced by~\cite{Galperin:1985de} and used throughout the paper can easily be generalised to give any Gibbons-Hawking metric~\cite{Hawking:1976jb,Gibbons:1978tef,Gibbons:1979xm}. In previous work, we gave a twistor construction for \emph{radiative} Gibbons-Hawking metrics, which can be defined in the complexified setting or for split signature~\cite{Adamo:2022mev}; the construction here will be valid for Euclidean signature and hence captures the non-radiative, gravitational instanton Gibbons-Hawking metrics.

Consider a complex deformation of $\PT$ of the form $\nbar=\dbar+\{\sh,\cdot\}$ for $\cO(2)$-valued Hamiltonian
\begin{equation}
    \sh=f(\eta,\lambda)\,\bar e^0\,,\label{eq:gibbons-hawking_hamiltonian}
\end{equation}
where $f$ is any function that has total homogeneity 4 in $(\mu^{\dot\alpha},\lambda_\alpha)$ but depends on $\mu^{\dot\alpha}$ only through $\eta$. Holomorphic rational curves $(\mu^{\dot\alpha}=\sF^{\dot\alpha}(x,\lambda),\lambda_{\alpha})$ in the complex structure of \eqref{eq:gibbons-hawking_hamiltonian} are defined by
\begin{equation}
    \left.\dbar\right|_{X}\sF^\pm=\pm\sF^\pm\,\dot f\,\bar e^0\,, \qquad \dot{f}:=\frac{\partial f}{\partial \eta}\,.
\end{equation}
These equations are solved by
\begin{equation}
\sF^\pm(x,\lambda)=\la\chi_\pm\,\lambda\ra\,\exp\!\left(\mp\frac{2\pi\im}{p}\,\tau\pm \sg(x,\lambda)\right)\,,
\end{equation}
where $\tau\sim\tau+p$ is a coordinate on a circle of radius $p/2\pi$ and 
\begin{equation}
\sg(x,\lambda):=\int_{\P^1}\d^2\lambda'\,\frac{\la\iota\,\lambda\ra}{\la\iota\,\lambda'\ra\la\lambda\,\lambda'\ra}\left.\dot f\right|_{X}\,,\qquad \d^2\lambda\equiv\frac{\la \lambda\,\d\lambda\ra\wedge\bar e^0}{2\pi\im}\,.
\end{equation}
In this integral formula, $\dot f|_{X}\equiv \dot f(\la\chi_+\,\lambda'\ra\la\chi_-\,\lambda'\ra,\lambda')$ is the restriction of $\dot f$ to the holomorphic rational curve $X$, where $\eta=-\im \vec{x}^{\alpha\beta}\lambda_\alpha\lambda_\beta$. The constant spinor $\iota^\alpha$ is arbitrarily chosen, and is needed since $\sh$ fixes $\sg$ only up to an element of $H^{0}(\P^1,\mathcal{O})$.

Let $\CM$ denote the real, 4-dimensional moduli space of these holomorphic curves (whose existence is locally guaranteed by theorems of Kodaira~\cite{Kodaira:1962,Kodaira:1963}). The pullback of the weighted symplectic form $\Sigma$ to $\CM\times\P^1$ given by the holomorphic curves is
\begin{equation}
p^*\Sigma=-2\lambda_\alpha\,\lambda_\beta\left(\d \chi_+^\alpha\wedge\d \chi_-^\beta-\frac{2\pi}{p}\,\d \tau\wedge\d \vec{x}^{\alpha\beta}-\im\,\d_{x}\sg\wedge\d \vec{x}^{\alpha\beta}\right) \mod\d\lambda_\alpha\,,
\end{equation}
where $\d_x\sg$ is the differential of $\sg$ with $\lambda$ held constant. At this stage, it is not immediately possible to identify a triplet of ASD 2-forms on $\CM$, as $\d_{x}\sg$ depends on $\lambda_{\alpha}$.

To proceed, write
\begin{equation}
    \lambda_\alpha\,\lambda_\beta\,\d \vec{x}^{\alpha\beta}\wedge\d_x\sg=\lambda_\alpha\,\lambda_\beta\,\vartheta^{\alpha\beta}(x)\,,
\end{equation}
where $\vartheta^{\alpha\beta}$ is the 2-form on $\CM$ defined by
 \begin{equation}
    \vartheta^{\alpha\beta}(x):=-\im\,\iota^{(\alpha}\d \vec{x}^{\beta)\gamma}\wedge\d \vec{x}^\delta{}_\gamma\int_{X}\d^2\lambda\,\frac{\lambda_\delta}{\langle\iota\,\lambda\rangle}\,\left.\ddot f\right|_{X}\,.
\end{equation}
The 2-form $\vartheta^{\alpha\beta}$ can be decomposed further as
\begin{equation}
    \vartheta^{\alpha\beta}=\im\, \tilde V\,\d \vec{x}^{\alpha\gamma}\wedge\d \vec{x}^\beta{}_\gamma+\im\, \d \vec{x}^{\alpha\beta}\wedge\tilde \sa\,,
\end{equation}
for the scalar $\tilde{V}$ and 1-form $\tilde{\sa}$
\begin{equation}
    \tilde V(x)=-\frac{1}{2}\int_{X}\d^2\lambda\,\left.\ddot f\right|_{X}\,,\qquad \tilde{\sa}=\d\vec{x}^{\alpha\beta}\int_{X}\d^2\lambda\,\frac{\iota_\alpha\lambda_\beta}{\langle\iota\,\lambda\rangle}\,\left.\ddot f\right|_{X}\,.
\end{equation}
In terms of these data, we can now write
\begin{equation}
    p^*\Sigma=\frac{2\pi}{p}\,\lambda_\alpha\,\lambda_\beta\,\Sigma^{\alpha\beta}(x)\quad\mod\d\lambda_\alpha\,,
\end{equation}
for
\begin{equation}
    \Sigma^{\alpha\beta}=2\left[\d \tau-\frac{p}{2\pi}\left(\tilde{\sa}-\frac{\sa}{2}\right)\right]\wedge\d \vec{x}^{\alpha\beta}+\frac{p}{\pi}\left(\tilde V+\frac{1}{2r}\right)\d \vec{x}^{\alpha\gamma}\wedge\d \vec{x}^\beta{}_\gamma\,.
\end{equation}
a triplet of ASD 2-forms on $\CM$.

Identifying a tetrad from $\Sigma^{\alpha\beta}=e^{\alpha\dot\alpha}\wedge e^{\beta}{}_{\dot\alpha}$, one sees that these are the ASD 2-forms associated to the Gibbons-Hawking metric
\begin{equation}
    \d s^2=\frac{(\d \tau+A)^2}{V}+V\,\d\vec x^2\,,
\end{equation}
with
\begin{equation}
    V=\frac{p}{2\pi}\left(\tilde V+\frac{1}{2r}\right)\,,\qquad A=\frac{p}{2\pi}\left(\frac{\sa}{2}-\tilde{\sa}\right)\,.
\end{equation}
It can be checked immediately that the pair $(V,A)$ satisfies the Bogomolny equation $\d V=\star_3\d A$ thanks to the identity
\begin{equation}
    \star_3\,\d \vec{x}^{\alpha\beta}=\d \vec{x}^{\alpha\gamma}\wedge\d \vec{x}^\beta{}_{\gamma}\,,
\end{equation}
so the metric is vacuum and self-dual (i.e., hyperk\"ahler) as required.

Note that the entire construction can be run in reverse: given a Gibbons-Hawking metric defined by $(V,A)$, we can reconstruct the associated $f(\eta,\lambda)$ and hence the complex structure on twistor space. The SDTN case studied throughout the paper is obtained as the special case where $p=8\pi M$ and $f=\eta^2/(4M)$, for which $\tilde V=1/(4M)$ is a constant and $\tilde{\sa}$ is pure gauge, hence it can be reabsorbed and set to zero by a shift in $\tau$. 


\section{2-point amplitude from the K\"ahler scalar}\label{app:2-point}

In previous work~\cite{Adamo:2023fbj}, we obtained the two-point amplitude for negative helicity gravitons scattering on SDTN by directly evaluating the quadratic part of the Plebanski action \eqref{MHVgen1} on quasi-momentum eigenstates:
\be\label{2pt0}
\begin{split}
\int_{\CM}\Sigma^{\alpha\beta}\wedge\gamma^{-}_{1\,\alpha}{}^{\gamma}\wedge\gamma^{+}_{2\,\beta\gamma}&\propto-\delta(\omega_1+\omega_2)\,M\,(4M\omega_2)!\,\frac{\omega_2^{4M\omega_2-2}\,\la1\,2\ra^{4+4M\omega_2}}{|\vec{k}_1+\vec{k}_2|^{8M\omega_2+2}} \\
&\propto\delta(\omega_1+\omega_2)\,M\,(4M\omega_2)!\,\frac{\omega_2^{4M\omega_2-2}\,\la1\,2\ra^{3}}{[1\,2]^{4M\omega_2+1}}\,,
\end{split}
\ee
ignoring overall numerical factors (which can be viewed as an overall normalisation of the generating functional).

Now, the MHV generating functional \eqref{MHVgen3} used in this paper is written in terms of the K\"ahler scalar -- or first Plebanski scalar -- of SDTN, and differs from the Plebanski action by two integrations-by-parts. In principle, this could lead to different two-point amplitudes due to contributions from boundary terms. While such boundary terms should vanish in this setting due to the assumptions of asymptotic flatness, it is hardly obvious that the object
\be\label{2pt1}
\la1\,2\ra^4\int_{\CM}\d^{4}x\,\sqrt{|g|}\,\Omega_{\mathrm{SDTN}}\,\phi_1^{-}\,\phi_2^{+}\,,
\ee
will give the same formula as \eqref{2pt0} for the 2-point amplitude.

Here, we show explicitly that \eqref{2pt1} does indeed match the previous result \eqref{2pt0} for the 2-point amplitude around the SDTN metric \cite{Adamo:2023fbj}. Using the boundary conditions \eqref{ybcond}, it can be shown that the on-shell twistor sigma model action for $h=0$ reduces to
\begin{equation}
    \Omega_{\mathrm{SDTN}}=8M\la\chi_-\,\chi_+\ra+2\,\frac{\la\chi_-\,1\ra^2\la\chi_+\,2\ra^2+\la\chi_+\,1\ra^2\la\chi_-\,2\ra^2}{\la1\,2\ra^2}\,.\label{eq:kahler_scalar_from_sigma_model}
\end{equation}
It can be verified that this object satisfies the first heavenly equation, and upon interpreting the quantities $\chi_-^\alpha\,\e^{\im t/4M}$ as holomorphic coordinates on $\C^2$, this matches the K\"ahler scalar \eqref{eq:kahler_potential} when the momenta of the two negative-helicity gravitons are antipodal. Equations \eqref{MHVgen3} -- \eqref{eq:kahler_scalar_from_sigma_model} lead to the following expression for the tree-level amplitude of two negative-helicity, minimal gravitons
\begin{multline}
    \mathcal{M}_2=4\pi\,\la 1\,2\ra^4\,\delta(\omega_1+\omega_2)\int\d^3\vec x\,\e^{\im(\vec k_1+\vec k_2)\cdot\vec x}\left(1+\frac{2M}{r}\right)\\\times\left(4Mr+\frac{\la \chi_-\,1\ra^2\la\chi_+\,2\ra^2+\la\chi_+\,1\ra^2\la\chi_-\,2\ra^2}{\la1\,2\ra^2}\right)\la\chi_-\,1\ra^{-4M\omega_1}\,\la\chi_+\,2\ra^{4M\omega_2}\,.
\end{multline}
Setting $\omega_2=\omega=-\omega_1>0$ and introducing the standard stereographic coordinates, we find
\begin{multline}
    \mathcal{M}_2=4\pi\,\la 1\,2\ra^4\,\delta(\omega_1+\omega_2)\int\d^3\vec x\,\e^{\im(\vec k_1+\vec k_2)\cdot\vec x}\left(1+\frac{2M}{r}\right)\left[\frac{r(\zeta-z_1)(\bar\zeta z_2+1)}{1+\zeta\bar\zeta}\right]^{4M\omega}\\\times\left(4Mr+r^2\frac{(\zeta-z_1)^2(\bar \zeta z_2+1)^2+(\zeta-z_2)^2(\bar\zeta z_1+1)^2}{\la1\,2\ra^2(1+\zeta\bar\zeta)^2}\right)\,.
\end{multline}
We can decompose this integral as
\begin{equation}
    \mathcal{M}_2=16\pi M\la 1\,2\ra^4\delta(\omega_1+\omega_2)[\mathcal{I}_0+\mathcal{I}_1+\mathcal{J}_0(z_1,z_2)+\mathcal{J}_1(z_1,z_2)+\mathcal{J}_0(z_2,z_1)+\mathcal{J}_1(z_2,z_1)]
\end{equation}
in terms of the basis integrals
\begingroup
\allowdisplaybreaks
\begin{align}
    \mathcal{I}_0&=2M\int\d^3\vec x\,\e^{\im\vec k\cdot\vec x}\left[\frac{r(\zeta-z_1)(\bar\zeta z_2+1)}{1+\zeta\bar\zeta}\right]^{4M\omega}\,,\\
    \mathcal{I}_1&=\int\d^3\vec x\,\e^{\im\vec k\cdot\vec x}\left[\frac{r(\zeta-z_1)(\bar\zeta z_2+1)}{1+\zeta\bar\zeta}\right]^{4M\omega}r\,,\\
    \mathcal{J}_0(a,b)&=\frac{1}{2\la1\,2\ra^2}\int\d^3\vec x\,\e^{\im\vec k\cdot\vec x}\left[\frac{r(\zeta-z_1)(\bar\zeta z_2+1)}{1+\zeta\bar\zeta}\right]^{4M\omega}\frac{r(\zeta-a)^2(\bar\zeta b+1)^2}{(1+\zeta\bar\zeta)^2}\,,\\
    \mathcal{J}_1(a,b)&=\frac{1}{4M\la1\,2\ra^2}\int\d^3\vec x\,\e^{\im\vec k\cdot\vec x}\left[\frac{r(\zeta-z_1)(\bar\zeta z_2+1)}{1+\zeta\bar\zeta}\right]^{4M\omega}\frac{r^2(\zeta-a)^2(\bar\zeta b+1)^2}{(1+\zeta\bar\zeta)^2}\,.
\end{align}
\endgroup
Each of these integrals can be evaluated following the procedure described in the Appendix of~\cite{Adamo:2023fbj}, leading to
\begin{equation}
\mathcal{M}_2=-4\pi\im\, M\,(4M\omega)!(-\omega)^{4M\omega-2}\,\frac{\la 1\,2\ra^3}{[1\,2]^{4M\omega+1}}\,\delta(\omega_1+\omega_2)\,.
\end{equation}
Up to an overall prefactor, this is exactly the amplitude \eqref{2pt0} found previously in~\cite{Adamo:2023fbj}.
    
    \bibliographystyle{JHEP}
	\bibliography{sdc}
\end{document}